\documentclass[a4paper,10pt]{elsarticle}
\usepackage{jcomp}
\usepackage{framed,multirow}
\usepackage[utf8]{inputenc}
\usepackage{bm}
\usepackage{amsmath}
\usepackage{mathrsfs}
\usepackage{graphicx}
\usepackage{epsfig, setspace}
\usepackage{url}
\usepackage{graphicx, transparent, color}
\usepackage{algorithm} 
\usepackage{algorithmicx}
\usepackage{etex}
\usepackage[bookmarks=true]{hyperref}
\usepackage{amsmath}
\usepackage{xcolor, soul}
\usepackage{rotating} % Rotating the figures but keeps the page in portrait mode
\usepackage{pdflscape} % Rotates single page into Landscape mode
\usepackage{silence}
\usepackage{ulem,cancel}
\newtheorem{remark}{\bf Remark}[section]
\newtheorem{example}{Example}[section]
\usepackage{graphicx}
\usepackage{float}
\usepackage{subfigure}% subcaptions for subfigures
\usepackage{subfigmat}% matrices of similar subfigures,

\usepackage{tikz}
\usepackage{capt-of}
\usepackage{setspace,lipsum}
\usepackage{lineno}
\usepackage{setspace,lipsum}
\usepackage{listings}
\usepackage{xcolor}
\definecolor{aquamarine}{rgb}{0.5, 1.0, 0.83}
\definecolor{OliveGreen}{rgb}{0,0.6,0}

\sethlcolor{aquamarine}

\definecolor{codegreen}{rgb}{0,0.6,0}
\definecolor{codegray}{rgb}{0.5,0.5,0.5}
\definecolor{codepurple}{rgb}{0.58,0,0.82}
\definecolor{backcolour}{rgb}{0.95,0.95,0.92}

\lstdefinestyle{mystyle}{
    backgroundcolor=\color{backcolour},   
    commentstyle=\color{codegreen},
    keywordstyle=\color{magenta},
    numberstyle=\tiny\color{codegray},
    stringstyle=\color{codepurple},
    basicstyle=\ttfamily\footnotesize,
    breakatwhitespace=false,         
    breaklines=true,                 
    captionpos=b,                    
    keepspaces=true,                 
    numbers=left,                    
    numbersep=5pt,                  
    showspaces=false,                
    showstringspaces=false,
    showtabs=false,                  
    tabsize=2
}
\begin{document}

\begin{frontmatter}
\title{A generalized adaptive central-upwind scheme for compressible flow simulations and preventing spurious vortices}

\author[AA_address]{Amareshwara Sainadh Chamarthi \cortext[cor1]{Corresponding author. \\ 
E-mail address: sainath@caltech.edu (Amareshwara Sainadh  Ch.).}}
\address[AA_address]{Division of Engineering and Applied Science, California Institute of Technology, Pasadena, CA, USA}

\begin{abstract}
This work introduces a novel adaptive central-upwind scheme designed for simulating compressible flows with discontinuities in the flow field. The proposed approach offers significant improvements in computational efficiency over the central gradient-based reconstruction approach presented in \cite{hoffmann2024centralized} (Hoffmann, Chamarthi and Frankel,  JCP 2024). By leveraging a combination of conservative and characteristic variable reconstruction, the proposed approach demonstrates oscillation-free results while effectively reducing computational costs and improving the results. Furthermore, the adaptive central-upwind algorithm is generalized to be compatible not only with the gradient-based reconstruction as in \cite{hoffmann2024centralized} but with other existing methods. In this regard, with the proposed algorithm, the standard fifth/sixth-order reconstruction scheme has also been shown to outperform existing schemes with a 20-30$\%$ reduction in computational expense with improved results. Notably, the proposed approach has successfully prevented the generation of spurious vortices in the double shear layer test cases, even with linear schemes, showcasing its robustness and effectiveness.
\end{abstract}


\end{frontmatter}
\section{Introduction}\label{sec:intro}

Adaptive central-upwind schemes are required to simulate compressible flows when both turbulence and discontinuities, like shock waves or contact discontinuities, are present. The challenge in these simulations is to capture discontinuities (where upwind schemes with dissipative properties are required) while maintaining low dissipation (an attribute of central schemes) in turbulent flow regions, which is a difficult balance to achieve. This is particularly crucial in large eddy simulations (LES) on coarse meshes, where resolving a wide range of turbulence scales with minimal dissipation is essential. Adaptive central-upwind schemes address this by combining the strengths of upwind and central schemes, allowing them to adapt to the complex demands of such simulations \cite{Pirozzoli2011}.

The adaptive central-upwind schemes are minimally dissipative, leading to sharp shocks and well-resolved broadband turbulence, but rely on an appropriate shock sensor. The shock or discontinuity sensor is the main challenge for the hybrid central/upwind method, as mentioned in \cite{johnsen2010assessment}. Various researchers have proposed different approaches to address this challenge. For example, Kim and Kwon suggested a hybrid scheme that combines a central scheme with the numerical dissipation of the upwind weighted essentially non-oscillatory (WENO) scheme using a weighing function \cite{kim2005high}. Hu, Wang, and Adams proposed the adaptive central-upwind WENO scheme, known as WENO-CU6, which leverages a low dissipation central scheme in smooth regions while maintaining strong shock-capturing capabilities through the WENO approach \cite{Hu2010}. Subsequently, Fu et al. developed the Targeted ENO family of schemes, which further improves upon the capabilities of WENO-CU6 \cite{Fu2016}. Other researchers, such as Liu et al. \cite{Liu2015}, Wong and Lele \cite{Wong2017}, and Subramaniam et al. \cite{subramaniam2019high}, have also made contributions by developing hybrid weighted interpolation schemes and improved versions of WENO-CU6 with localized dissipative interpolation. Subramaniam et al. \cite{subramaniam2019high} proposed an explicit-compact interpolation along with compact finite differences that provide higher resolution and more localized dissipation compared to that of Wong and Lele \cite{Wong2017}. Furthermore, Chamarthi and Frankel \cite{chamarthi2021high} introduced an adaptive central-upwind scheme, which employs a sixth-order linear-compact scheme and the fifth-order Monotonicity Preserving scheme (MP) using the Boundary Variation Diminishing (BVD) algorithm. Hoffmann, Chamarthi, and Frankel \cite{hoffmann2024centralized} have proposed a wave-appropriate central-upwind scheme by taking advantage of the wave structure of the Euler equations \cite{chamarthi2023wave}. The proposed scheme successfully simulates the hypersonic transitional flows on coarse grids. While it outperformed existing methods for benchmark and hypersonic flow test cases, the wave-appropriate centralization approach is still expensive as it requires the reconstruction of the cell-interface values to be carried out for characteristic variables. It is well known that transforming conservative variables to characteristic variables, while necessary for shock-capturing, is expensive and is not required for regions with no discontinuities. On a different note, Feng et al. have employed optimization techniques to develop an adaptive central-upwind scheme that balances dissipation and shock-capturing \cite{feng2022multi}. Their approach involves using multi-objective Bayesian optimization to construct a WENO-based subgrid-scale model for implicit Large Eddy Simulations of compressible flows. Feng et al. further improved the approach by using deep reinforcement learning \cite{feng2023deep}  and Bayesian optimization \cite{feng2024general} techniques. In their papers, Feng et al. simulated the double periodic shear layer test case and observed that the proposed optimized schemes are free of spurious vortices.

 The objective of the current paper is to propose a generalized and efficient adaptive central-upwind scheme based on the wave-appropriate scheme proposed by Hoffmann, Chamarthi, and Frankel \cite{hoffmann2024centralized}. The proposed algorithm aims to reduce computational expenses by reconstructing conservative variables in regions without discontinuities and to prevent spurious vortices in the double periodic shear layer test case.

The rest of the manuscript is organized as follows: In Section \ref{sec:governingEquations}, the governing equations are presented. Section \ref{sec:num} presents the numerical methods, including the novel adaptive central-upwind approach with conservative and characteristic variable reconstruction. Section \ref{sec:results} consists of the numerical results with the proposed algorithm and the conclusions are presented in Section \ref{sec:conclusions}.

\section{Governing Equations} \label{sec:governingEquations}

In this study, the three-dimensional compressible Navier-Stokes equations are solved in Cartesian coordinates:
\begin{equation}\label{eqn:cns}
\frac{\partial \mathbf{U}}{\partial t}+\frac{\partial \mathbf{F^c}}{\partial x}+\frac{\partial \mathbf{G^c}}{\partial y}+\frac{\partial \mathbf{H^c}}{\partial z}=\frac{\partial \mathbf{F^v}}{\partial x}+\frac{\partial \mathbf{G^v}}{\partial y}+\frac{\partial \mathbf{H^v}}{\partial z},
\end{equation}
where  $\textbf{U}$ is the conservative variable vector, $\mathbf{F^c}$, $\mathbf{G^c}$,  $\mathbf{H^c}$ and $\mathbf{F^v}$, $\mathbf{G^v}$, $\mathbf{H^v}$, are the convective (superscript $c$) and viscous (superscript $v$) flux vectors in each coordinate direction, respectively.  The conservative variable, convective, and viscous flux vectors are given as:

\begin{subequations}
    \centering
    \begin{gather}
        \mathbf{U} = \begin{pmatrix}
        \rho \\
        \rho u \\
        \rho v \\
        \rho w \\
        \rho E
        \end{pmatrix},
        \quad
        \mathbf{F}^c = \begin{pmatrix}
        \rho u \\
        \rho u u + p \\
        \rho u v + p \\
        \rho u w + p \\
        \rho u H 
        \end{pmatrix},
        \quad
        \mathbf{G}^c = \begin{pmatrix}
        \rho v \\
        \rho v u + p \\
        \rho v v + p \\
        \rho v w + p \\
        \rho v H 
        \end{pmatrix},
        \quad
        \mathbf{H}^c = \begin{pmatrix}
        \rho w \\
        \rho w u + p \\
        \rho w v + p \\
        \rho w w + p \\
        \rho w H 
        \end{pmatrix},
        \tag{\theequation a--\theequation d}
    \end{gather}
\end{subequations}

\begin{equation}
\begin{array}{l}\label{eqn-visc}
\mathbf{F^v}=\left[0, \tau_{x x}, \tau_{x y}, \tau_{x z}, u \tau_{x x}+v \tau_{x y}+w \tau_{x z}-q_{x}\right]^{T}, \\
\mathbf{G^v}=\left[0, \tau_{x y}, \tau_{y y}, \tau_{y z}, u \tau_{y x}+v \tau_{y y}+w \tau_{y z}-q_{y}\right]^{T}, \\
\mathbf{H^v}=\left[0, \tau_{x z}, \tau_{y z}, \tau_{z z}, u \tau_{z x}+v \tau_{z y}+w \tau_{z z}-q_{z}\right]^{T},
\end{array}
\end{equation}
\noindent where $\rho$ is density, $u$, $v$, and $w$ are the velocities in the $x$, $y$, and $z$ directions, respectively, $p$ is the pressure, $E = e + \left(u^2 + v^2 + w^2 \right)/2$ is the specific total energy, and $H = E + p/\rho$ is the specific total enthalpy. The equation of state is for a calorically perfect gas so that $e = p/ \left[ \rho (\gamma-1) \right]^{-1}$ is the internal energy, where $\gamma$ is the ratio of specific heats. The components of the viscous stress tensor $\tau$ and the heat flux $q$ are defined in tensor notations as:
\begin{equation}\label{eqn:5-stress}
\tau_{i j}=\frac{\mu}{\operatorname{Re}}\left(\frac{\partial u_{i}}{\partial x_{j}}+\frac{\partial u_{j}}{\partial x_{i}}-\frac{2}{3} \frac{\partial u_{k}}{\partial x_{k}} \delta_{i j}\right),
\end{equation}
\begin{equation}\label{eqn:6-heat}
\begin{aligned}
\mathrm{q}_{i}=-\frac{\mu}{\operatorname{Re Pr Ma}(\gamma-1)} \frac{\partial T}{\partial x_{i}}, \quad T= \text{Ma}^{2} \gamma \frac{p}{\rho},
\end{aligned}
\end{equation}
where $\mu$ is the dynamic viscosity, $T$ is the temperature, $Ma$ and $Re$ are the Mach number and Reynolds number, and Pr is the Prandtl number.

\section{Numerical methods}\label{sec:num}

Using a conservative numerical method, the governing equations cast in semi-discrete form for a Cartesian cell $I_{i,j,k} = \left[ x_{i-\frac{1}{2}}, x_{i+\frac{1}{2}} \right] \times \left[ y_{i-\frac{1}{2}}, y_{i+\frac{1}{2}} \right] \times \left[ z_{i-\frac{1}{2}}, z_{i+\frac{1}{2}} \right]$ can be expressed via the following ordinary differential equation: 

\begin{align}
    \begin{aligned}
        \frac{\text{d}}{\text{d} t} \check{\mathbf{U}}_{i,j,k} = \mathbf{Res}_{i,j,k} = &- \left. \frac{\text{d} \check{\mathbf{F}}^c}{\text{d} x} \right|_{i,j,k} - \left. \frac{\text{d} \check{\mathbf{G}}^c}{\text{d} y} \right|_{i,j,k} - \left. \frac{\text{d} \check{\mathbf{H}}^c}{\text{d} z} \right|_{i,j,k} \\ 
        &+ \left. \frac{\text{d} \check{\mathbf{F}}^v}{\text{d} x} \right|_{i,j,k} + \left. \frac{\text{d} \check{\mathbf{G}}^v}{\text{d} y} \right|_{i,j,k} + \left. \frac{\text{d} \check{\mathbf{H}}^v}{\text{d} z} \right|_{i,j,k},
    \end{aligned}
\end{align}

\noindent where the check accent, $\check{(\cdot)}$, indicates a numerical approximation of physical quantities, $\mathbf{Res}_{i,j,k}$ is the residual function, and the remaining terms are cell centre numerical flux derivatives of the physical fluxes in Equation (\ref{eqn:cns}). For brevity, we continue with only the $x$-direction as it is straightforward to extend to all three dimensions in a dimension-by-dimension manner. The indices  $j$ and $k$ are also dropped for simplicity.

\subsection{Viscous Flux Spatial Discretization Scheme}\label{sec:visc}

Viscous fluxes, $\check{\mathbf{F}^v}$, are computed using the fourth-order $\alpha$-damping scheme of Chamarthi \cite{chamarthi2023gradient}, which is based on the $\alpha$-damping approach of Nishikawa \cite{Nishikawa2010}. In the one-dimensional scenario, the cell center numerical viscous flux derivative is:

\begin{equation}
    \left. \frac{\mathrm{d} \mathbf{F}^v}{\mathrm{d} x} \right|_{i} = \frac{1}{\Delta x} \left( \mathbf{F}^{v}_{i+\frac{1}{2}} - \mathbf{F}^{v}_{i-\frac{1}{2}} \right).
\end{equation}

\noindent The cell interface numerical viscous flux is computed as,

\begin{equation}
    \mathbf{F}^v_{i+\frac{1}{2}} = 
    \begin{pmatrix}
        0 \\
        -\tau_{i+\frac{1}{2}} \\
        -\tau_{i+\frac{1}{2}} u_{i+\frac{1}{2}} + q_{i+\frac{1}{2}}, \\
    \end{pmatrix},         \tau_{i+\frac{1}{2}} = \frac{4}{3} {\mu}_{i+\frac{1}{2}} \left. \frac{\partial u}{\partial x} \right|_{i+\frac{1}{2}},
        \quad
        q_{i+\frac{1}{2}} = -{\kappa}_{i+\frac{1}{2}} \left. \frac{\partial T}{\partial x} \right|_{i+\frac{1}{2}}.
\end{equation}

\noindent For an arbitrary variable, $\phi$, the $\alpha$-damping approach computes cell interface gradients as:

\begin{equation}
    \left. \frac{\partial \phi}{\partial x} \right|_{i+ \frac{1}{2}} = \frac{1}{2} \left( \left. \frac{\partial \phi}{\partial x} \right|_{i} + \left. \frac{\partial \phi}{\partial x} \right|_{i+1} \right) + \frac{\alpha}{2 \Delta x} \left( \phi_R - \phi_L \right),         \phi_L = \phi_i + \left. \frac{\partial \phi}{\partial x} \right|_{i} \frac{\Delta x}{2}, \phi_R = \phi_{i+1} - \left. \frac{\partial \phi}{\partial x} \right|_{i+1} \frac{\Delta x}{2},
\end{equation}

\noindent where, in this work, $\alpha = 4$. The gradients at cell centers are computed using the fourth order compact finite-difference scheme, as in \cite{chamarthi2023gradient}, which is as follows:
\begin{equation}
    \frac{5}{14} \left. \frac{\partial \phi}{\partial x} \right|_{i-1} + \left. \frac{\partial \phi}{\partial x} \right|_{i} + \frac{5}{14} \left. \frac{\partial \phi}{\partial x} \right|_{i+1} = \frac{1}{28 \Delta x} \left( \phi_{i+2} - \phi_{i-2} \right) + \frac{11}{14 \Delta x} \left( \phi_{i+1} - \phi_{i-1} \right).
    \label{eqn:firstDerivative}
\end{equation}

\subsection{Convective Flux Spatial Discretization Scheme}
Similar to the viscous flux discretization, the cell centre numerical convective flux derivative is expressed as:
\begin{equation}
    \left. \frac{\text{d} \check{\mathbf{F}}^c}{\text{d} x} \right|_{i} = \frac{1}{\Delta x} \left( \check{\mathbf{F}}^{c}_{i+\frac{1}{2}} - \check{\mathbf{F}}^{c}_{i-\frac{1}{2}} \right),
\end{equation}

\noindent where $i \pm \frac{1}{2}$ indicates right and left cell interface values, respectively. $\check{\mathbf{F}}^c_{i \pm \frac{1}{2}}$ are computed using an approximate Riemann solver. This work uses the componentwise local Lax-Friedrichs (cLLF) \cite{fleischmann2019numerical} approximate Riemann solver unless otherwise explicitly stated. The numerical fluxes at cell boundaries computed using a Riemann solver can be expressed in the following standard form:

\begin{equation}
    \check{\mathbf{F}}^c_{i \pm \frac{1}{2}} = \frac{1}{2} \left[ \check{\mathbf{F}}^c \left( {\mathbf{U}}^{L}_{i \pm \frac{1}{2}} \right) + {\mathbf{F}}^c \left( {\mathbf{U}}^{R}_{i \pm \frac{1}{2}} \right) \right] - \frac{1}{2} \left| \mathbf{A}_{i \pm \frac{1}{2}} \right| \left( {\mathbf{U}}^{R}_{i \pm \frac{1}{2}} - {\mathbf{U}}^{L}_{i \pm \frac{1}{2}} \right),
\end{equation}

\noindent where the $L$ and $R$ superscripts denote the left- and right-biased states, respectively, and $\left| \mathbf{A}_{i \pm \frac{1}{2}} \right|$ denotes the convective flux Jacobian. The objective is to obtain the left- and right-biased states, $\mathbf{U}^{L}_{i \pm \frac{1}{2}}$ and $\mathbf{U}^{R}_{i \pm \frac{1}{2}}$. The procedure to obtain these interface values is elucidated in the following sections.

\subsubsection{Linear Schemes:}

In this subsection, we provide the details of the calculations of candidate polynomials that can be used to approximate the values of $\mathbf{U}^{L}_{i \pm \frac{1}{2}}$ and $\mathbf{U}^{R}_{i \pm \frac{1}{2}}$. While the adaptive central-upwind approach to be presented later is a general approach that might potentially be used with most of the schemes, we consider the standard fifth-order upwind scheme \cite{Hu2010, Jiang1995,suresh1997accurate} and the corresponding sixth-order scheme, and the Gradient-Based Reconstruction (GBR) schemes proposed in \cite{chamarthi2023gradient,chamarthi2023efficient}.

\textbf{Fifth-order upwind scheme:} The well-known fifth-order upwind schemes for obtaining the values of the left and right interfaces are as follows:
\begin{equation}
\begin{aligned}
& \phi_{i+1 / 2}^{L5, Linear}=\frac{1}{30} \phi_{i-2}-\frac{13}{60} \phi_{i-1}+\frac{47}{60} \phi_{i+0}+\frac{9}{20} \phi_{i+1}-\frac{1}{20} \phi_{i+2}, \\
& \phi_{i+1 / 2}^{R5, Linear}=\frac{1}{30} \phi_{i+3}-\frac{13}{60} \phi_{i+2}+\frac{47}{60} \phi_{i+1}+\frac{9}{20} \phi_{i+0}-\frac{1}{20} \phi_{i-1},
\end{aligned}
\label{eqn:fifth-linear}
\end{equation}
where $\phi$ is an arbitrary variable, either conservative ($\mathbf{U}$) or characteristic ($\mathbf{C}$) variables are used in the present paper. The superscripts $L5$ and $R5$ represent left and right-biased fifth-order formulas, and the $Linear$ superscript represents the scheme is linear. Linear schemes cannot be used for flows with discontinuities as they will lead to oscillations. The details of the non-linear scheme will be presented later.

\textbf{Upwind-biased GBR scheme:} The GBR schemes differ slightly from the fifth-order mentioned above scheme. As such, the interface values depend not only on the cell centre values but also on the gradients at the cell centres. The advantage of having gradients is that they can be used in the parts of the subroutine, such as viscous flux discretization, shock capturing, post-processing quantities, and so on \cite{chamarthi2023wave,chamarthi2023gradient}. GBR method employs the first two moments of the Legendre polynomial evaluated on $x_{i-1 / 2} \leq x \leq x_{i+1 / 2}$ and written for a general variable, $\phi$, as:

\begin{equation}\label{eqn:legendre}
    \textcolor{black}{\phi(x) = \phi_{i} + \phi'_{i} \left(x-x_{i}\right) + \frac{3 \phi''_{i}}{2}  \kappa \left[\left(x-x_{i}\right)^{2}-\frac{\Delta x_i^{2}}{12}\right],}
\end{equation}

\noindent where $\phi'_{i}$ and $\phi''_{i}$ respectively represent the first and second derivatives of $\phi_{i}$. If $x = x_i \pm \Delta x/2$ and $\kappa = 1/3$, the following equations for the left- and right-biased states are obtained:
    
\begin{subequations}
    \begin{align}
\phi^{L,GBR, Linear}_{i+\frac{1}{2}} &= \phi_{i+0} + \frac{\Delta x}{2} \phi'_{i+0} + \frac{\Delta x^2}{12} \phi''_{i+0},\\
\quad
\phi^{R,GBR, Linear}_{i+\frac{1}{2}} &= \phi_{i+1} - \frac{\Delta x}{2} \phi'_{i+1} + \frac{\Delta x^2}{12} \phi''_{i+1}.
    \end{align}
    \label{eqn:legendreInterpolation}
\end{subequations}
Like the fifth-order upwind scheme, the superscripts $L, GBR$ and $R, GBR$ represent left and right-biased gradient-based reconstruction formulas. In this work, $\phi'_{i}$ was computed using eighth order explicit central differences \cite{hoffmann2024centralized}:
    
\begin{equation}
    \phi'_{i} = \frac{1}{\Delta x} \left( \frac{1}{280} \phi_{i-4} - \frac{4}{105} \phi_{i-3} + \frac{1}{5} \phi_{i-2} - \frac{4}{5} \phi_{i-1} + \frac{4}{5} \phi_{i+1} - \frac{1}{5} \phi_{i+2} +\frac{4}{105} \phi_{i+3} - \frac{1}{280} \phi_{i+4} \right).
    \label{eqn:firstDerivative-eighth}
\end{equation}

\noindent $\phi''_{i}$ was computed by the following formula \cite{chamarthi2023gradient}:

\begin{equation}
    \phi''_{i} = \frac{2}{\Delta x} \left( \phi_{i+1} - 2 \phi_{i} + \phi_{i-1} \right) - \frac{1}{2 \Delta x} \left( \phi'_{i+1} - \phi'_{i-1} \right).
    \label{eqn:secondDerivative}
\end{equation}

\textbf{Central schemes:} The abovementioned schemes are left and right-biased upwind schemes. They provide necessary dissipation in some regions but are unsuitable for turbulent simulations as they require a low dissipation central scheme. A central scheme can be obtained by averaging the left- and right-biased upwind reconstructions (they can also be derived in other possible ways as well).
\begin{subequations}\label{eqn:centralScheme}
    \begin{gather}
        \phi^{L}_{i+\frac{1}{2}} = \phi^{C}_{i+\frac{1}{2}} = \left( 1 - \eta \right) \phi^{L, Linear}_{i+\frac{1}{2}} + \eta  \phi^{R, Linear}_{i+\frac{1}{2}},
        \tag{\theequation a}
        \\[5pt]
        \phi^{R}_{i+\frac{1}{2}} = \phi^{C}_{i+\frac{1}{2}} = \eta  \phi^{L, Linear}_{i+\frac{1}{2}} + \left( 1 - \eta  \right) \phi^{R, Linear}_{i+\frac{1}{2}},
        \tag{\theequation b}
    \end{gather}
\end{subequations}
where $\eta = 0.5$ and $\left( \cdot \right)^C$ denotes the centralized reconstruction. The well-known sixth-order central scheme can be obtained by averaging the left and right-biased reconstruction formulas given by Equation (\ref{eqn:fifth-linear}), and is as follows:
\begin{equation}
\phi^{C6,Linear}_{i+\frac{1}{2}}=\frac{1}{2}\left(\phi_{i+1 / 2}^{L5, Linear} + \phi_{i+1 / 2}^{R5, Linear}\right)=\frac{1}{60}\left(\phi_{i-2}-8 \phi_{i-1}+37 \phi_i+37 \phi_{i+1}-8 \phi_{i+2}+\phi_{i+3}\right).
\end{equation}
Similarly, the centralized GBR scheme can be obtained by averaging the Equations (\ref{eqn:legendreInterpolation}), and is as follows:

\begin{equation}
\phi^{C,GBR,Linear}_{i+\frac{1}{2}}=\frac{1}{2}\left(\phi_{i+1 / 2}^{L,GBR, Linear} + \phi_{i+1 / 2}^{R,GBR, Linear}\right).
\end{equation}

\subsubsection{Monotonicity Preserving approach}

The linear reconstruction schemes represented by Equations (\ref{eqn:fifth-linear}) and (\ref{eqn:legendreInterpolation}) are susceptible to oscillations when there are discontinuities present. To address this issue, we utilized MP limiting. The following details the MP limiting procedure specifically for the left-biased state, although the procedure is identical for the right-biased state. The limiting procedure is typically carried out for the characteristic variables, $\mathbf{C}$, for oscillation-free results, as explained in \cite{suresh1997accurate,van2006upwind}. The transformation from conservative variables to characteristic variables will be explained later. The MP limiting criterion for the GBR scheme \cite{chamarthi2023gradient} is:

\begin{equation}
    \mathbf{C}^{L}_{i+\frac{1}{2}} = 
    \begin{cases}
        \mathbf{C}^{L,Linear}_{i+\frac{1}{2}} & \text{if } \left( \mathbf{C}^{L,Linear}_{i+\frac{1}{2}} - \mathbf{C}_i \right) \left( \mathbf{C}^{L,Linear}_{i+\frac{1}{2}} - \mathbf{C}^{L,MP}_{i+\frac{1}{2}} \right) \leq 10^{-40}, \\[5pt]
        \mathbf{C}^{L,Non-Linear}_{i+\frac{1}{2}} & \text{otherwise},
    \end{cases}
    \label{eqn:mpLimitingCriterion}
\end{equation}

\noindent where $\mathbf{C}^{L,Linear}_{i+\frac{1}{2}}$ corresponds to Eqn. \ref{eqn:legendreInterpolation} and the remaining terms are:

\begin{equation}
\begin{aligned}
\mathbf{C}^{L,Non-Linear}_{i+\frac{1}{2}} &= \mathbf{C}^{L,Linear}_{i+\frac{1}{2}} + \text{minmod} \left( \mathbf{C}^{L,MIN}_{i+\frac{1}{2}} - \mathbf{C}^{L,Linear}_{i+\frac{1}{2}}, \mathbf{C}^{L,MAX}_{i+\frac{1}{2}} - \mathbf{C}^{L,Linear}_{i+\frac{1}{2}} \right),\\
\mathbf{C}^{L,MP}_{i+\frac{1}{2}} &= \mathbf{C}^{L,Linear}_{i+\frac{1}{2}} + \text{minmod} \left[ \mathbf{C}_{i+1}-\mathbf{C}_{i}, \mathscr{A} \left( \mathbf{C}_{i}-\mathbf{C}_{i-1} \right) \right], \\
\mathbf{C}^{L,MIN}_{i+\frac{1}{2}} &= \max \left[ \min \left( \mathbf{C}_{i}, \mathbf{C}_{i+1}, \mathbf{C}^{L,MD}_{i+\frac{1}{2}} \right), \min \left( \mathbf{C}_{i}, \mathbf{C}^{L,UL}_{i+\frac{1}{2}}, \mathbf{C}^{L,LC}_{i+\frac{1}{2}} \right) \right],\\
\mathbf{C}^{L,MAX}_{i+\frac{1}{2}} &= \min \left[ \max \left( \mathbf{C}_{i}, \mathbf{C}_{i+1}, \mathbf{C}^{L,MD}_{i+\frac{1}{2}} \right), \max \left( \mathbf{C}_{i}, \mathbf{C}^{L,UL}_{i+\frac{1}{2}}, \mathbf{C}^{L,LC}_{i+\frac{1}{2}} \right) \right], \\
\mathbf{C}^{L,MD}_{i+\frac{1}{2}} &= \frac{1}{2} \left( \mathbf{C}_{i} + \mathbf{C}_{i+1} \right) - \frac{1}{2} d^{L,M}_{i+\frac{1}{2}}, \mathbf{C}^{L,UL}_{i+\frac{1}{2}} = \mathbf{C}_{i} + 4 \left( \mathbf{C}_{i} - \mathbf{C}_{i-1} \right), \\
\mathbf{C}^{L,LC}_{i+\frac{1}{2}} &= \frac{1}{2} \left( 3 \mathbf{C}_{i} - \mathbf{C}_{i-1} \right) + \frac{4}{3} d^{L,M}_{i-\frac{1}{2}}, \ \quad      d^{L,M}_{i+\frac{1}{2}} = \text{minmod} \left( 0.5(d_i + d_{i+1}),2d_i, 2d_{i+1} \right), \\
        d_i &= 2 \left( \mathbf{C}_{i+1} - 2\mathbf{C}_{i} + \mathbf{C}_{i-1} \right) - \frac{\Delta x}{2} \left( \mathbf{C}'_{i+1} - \mathbf{C}'_{i-1} \right),
\end{aligned}
\end{equation}

\noindent where $\mathscr{A} = 4$ and $\text{minmod} \left( a,b \right) = \frac{1}{2} \left[ \text{sgn}(a) + \text{sgn}(b) \right] \min \left( \left| a \right|, \left| b \right| \right)$. \textcolor{black}{The GBR method that employs explicit finite differences and the above MP limiter is called MEG (\textbf{M}onotonicity Preserving \textbf{E}xplicit \textbf{G}radient) as in \cite{chamarthi2023gradient}. As it uses the eighth-order explicit gradients, it is denoted as MEG8.} For the fifth-order upwind scheme, the computations of the $d_{i+1 / 2}^M$ and $d_i$ differ from that of the GBR scheme and are as follows \cite{suresh1997accurate}:

\begin{equation}\label{eqn:mp_improve}
    \begin{aligned}
    d_{i+1 / 2}^M & =\operatorname{minmod}\left(4 \mathbf{C}_i-\mathbf{C}_{i+1}, 4 \mathbf{C}_{i+1}-\mathbf{C}_i, \mathbf{C}_i, \mathbf{C}_{i+1}\right),\\
    d_i  &=\mathbf{C}_{i-1}-2 \mathbf{C}_i+\mathbf{C}_{i+1}.
    \end{aligned}
\end{equation}
The fifth-order upwind method with MP limiting is known as MP5, as in \cite{suresh1997accurate}, in this paper. Only the adaptive central-upwind scheme is considered in this paper and is denoted as MP6.

\subsubsection{Novel adaptive central-upwind and conservative-characteristic variable reconstruction}

This section presents the novel central-upwind scheme with adaptive conservative-characteristic variable reconstruction. Before proceeding to the novel algorithm, a brief background of the approach is presented. 

\begin{itemize}
	\item In Ref. \cite{chamarthi2023wave} authors took advantage of the wave structure of the Euler equations. Once the variables are transformed from physical to characteristic space, the characteristic variables have specific properties. The first and last variable is known as the acoustic waves, the second variable is the entropy wave, and the rest are known as shear waves. Together, the entropy and shear waves are known as linearly degenerate waves. The density varies across the entropy wave in characteristic space, and the rest of the variables remain unchanged. Ref. \cite{chamarthi2023wave} took advantage of this and reduced the frequent activation of the MP criterion, Equation (\ref{eqn:mpLimitingCriterion}). Shockwaves are detected by the Ducros sensor, which is used for waves (characteristic variables) other than entropy waves. However, all the waves are computed using the upwind, inherently dissipative schemes.
	\item The approach is improved in \cite{hoffmann2024centralized} that all the linearly degenerate waves are computed using a central scheme if the necessary discontinuity detection criteria are met. The acoustic waves are still computed using an upwind scheme for stability, and the proposed approach significantly improved benchmark test cases and was able to predict hypersonic transitional flows.
\end{itemize}

One disadvantage of Ref. \cite{chamarthi2023wave} and \cite{hoffmann2024centralized} is that the variables are transformed from physical (conservative) to characteristic variables, which is computationally expensive. It would be efficient to convert to characteristic variables only in regions where there are discontinuities, which are, strictly speaking, localized to some regions of the flow, and the rest of the regions are computed using conservative variables. While the Ducros sensor detects shockwaves, the contact discontinuities must be detected by a different approach and done in physical space. In physical space, across the contact discontinuity, only density varies, and the rest of the variables remain continuous; therefore, if a contact sensor coupled with the Ducros sensor can identify the discontinuities, then the regions without discontinuities can be computed using conservative variables thereby avoiding the variable transformation. The following steps describe the novel algorithm that avoids this expensive transformation procedure:

\begin{description}
\item[Step 1.] Compute the linear left and right-biased reconstructions of the variable density either by the fifth-order upwind schemes or the GBR schemes (depending on the linear scheme of choice the user has decided upon, either MEG8 and MP). 
\item [Step 2.] Evaluate the MP limiting criterion (Equation (\ref{eqn:mpLimitingCriterion}))  for the variable density ($\rho$ = $\mathbf{U}_{1}$) using the following equations
\begin{equation}\label{con:den}
    \begin{aligned}
L_{\rho}&=\left( \mathbf{U}^{L,Linear}_{i+\frac{1}{2},1} - \mathbf{U}_{i+0,1} \right) \left( \mathbf{U}^{L,Linear}_{i+\frac{1}{2},1} - \mathbf{U}^{L,MP}_{i+\frac{1}{2},1} \right),
 \\[10pt]
R_{\rho}&= \left( \mathbf{U}^{R,Linear}_{i+\frac{1}{2},1} - \mathbf{U}_{i+1,1} \right) \left( \mathbf{U}^{R,Linear}_{i+\frac{1}{2},1} - \mathbf{U}^{R,MP}_{i+\frac{1}{2},1} \right).
    \end{aligned}
\end{equation}

The above equations can be considered as contact discontinuity detectors in physical space. Also evaluate the Durcos sensor, ${\Omega_{d}}$, which is computed as follows: 
\begin{equation}
    \Omega_d = \frac{\left| -p_{i-2} + 16 p_{i-1} - 30 p_{i} + 16 p_{i+1} - p_{i+2} \right|}{\left| +p_{i-2} + 16 p_{i-1} + 30 p_{i} + 16 p_{i+1} + p_{i+2} \right|} \frac{ \left( \nabla \cdot \mathbf{u} \right)^2}{ \left( \nabla \cdot \mathbf{u} \right)^2 + \left| \nabla \times \mathbf{u} \right|^2},
    \label{eqn:ducros}
\end{equation}
\noindent where $\mathbf{u}$ is the velocity vector, and the derivatives of velocities are computed by the Equation (\ref{eqn:firstDerivative}) as in \cite{chamarthi2023gradient}.  We modify $\Omega_{d}$ by using its maximum value in a three-cell neighborhood:

\begin{equation}
    \Omega_d = \max \left( \Omega_{i+m} \right), \quad \text{for } m = -1,0,1.    
\end{equation}

\item[Step 3.] If both the Ducros sensor and the contact discontinuity detectors identify that there are no discontinuities in the region as per the following conditions:

\begin{equation}\label{con:check}
\text{if } \left({\Omega_{d}} > 0.01\right) \ \& \  \left(L_{\rho} \leq 10^{-40}\right)  \& \  \left(R_{\rho} \leq 10^{-40}\right).  \\
\end{equation}

One can reconstruct the conservative variables according to the following algorithm:

In all directions:

\begin{equation}\label{eqn:centralScheme_d}
    \mathbf{U}^{L,R}_{i+\frac{1}{2},b} = 
    \left\{
    \begin{array}{ll}
        \text{if } b = 1,5\text{:} & \mathbf{U}^{C,Linear}_{i+\frac{1}{2},b} \quad \text{i.e.}\ \rho \ \& \  \rho E \ (\eta =0.5, \text{in Eqn. \ref{eqn:centralScheme}})
    \end{array}
    \right.
\end{equation}

In $x$-direction:

\begin{equation}\label{eqn:centralScheme_x}
    \mathbf{U}^{L,R}_{i+\frac{1}{2},b} = 
    \left\{
    \begin{array}{ll}
        \text{if } b = 2\text{:} & \mathbf{U}^{L,R,Linear}_{i+\frac{1}{2},b} \quad \text{i.e.}\ \rho u \ (\eta =1, \text{in Eqn. \ref{eqn:centralScheme}})
        \\[15pt]
        \text{if } b = 3,4\text{:} & \mathbf{U}^{C,Linear}_{i+\frac{1}{2},b} \quad \text{i.e.}\ \rho v \ \& \  \rho w \ (\eta =0.5, \text{in Eqn. \ref{eqn:centralScheme}})
    \end{array}
    \right.
\end{equation}

In $y$-direction:

\begin{equation}\label{eqn:centralScheme_y}
    \mathbf{U}^{L,R}_{i+\frac{1}{2},b} = 
    \left\{
    \begin{array}{ll}
        \text{if } b = 3\text{:} & \mathbf{U}^{L,R,Linear}_{i+\frac{1}{2},b} \quad \text{i.e.}\ \rho v \ (\eta =1, \text{in Eqn. \ref{eqn:centralScheme}})
        \\[15pt]
        \text{if } b = 2,4\text{:} & \mathbf{U}^{C,Linear}_{i+\frac{1}{2},b} \quad \text{i.e.}\ \rho u \ \& \  \rho w \ (\eta =0.5, \text{in Eqn. \ref{eqn:centralScheme}})
    \end{array}
    \right.
\end{equation}

In $z$-direction:

\begin{equation}\label{eqn:centralScheme_z}
    \mathbf{U}^{L}_{i+\frac{1}{2},b} = 
    \left\{
    \begin{array}{ll}
        \text{if } b = 4\text{:} & \mathbf{U}^{L,R,Linear}_{i+\frac{1}{2},b} \quad \text{i.e.}\ \rho w \ (\eta =1, \text{in Eqn. \ref{eqn:centralScheme}})
        \\[15pt]
        \text{if } b = 2,3\text{:} & \mathbf{U}^{C,Linear}_{i+\frac{1}{2},b} \quad \text{i.e.}\ \rho u \ \& \  \rho v \ (\eta =0.5, \text{in Eqn. \ref{eqn:centralScheme}})
    \end{array}
    \right.
\end{equation}

Only linear schemes are required in the above procedure, as there are no discontinuities in the computational region.

\item[Step 4.] If the criterion in Equation (\ref{con:check}) is not met, then one can proceed with the characteristic variable reconstruction as in Ref. \cite{hoffmann2024centralized}. Compute Roe-averaged variables at the interface to construct the left, $\mathbf{L}_n$, and right, $\mathbf{R}_n$, eigenvectors of the normal convective flux Jacobian. \\
    
\item[Step 5.] For the GBR scheme transform $\mathbf{U}_{i}$, $\mathbf{U}'_{i}$, and $\mathbf{U}''_{i}$ to characteristic space by multiplying them by $\mathbf{L}_n$:

    \begin{subequations}
        \begin{gather}
            \mathbf{C}_{i+m,b} = \mathbf{L}_{n,i+\frac{1}{2}} \mathbf{U}_{i+m}, 
            \quad
            \mathbf{C}'_{i+m,b} = \mathbf{L}_{n,i+\frac{1}{2}} \mathbf{U}'_{i+m},
            \quad
            \mathbf{C}''_{i+m,b} = \mathbf{L}_{n,i+\frac{1}{2}} \mathbf{U}''_{i+m}.
            \tag{\theequation a--\theequation c}
        \end{gather}
    \end{subequations}
    
 For the fifth-order upwind scheme transform only $\mathbf{U}_{i}$ to characteristic space by multiplying them by $\mathbf{L}_n$
    
        \begin{subequations}
        \begin{gather}
            \mathbf{C}_{i+m,b} = \mathbf{L}_{n,i+\frac{1}{2}} \mathbf{U}_{i+m},
        \end{gather}
    \end{subequations}

    for $m = -2,-1,0,1,2,3$ and $b = 1,2,3,4,5$, representing the vector of characteristic variables which are defined as follows in the current implementation: 

    \begin{table}[H]
        \centering
        \caption{Characteristic wave types.}
        \begin{tabular}{c c c}
            \hline
            \hline
            $b = 1,5$ & $b = 2$ & $b = 3,4$ \\
            \hline
            Acoustic & Entropy/Contact & Shear/Vortical  \\
            \hline
            \hline
        \end{tabular}
        \label{tab:characteristicWaveStructure}
    \end{table}
        
 \item[Step 6.] Using Equations (\ref{eqn:legendreInterpolation}), obtain the unlimited reconstruction to cell interfaces in characteristic space via:

\begin{subequations}
    \begin{align}
        \mathbf{C}^{L, GBR, Linear}_{i+\frac{1}{2},b} &= \mathbf{C}_{i+0,b} + \frac{\Delta x}{2} \mathbf{C}'_{i+0,b} + \frac{\Delta x^2}{12} \mathbf{C}''_{i+0,b},\\
 \quad
        \mathbf{C}^{R, GBR, Linear}_{i+\frac{1}{2},b} &= \mathbf{C}_{i+1,b} - \frac{\Delta x}{2} \mathbf{C}'_{i+1,b} + \frac{\Delta x^2}{12} \mathbf{C}''_{i+1,b}.
    \end{align}
    \label{eqn:unlimitedCharacteristicInterpolation-gbr}
\end{subequations}

For the fifth-order reconstruction schemes Equations (\ref{eqn:fifth-linear}) are used for the interface values  and are as follows:
\begin{subequations}
    \begin{align}
 \mathbf{C}_{i+\frac{1}{2},b}^{L5, Linear}&=\frac{1}{30} \mathbf{C}_{i-2,b}-\frac{13}{60} \mathbf{C}_{i-1,b}+\frac{47}{60} \mathbf{C}_{i+0,b}+\frac{9}{20} \mathbf{C}_{i+1,b}-\frac{1}{20} \mathbf{C}_{i+2,b},\\
 \quad
\mathbf{C}_{i+\frac{1}{2},b}^{R5, Linear}&=\frac{1}{30} \mathbf{C}_{i+3,b}-\frac{13}{60} \mathbf{C}_{i+2,b}+\frac{47}{60} \mathbf{C}_{i+1,b}+\frac{9}{20} \mathbf{C}_{i+0,b}-\frac{1}{20} \mathbf{C}_{i-1,b}.
    \end{align}
    \label{eqn:unlimitedCharacteristicInterpolation-mp5}
\end{subequations}

\noindent The left-biased reconstruction is then treated by the following algorithm:

\begin{equation}
    \mathbf{C}^{L}_{i+\frac{1}{2},b} = 
    \left\{
    \begin{array}{ll}
        \text{if } b = 1,5\text{:} & \begin{cases}
            \mathbf{C}^{L,Non-Linear}_{i+\frac{1}{2},b} & \text{if } {\Omega_{d}} > 0.01, 
            \\[10pt]
            \mathbf{C}^{L,Linear}_{i+\frac{1}{2},b} & \text{otherwise},
        \end{cases} 
        \\[30pt]
        \text{if } b = 2\text{:} & \begin{cases}
            \mathbf{C}^{L,Non-Linear}_{i+\frac{1}{2},b} & \text{if } \left( \mathbf{C}^{L,Linear}_{i+\frac{1}{2}} - \mathbf{C}_i \right) \left( \mathbf{C}^{L,Linear}_{i+\frac{1}{2}} - \mathbf{C}^{L,MP}_{i+\frac{1}{2}} \right) \geq 10^{-40}, 
            \\[10pt]
            \mathbf{C}^{C,Linear}_{i+\frac{1}{2},b} & \text{otherwise},
        \end{cases} 
        \\[30pt]
        \text{if } b = 3,4\text{:} & \begin{cases}
            \mathbf{C}^{L,Non-Linear}_{i+\frac{1}{2},b} & \text{if } {\Omega_{d}} > 0.01, 
            \\[10pt]
            \mathbf{C}^{C,Linear}_{i+\frac{1}{2},b} & \text{otherwise}.
        \end{cases}
    \end{array}
    \right.
    \label{eqn:centralizedWaveSensorCriterion}
\end{equation}

A similar procedure is carried out for the right-biased reconstruction.

\item[Step 7.] After obtaining $\mathbf{C}^{L,R}_{i+\frac{1}{2},b}$, the variables are transformed back to physical fields:

\begin{equation}\label{eqn:characteristicToPhysical}
    \mathbf{U}^{L,R}_{i+\frac{1}{2}} = \mathbf{R}_{n,i+\frac{1}{2}} \mathbf{C}^{L,R}_{i+\frac{1}{2}}.
\end{equation}

\end{description}

The nonlinear scheme with the above procedure along with the GBR schemes (Equations (\ref{eqn:legendreInterpolation})) as linear schemes is denoted as MEG8-CC (CC for conservative-characteristic variables). Similarly, if fifth-order schemes (Equations (\ref{eqn:fifth-linear})) are used as linear schemes, then the nonlinear scheme is denoted as MP6-CC in this paper. The approach is generic, so even higher-order schemes like the seventh-order upwind schemes or implicit gradient schemes as in \cite{chamarthi2023gradient,chamarthi2023wave} can also be used. For simplicity, only the MEG8-CC and MP6-CC schemes are presented in this paper. Steps 4-7 are the same as that of the wave-appropriate centralization approach presented in \cite{hoffmann2024centralized}, and such scheme is denoted as MEG8-C as in \cite{hoffmann2024centralized}. The novel approach here is steps 1-3, where the conservative variables are reconstructed in a unique central-upwind approach if the shock and contact discontinuity sensors are satisfied.

  \begin{remark}
 \normalfont It is essential to note that a complete central scheme is not used even though the criteria in Equation (\ref{con:check}) are satisfied. In Equations (\ref{eqn:centralScheme_d}), (\ref{eqn:centralScheme_x}), (\ref{eqn:centralScheme_y}), and (\ref{eqn:centralScheme_z}) not all variables are \textit{centralized}. While $\rho$ and  $\rho E$ are centralized in all directions, the variables $\rho u$, $\rho v$ and $\rho w$ are upwind or centralized depending on the direction. The characteristic variable centralization explained in steps 4-7 and the conservative variable reconstruction in steps 1-3 above have \textit{some similar characteristics} and will be explained below. A complete central scheme would still give spurious vortices for the double shear layer test case, and the proposed algorithm would avoid spurious vortices.
 \end{remark}

 \noindent \textbf{Relation between centralization in characteristic and conservative variable space:}

In \cite{hoffmann2024centralized}, it was found that using a centralized reconstruction for all but the characteristic acoustic waves, i.e. all the linearly degenerate waves; entropy wave and the shear waves, was a robust and superior solution. The similarity between the central-upwind scheme in conservative variable space and the central-upwind scheme in characteristic space is explained below. In order to obtain charactertic variables, $\mathbf{C}$, the conservative variables,  $\mathbf{U}$, are multiplied by the left eigenvectors, $\mathbf{L}_n$. For simplicity we consider the two-dimensional scenario:
    
        \begin{equation}
            \mathbf{C}_{b} = \mathbf{L}_{n} \mathbf{U}_{b},
        \end{equation}

\begin{equation} \label{matrix}
        \left( \begin{array}{c}
            {\mathbf{C}}_{1} \\[5pt]
            {\mathbf{C}}_{2} \\[5pt]
            {\mathbf{C}}_{3} \\[5pt]
            {\mathbf{C}}_{4} \\[5pt]
        \end{array} \right) 
        = \left(\begin{array}{cccc}
\frac{1}{2}\left(\frac{\gamma-1}{2 c^2} q^2+\frac{q_n}{c}\right) & -\frac{1}{2}\left(\frac{\gamma-1}{c^2} u+\frac{n_x}{c}\right) & -\frac{1}{2}\left(\frac{\gamma-1}{c^2} v+\frac{n_y}{c}\right) & \frac{\gamma-1}{2 c^2} \\
\\[5pt]
1-\frac{\gamma-1}{2 c^2} q^2 & \frac{\gamma-1}{c^2} u & \frac{\gamma-1}{c^2} v & -\frac{\gamma-1}{c^2} \\
\\[5pt]
-q_{\ell} & \ell_x & \ell_y & 0\\
\\[5pt]
\frac{1}{2}\left(\frac{\gamma-1}{2 c^2} q^2-\frac{q_n}{c}\right) & -\frac{1}{2}\left(\frac{\gamma-1}{c^2} u-\frac{n_x}{c}\right) & -\frac{1}{2}\left(\frac{\gamma-1}{c^2} v-\frac{n_y}{c}\right) & \frac{\gamma-1}{2 c^2}
\end{array}\right)
        \left( \begin{array}{c}
            {\mathbf{U}}_{1} \\[5pt]
            {\mathbf{U}}_{2} \\[5pt]
            {\mathbf{U}}_{3} \\[5pt]
            {\mathbf{U}}_{4}
        \end{array} \right),
    \end{equation}
    where $\bm{n}$ = $[n_x \ n_y]^t$ and $[l_x \ l_y]^t$ is a tangent vector (perpendicular to $\bm{n}$) such as $[l_x \ l_y]^t$ = $[-n_y \ n_x]^t$. By taking $\bm{n}$ = $[1, 0]^t$ and $[0, 1]^t$ we obtain the corresponding eigenvectors in $x$ and $y$ directions (in the $x$-direction, $n_x = 1$, whereas $n_y=0$. The $y$-direction is analogous.). Furthermore,  $q^2 = u^2 + v^2 $, $q_l = u l_x + v l_y$, and $q_n = u n_x + v n_y$. In the matrix (\ref{matrix}), the $\mathbf{C}_1$ and $\mathbf{C}_4$ are acoustic waves, $\mathbf{C}_2$ is the entropy wave, and the $\mathbf{C}_3$ is the shear wave. The characteristic variable $\mathbf{C}_3$ in the characteristic space is as follows:
    \begin{equation}
    \mathbf{C}_3 = -q_l \mathbf{U}_{1} + l_x\mathbf{U}_{2} + l_y\mathbf{U}_{3}+0\mathbf{U}_{4}.
    \end{equation}
    
    In $x-$ direction, $[l_x \ l_y]^t$ = $[-n_y \ n_x]^t$= $[0 \ 1]$  which means $q_l = u l_x + v l_y$ = v. Therefore,
    
        \begin{equation}
    \mathbf{C}_3 = -v \mathbf{U}_{1} + 1\mathbf{U}_{3} = -\rho v + \rho v \approx 0 \ (\text{a small value in computations}).
    \end{equation}
    
        In $y-$ direction, $[l_x \ l_y]^t$ = $[-n_y \ n_x]^t$= $[-1 \ 0]$  which means $q_l = u l_x + v l_y$ = -u. Therefore,
    
        \begin{equation}
    \mathbf{C}_3 = u \mathbf{U}_{1} + (-1)\mathbf{U}_{2} = \rho u - \rho u \approx 0 \ (\text{a small value in computations}).
    \end{equation}
    
   The critical aspect is that in $x-$ direction $\rho v$ is being reconstructed in the characteristic space and centralized if the Ducros sensor criterion is satisfied. Likewise, in the $y-$ direction, $\rho u$ is reconstructed in the characteristic space and centralized if the Ducros sensor criterion is satisfied. Similarly, the variable $\rho v$ is centralized in conservative variable space in $x-$ direction (Equations (\ref{eqn:centralScheme_x})), and  $\rho u$ is centralized in conservative variable space in $y-$ direction (Equations (\ref{eqn:centralScheme_y})), respectively. The analysis straightforwardly extends to the three-dimensional scenario.
       
    Similarly, the characteristic variable $\mathbf{C}_2$ represents the contact discontinuity in characteristic space and across a contact discontinuity density changes in physical space. The corresponding variables are centralized in their respective computational spaces if there is no contact discontinuity in the computation region. It explains the relation between the centralization in characteristic and conservative variable space.
    
\section{Results and discussion}\label{sec:results}

This section tests the proposed numerical schemes for various benchmark test cases. For time integration, we use the explicit third-order TVD Runge-Kutta method \cite{Jiang1995}. Time integration is performed with a CFL = 0.4 for all the problems. The advantages of the proposed algorithm are compared with the TENO5 scheme \cite{Fu2016}. All the computations are carried out on the authors' Mac mini with an M1 processor.

\begin{example}{Double periodic Shear layer}\label{ex:dsl}
\end{example}
\textcolor{black}{In the first test case, we demonstrate that the proposed algorithm will prevent spurious vortices for the double periodic shear layer test case \cite{minion1997performance}. The test involves two initially parallel shear layers that develop into two significant vortices at $t = 1$. The non-dimensional parameters for this test case are presented in Table \ref{tab:shearLayerNondimensionalParameters}}.

\begin{figure}[H]
\centering\offinterlineskip
\subfigure[Reference, $\theta  = 120$, $800^2$]{\includegraphics[width=0.3\textwidth]{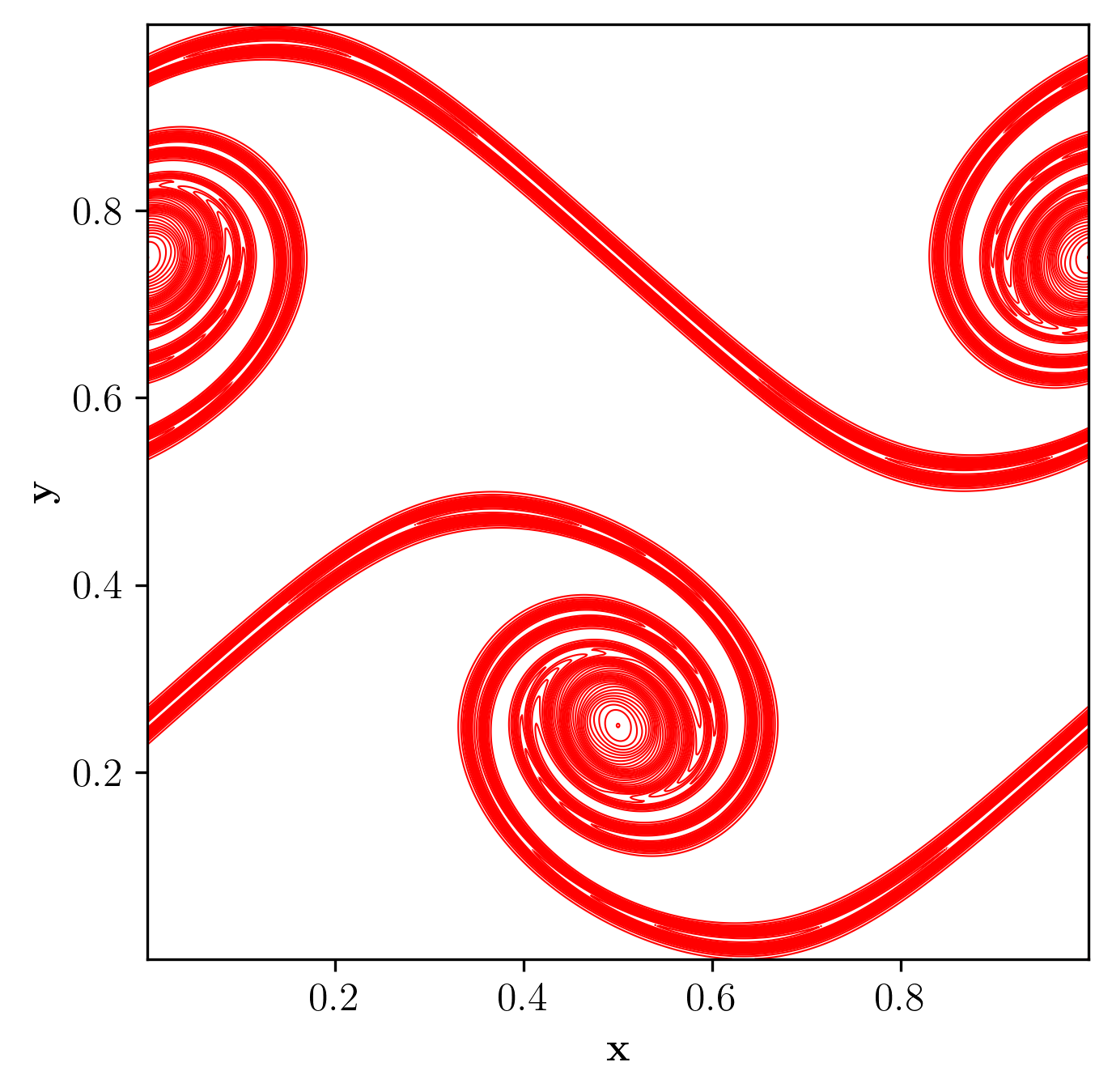}
\label{fig:fin-dsl_120}}
\subfigure[Reference, $\theta  = 80$, $512^2$]{\includegraphics[width=0.3\textwidth]{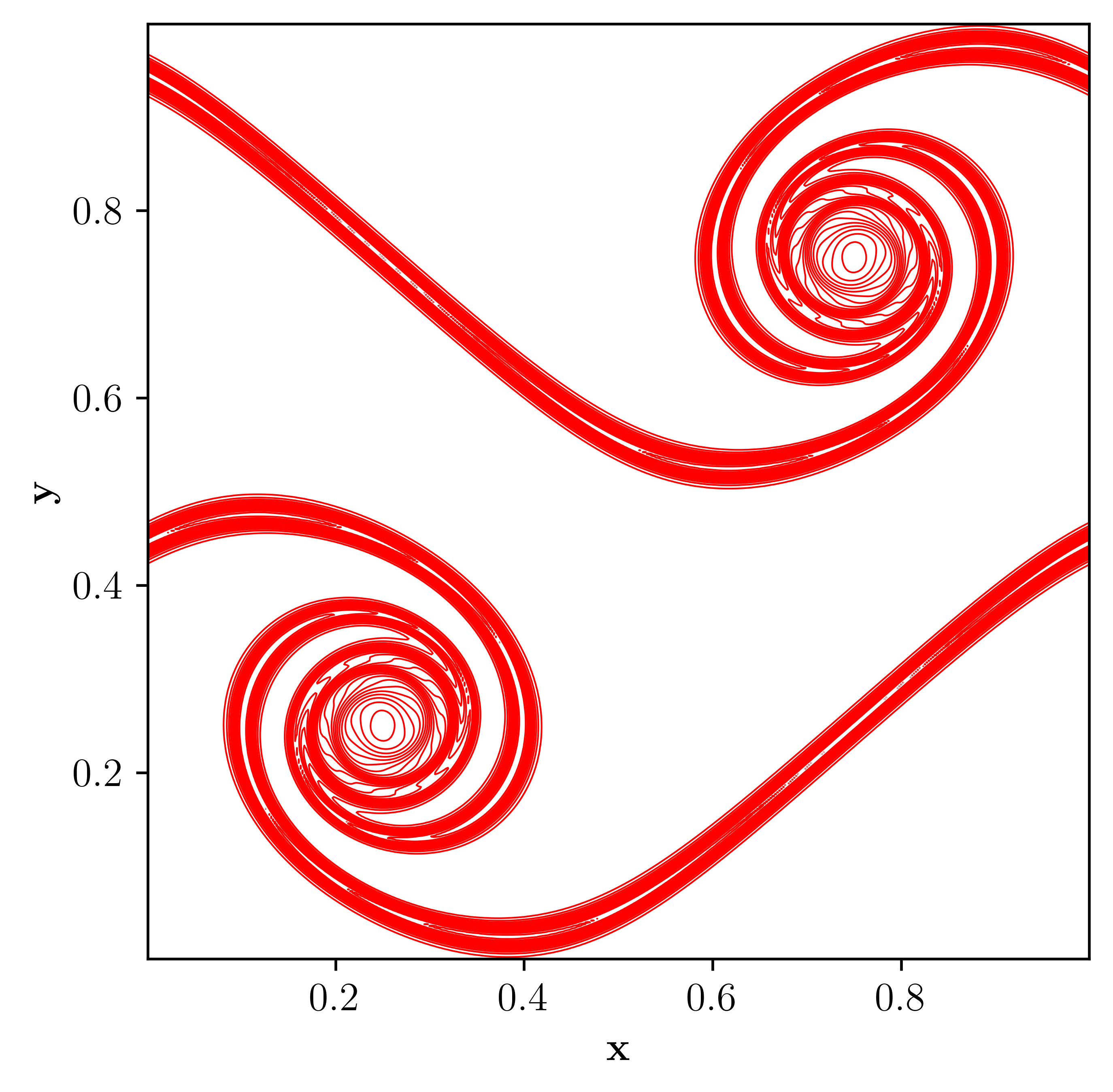}
\label{fig:fin-dsl_80}}
    \caption{$z$-vorticity contours of the test case Example \ref{ex:dsl} on fine grids.}
    \label{fig:dpsl_ref}
\end{figure}

\begin{table}[h!]
    \centering
    \caption{\textcolor{black}{Parameters of the periodic double shear layer test case.}}
    \begin{tabular}{c c c c}
        \hline
        \hline
        $\mathrm{Ma}$ & $\mathrm{Re}$ & Pr & $\gamma$ \\
        \hline
        0.1 & 10,000 & 0.73 & 1.4 \\
        \hline
        \hline
    \end{tabular}
    \label{tab:shearLayerNondimensionalParameters}
\end{table}
\noindent \textcolor{black}{The initial conditions were:}
\begin{subequations}
   \textcolor{black}{ \begin{align}
        p &= \frac{1}{\gamma \mathrm{Ma}^2}, \rho=1, \ u= 
        \begin{cases}
            \tanh \left[ \theta (y-0.25) \right], & \text{ if } (y \leq 0.5), \\
            \tanh \left[ \theta (0.75-y) \right], & \text{ if } (y > 0.5),
        \end{cases} 
        \\[10pt]
        v &= 0.05 \sin \left[ 2 \pi(x+0.25) \right] \text{for}\ \theta  = 80, \text{and}\ v = 0.05 \sin \left[ 2 \pi(x) \right] \text{for}\ \theta  = 120, 
    \end{align}}
\end{subequations}
The parameter $\theta$ defines the initial shear layer width. Two configurations are considered with $\theta$ =80 and $\theta$ =120, as shown in the above initial conditions. The reference solutions for $\theta$ =80 and $\theta$ =120, shown in Fig. \ref{fig:dpsl_ref}, are computed with MEG8-C scheme on a grid sizes of of $800 \times 800$  and $ 512 \times 512$, respectively. For this test case, if the grid is under-resolved, unphysical braid vortices and oscillations can occur on the shear layers according to the literature \cite{feng2022multi,feng2024general,minion1997performance,chamarthi2023role,clausen2013entropically,schranner2013physically}. Observing the well-resolved reference solution, there are no braid vortices or oscillations that form on the shear layers. 

\begin{figure}[H]
\centering\offinterlineskip
\subfigure[MEG8-CC]{\includegraphics[width=0.3\textwidth]{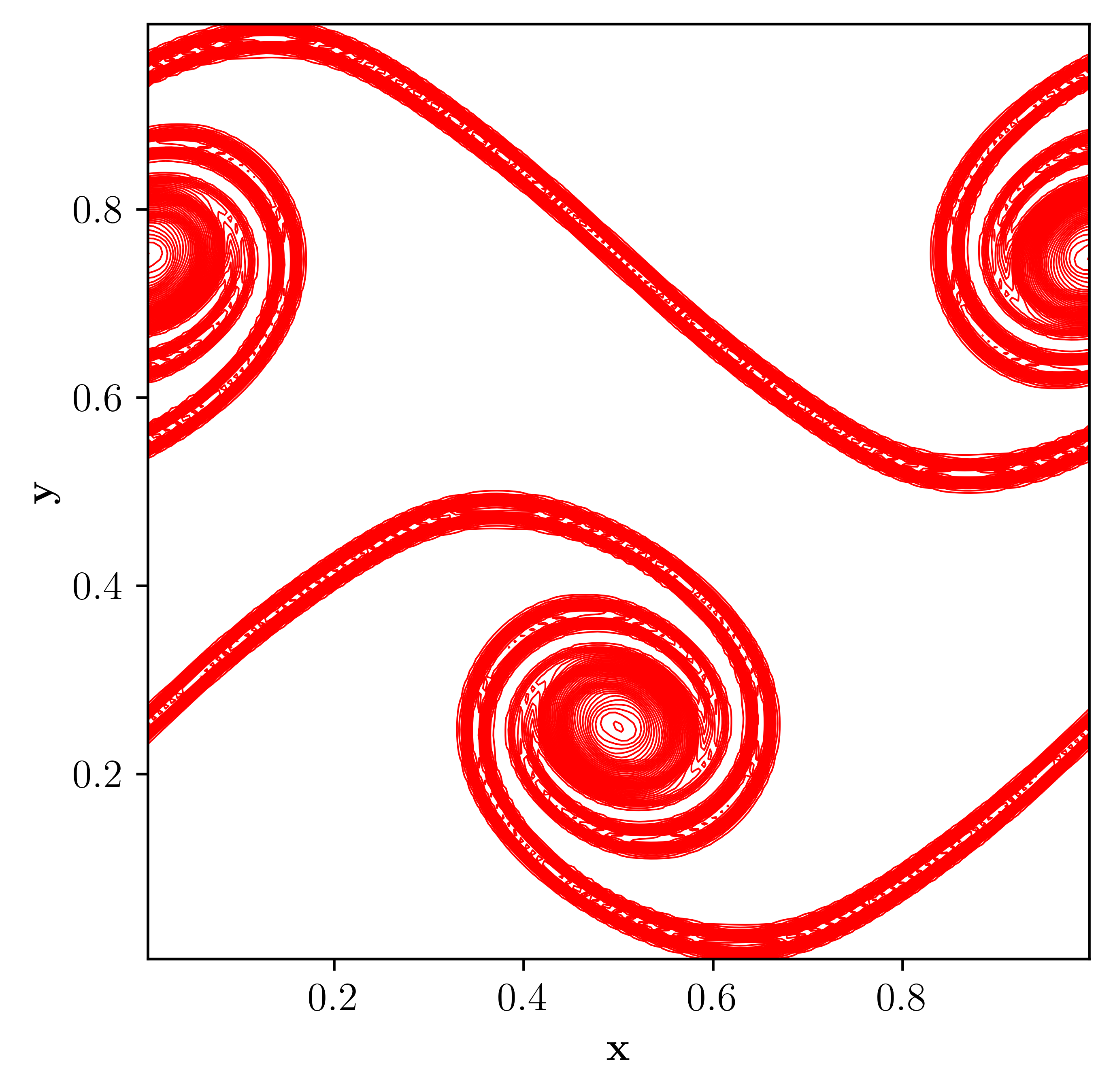}
\label{fig:mcc_dsl_120}}
\subfigure[MEG8-C]{\includegraphics[width=0.3\textwidth]{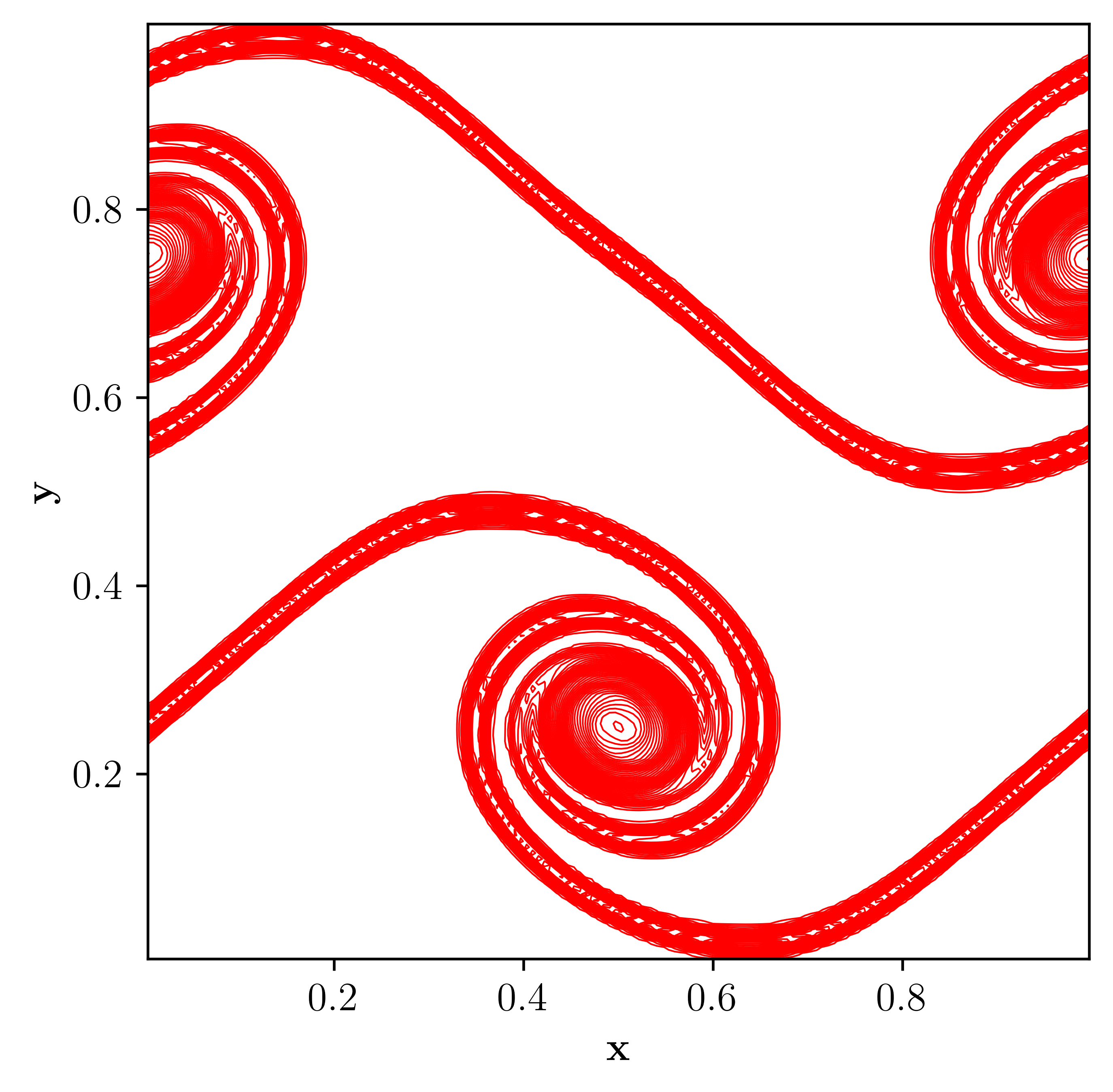}
\label{fig:mc_dsl_120}}
\subfigure[MEG8-Central]{\includegraphics[width=0.3\textwidth]{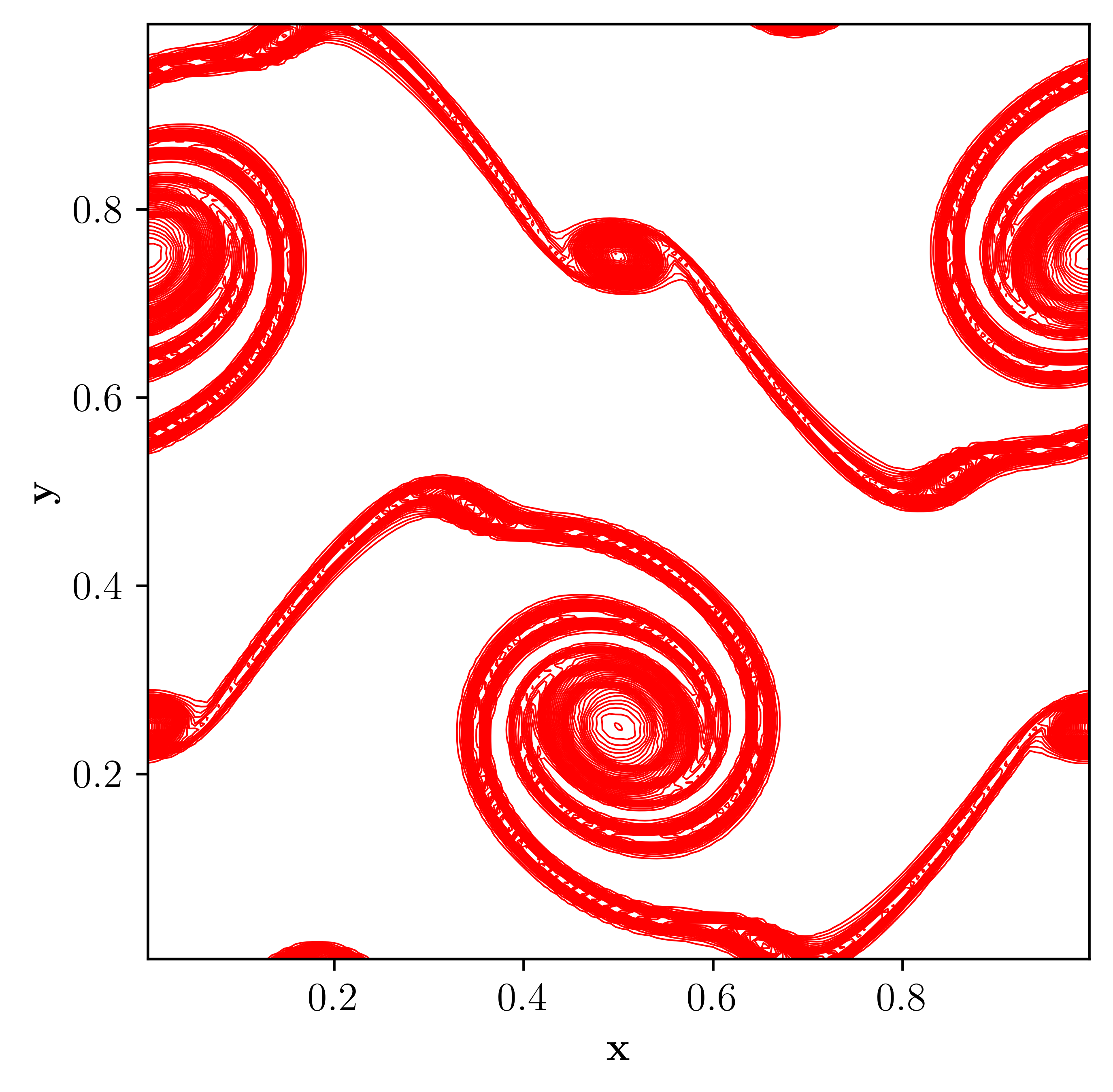}
\label{fig:m_cen_dsl_120}}
\subfigure[MP6-CC]{\includegraphics[width=0.3\textwidth]{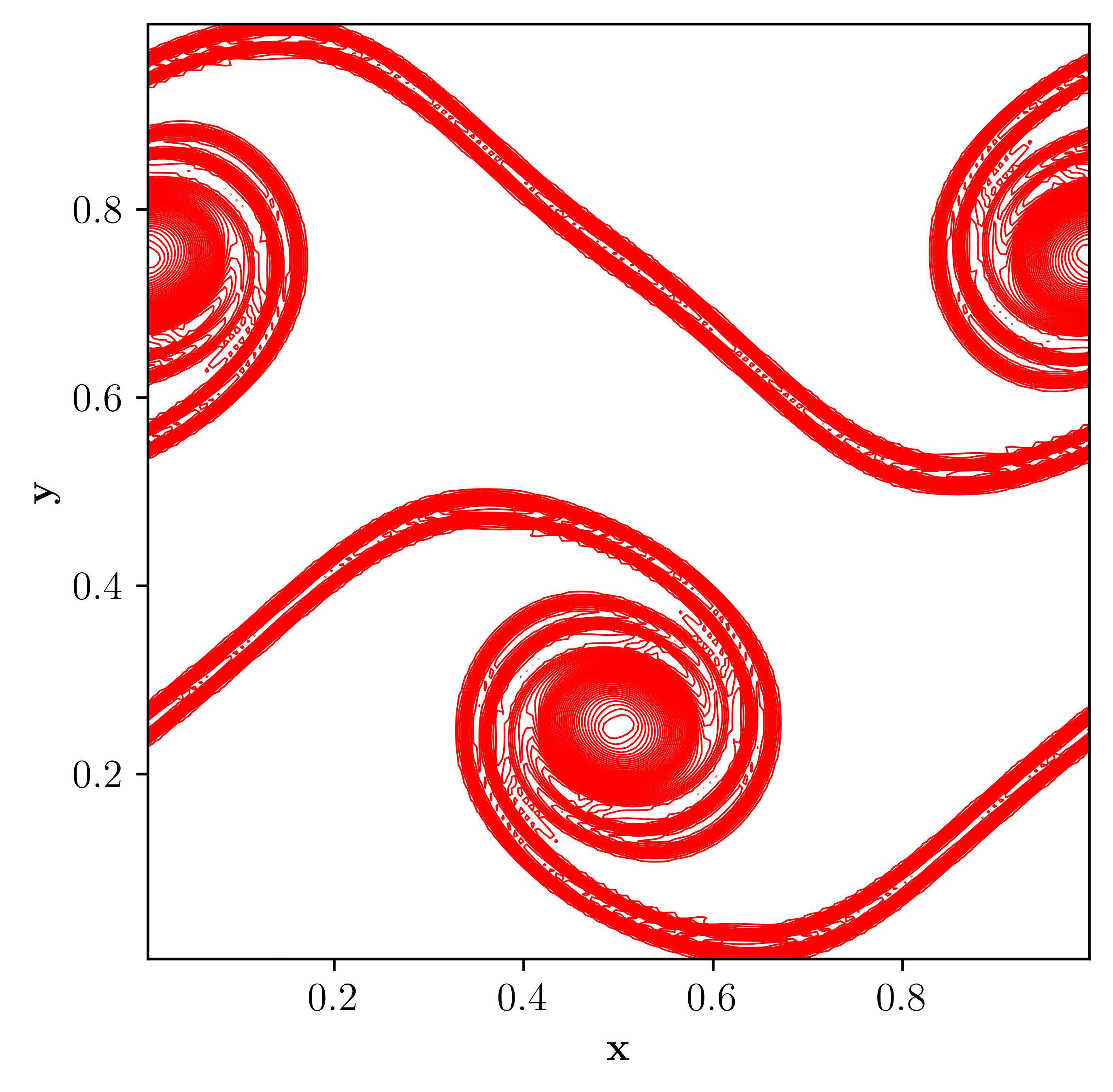}
\label{fig:MPcc_dsl_120}}
\subfigure[TENO5, 160 $\times$ 160]{\includegraphics[width=0.3\textwidth]{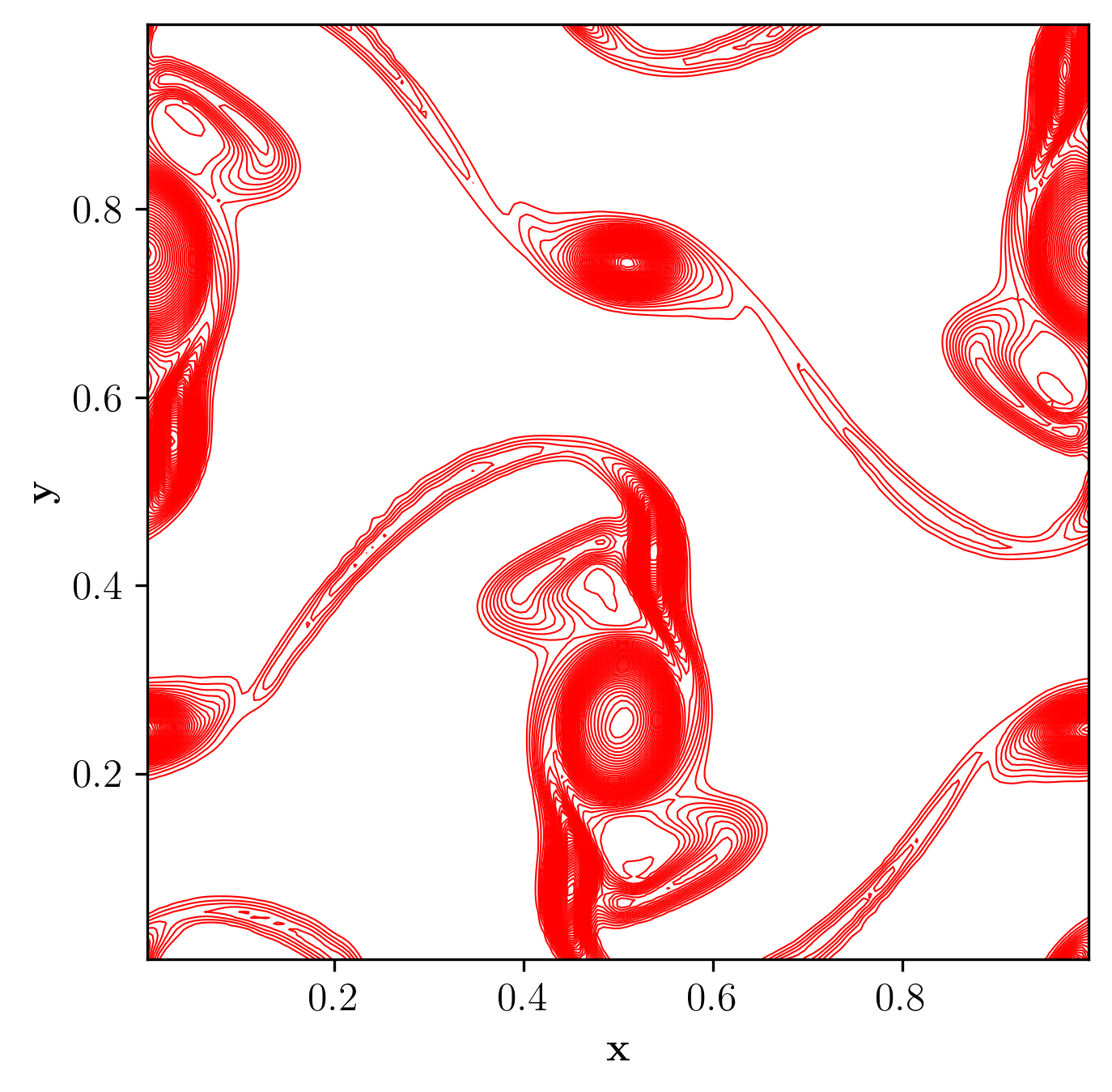}
\label{fig:teno_dsl_120}}
\subfigure[TENO5, 320 $\times$ 320]{\includegraphics[width=0.3\textwidth]{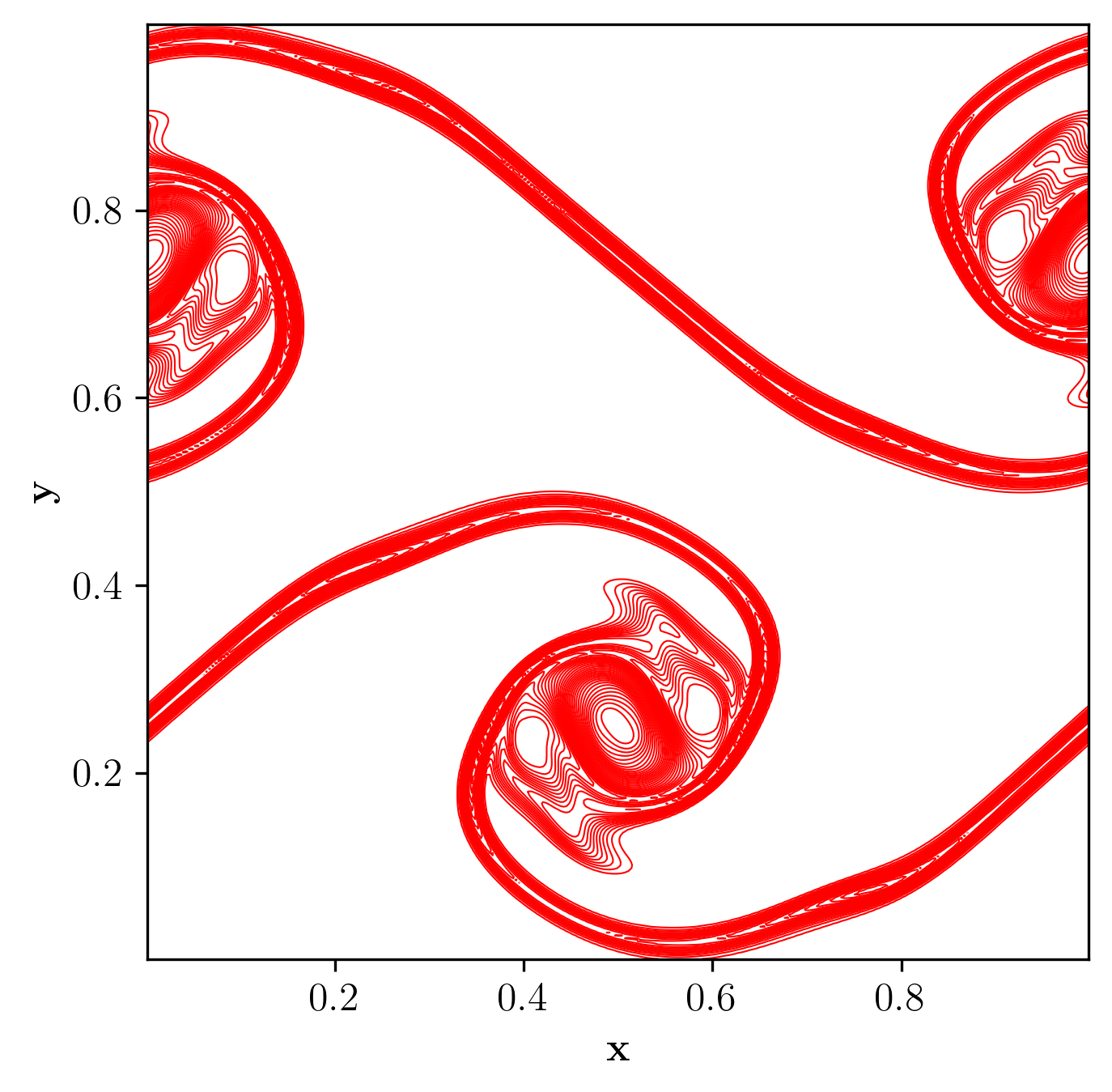}
\label{fig:teno_dsl_320}}
    \caption{$z$-vorticity contours of the considered schemes using a grid size of $160^2$ for $\theta$ =120.}
    \label{fig:dpsl_120}
\end{figure}

Fig. \ref{fig:dpsl_120} displays the $z$-vorticity computed for MEG8-CC, MEG8-C, MEG8-central, MP6-CC, and TENO5 schemes on a grid size of 160 $\times$ 160 (except for Fig. \ref{fig:teno_dsl_320}). From Fig. \ref{fig:mcc_dsl_120}, it can be seen that MEG8-CC best approximates the reference solution, as there are no unphysical braid vortices and oscillations. MEG8-C, Fig. \ref{fig:mc_dsl_120} is also relatively free of spurious vortices. MEG8-CC and MEG8-C use the unique adaptive central-upwind algorithm proposed, i.e. not all variables are reconstructed using the central scheme. Fig. \ref{fig:m_cen_dsl_120} shows the results obtained by the fully central scheme for all the variables if both Ducros and contact discontinuity sensors criteria are satisfied (Equation \ref{con:check}). The fully central scheme produced unphysical vortical structures unlike the adaptive central-upwind scheme (Equations (\ref{eqn:centralScheme_d}), (\ref{eqn:centralScheme_x}), (\ref{eqn:centralScheme_y}), and (\ref{eqn:centralScheme_z}) in conservative variable computational space). Likewise, the MP6-CC scheme, vorticity contours shown in Fig. \ref{fig:MPcc_dsl_120}, also has no spurious vortices. The stencils of the MEG8-CC and MP6-CC schemes are significantly different, yet both schemes resulted in results free of spurious vortices.

Figs \ref{fig:teno_dsl_120} and \ref{fig:teno_dsl_320} show the vorticity contours computed by the TENO5 scheme on grid sizes of 160 $\times$ 160 and 320 $\times$ 320, respectively. Even on a grid size of 320 $\times$ 320, the TENO5 scheme produced spurious results in the primary vortex. It is important to note that the MP6-CC and TENO5 schemes have the same stencil for reconstructing cell-interface values. It indicates that the proposed algorithm produces results free of oscillations and spurious vortices for this test case. Feng et al. \cite{feng2024general} simulated this test case on a grid size of 320 $\times$ 320 using TENO5 and TENO8. Even on that fine grid resolution, the TENO8 scheme produced spurious vortices along the shear layer (readers can refer to Fig. 13 of \cite{feng2024general}). These findings suggest that the proposed algorithm with a unique central-upwind scheme can prevent spurious vortices for this test case, which has been the subject of many papers in the literature. 

Table \ref{tab:dsl_cost} shows the computational cost of the proposed algorithms, MEG8-CC and MP6-CC, and MEG8-C and TENO5. The MP6-CC and MEG8-CC are almost similar in computational cost, whereas MEG8-C is twice as expensive due to the reconstruction of characteristic variables only. The TENO5 scheme will require four times more grid points and is 13 times more costly than the proposed schemes.

\begin{table}[H]
    \centering
    \caption{Comparison of computational costs and efficiency for the evaluated schemes for Example \ref{ex:dsl}.}
    \begin{tabular}{c c c c c}
        \hline
        \hline
        MEG8-CC & MEG8-C & MP6-CC & TENO5 (Coarse) & TENO5 (Fine) \\
        \hline
        110 s  & 205 s  & 100 s  & 250 s (160 $\times$ 160) & 1297 s (320 $\times$ 320)  \\
        \hline
        \hline
    \end{tabular}
    \label{tab:dsl_cost}
\end{table}

This test case does not involve shocks or contact discontinuities; therefore, the Ducros sensor or contact sensor was unnecessary. Therefore, a linear combination of the central-upwind scheme algorithm given by Equations (\ref{eqn:centralScheme_d}), (\ref{eqn:centralScheme_x}), (\ref{eqn:centralScheme_y}), and (\ref{eqn:centralScheme_z})  alone was sufficient. Central reconstruction scheme for all the variables except for $\rho u$ in $x-$ direction and $\rho v$ in $y-$ direction will suffice. In this regard, simulation is also carried out using the third-order upwind and fourth-order central scheme combination to show the uniqueness of the proposed algorithm. The third-order upwind scheme for left and right interfaces is as follows:

\begin{equation}
\begin{aligned} 
\phi_{i+1 / 2}^{L3, Linear} &= \frac{1}{6}\left(-\phi_{i-1} + 5\phi_{i} + 2\phi_{i+1} \right), \\
\phi_{i+1 / 2}^{R3, Linear} &= \frac{1}{6}\left(2\phi_{i} + 5\phi_{i+1} -\phi_{i+2} \right),
\end{aligned}
\label{eqn:3linear}
\end{equation}
and the fourth-order central scheme is:

\begin{equation}
\phi^{C4,Linear}_{i+\frac{1}{2}}=\frac{1}{2}\left(\phi_{i+1 / 2}^{L3, Linear} + \phi_{i+1 / 2}^{R3, Linear}\right)=\frac{1}{12}\left(- \phi_{i-1}+7 \phi_i+7 \phi_{i+1}- \phi_{i+2}\right).
\end{equation}

\begin{figure}[H]
\centering\offinterlineskip
\subfigure[Linear 3rd/4th]{\includegraphics[width=0.3\textwidth]{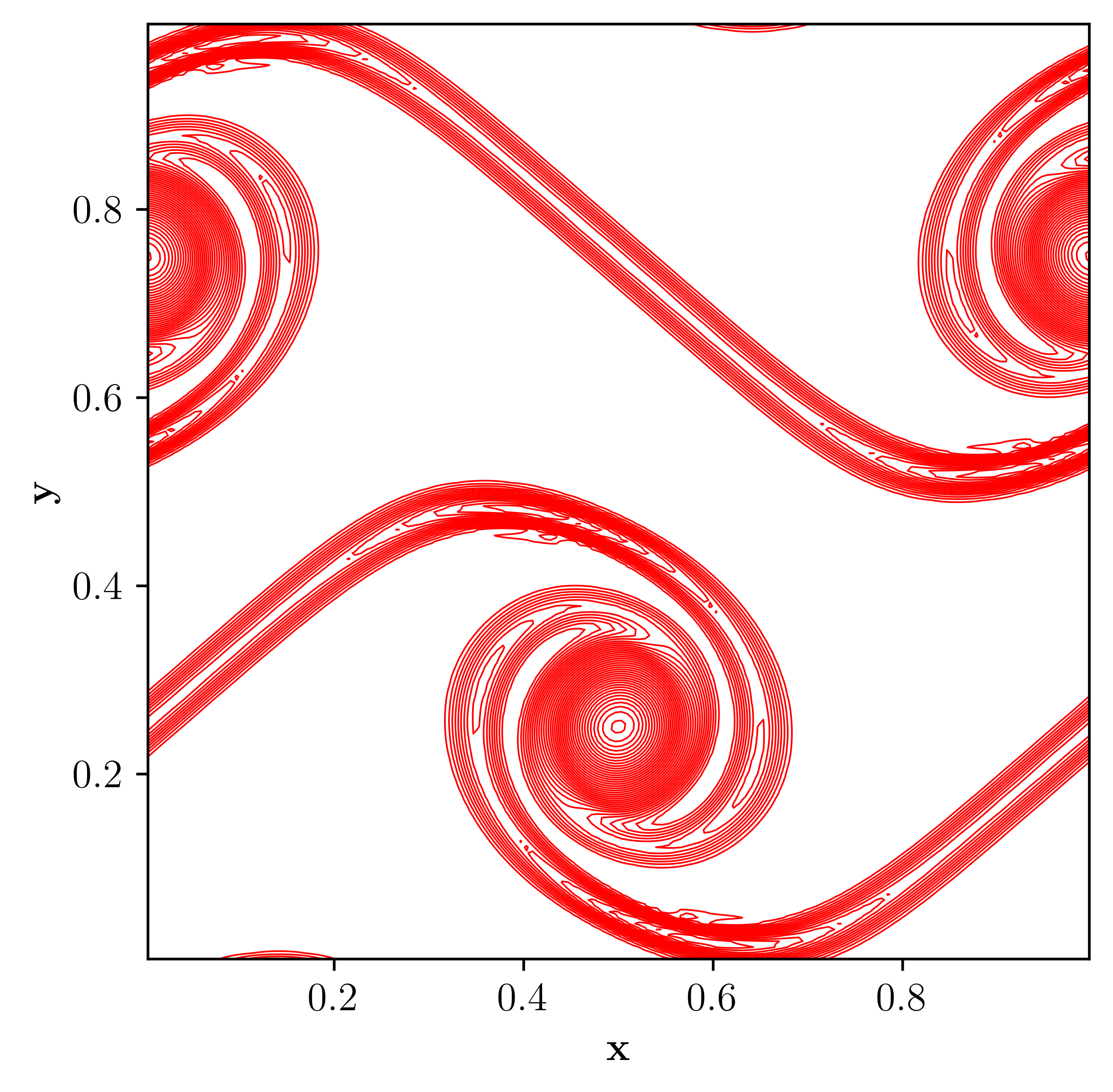}
\label{fig:34_dsl_120}}
\subfigure[Linear 5th/6th]{\includegraphics[width=0.3\textwidth]{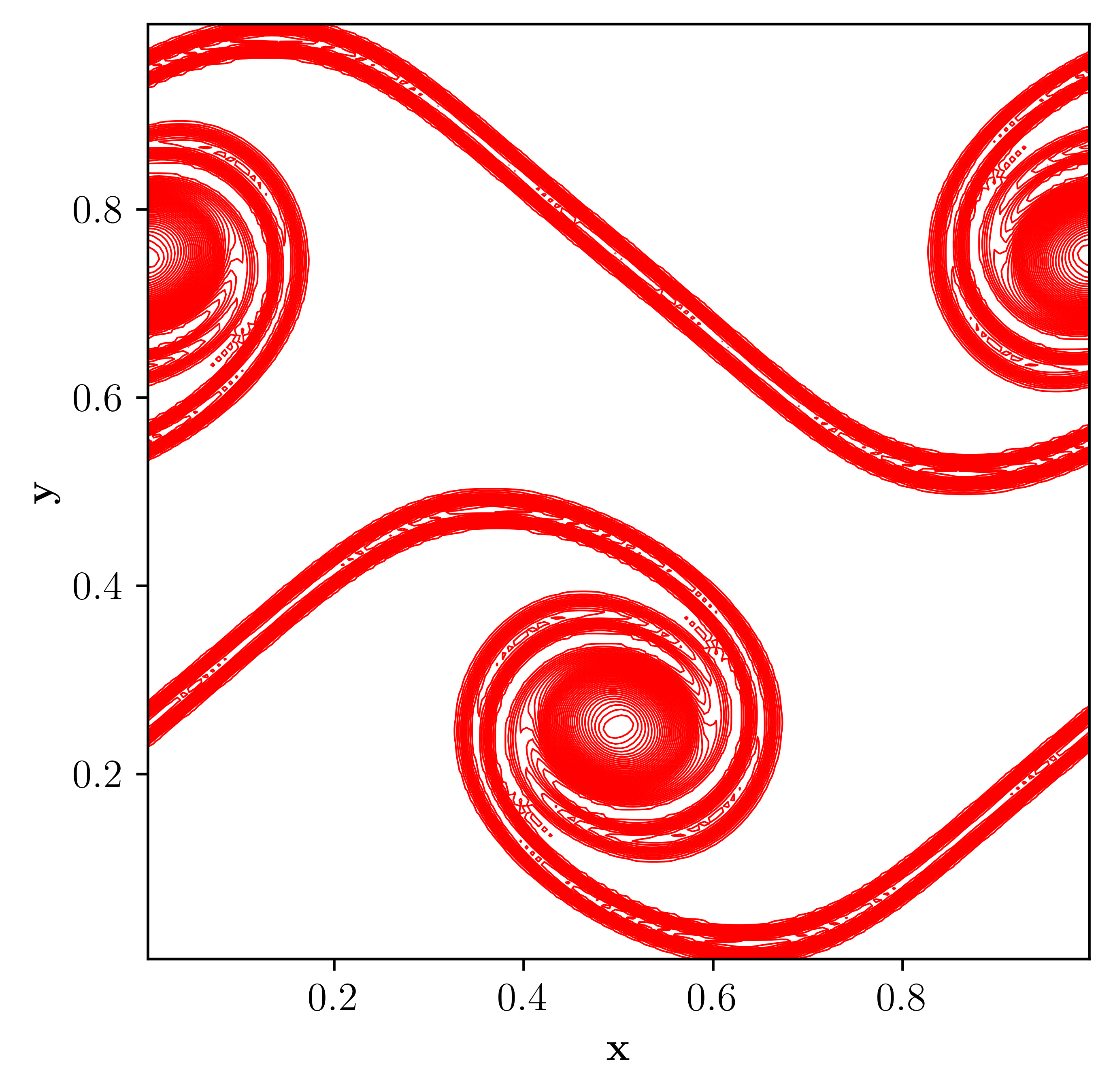}
\label{fig:56_dsl_120}}
    \caption{$z$-vorticity contours of the considered schemes using a grid size of $160^2$ for $\theta$ =120.}
    \label{fig:dpsl_160_linear}
\end{figure}

Figs \ref{fig:34_dsl_120} and \ref{fig:56_dsl_120} show the vorticity contours computed by the linear third-fourth order scheme and linear fifth-sixth-order scheme using the proposed central-upwind algorithm on grid size of 160 $\times$ 160, respectively. Even the linear schemes are free of spurious vortices for this test case using the proposed algorithm. The only difference between the third-fourth-order scheme and the fifth-sixth-order scheme is the thickness of the shear layer. The third-fourth-order scheme is thicker than the fifth-sixth-order scheme, which can be attributed to the difference in spectral properties and order of accuracy. With the proposed algorithm, even a third-fourth-order linear scheme prevented spurious vortices, unlike the TENO5 scheme. Furthermore, for the simulations using the linear schemes (it is applicable also for the MEG8-CC and MP6-CC schemes), the fourth-order alpha-damping scheme of Nishikawa \cite{Nishikawa2010} was used instead of the compact viscous scheme proposed in \cite{chamarthi2023gradient} (described in Section \ref{sec:visc}) and yet there are no spurious vortices. It indicates that the viscous flux discretization had little impact on this test case as long as the discretization has good spectral properties \cite{chamarthi2022}.

\begin{figure}[H]
\centering\offinterlineskip
\subfigure[MEG8-CC]{\includegraphics[width=0.3\textwidth]{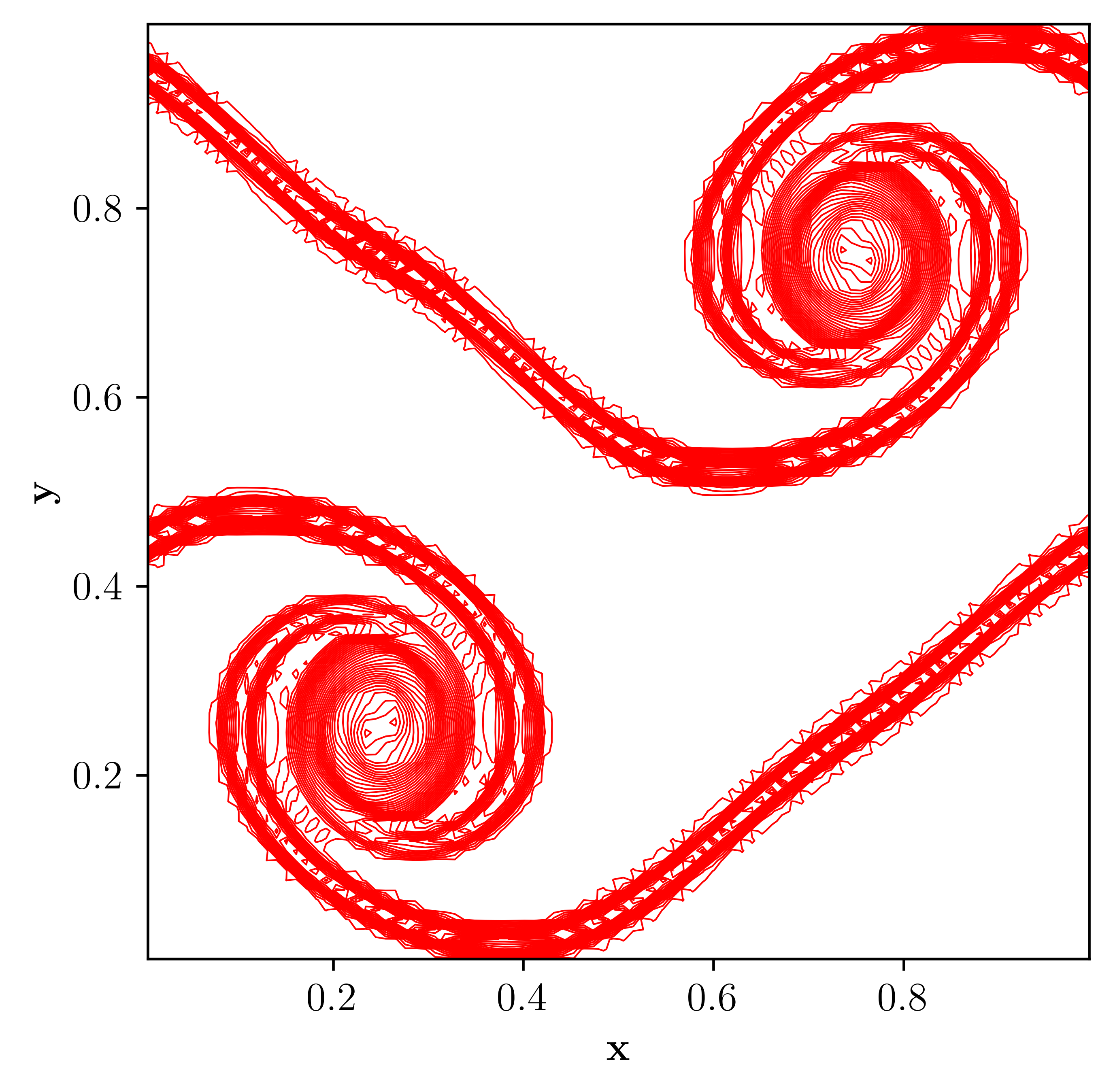}
\label{fig:mcc_dsl_80}}
\subfigure[Linear 3rd/4th]{\includegraphics[width=0.3\textwidth]{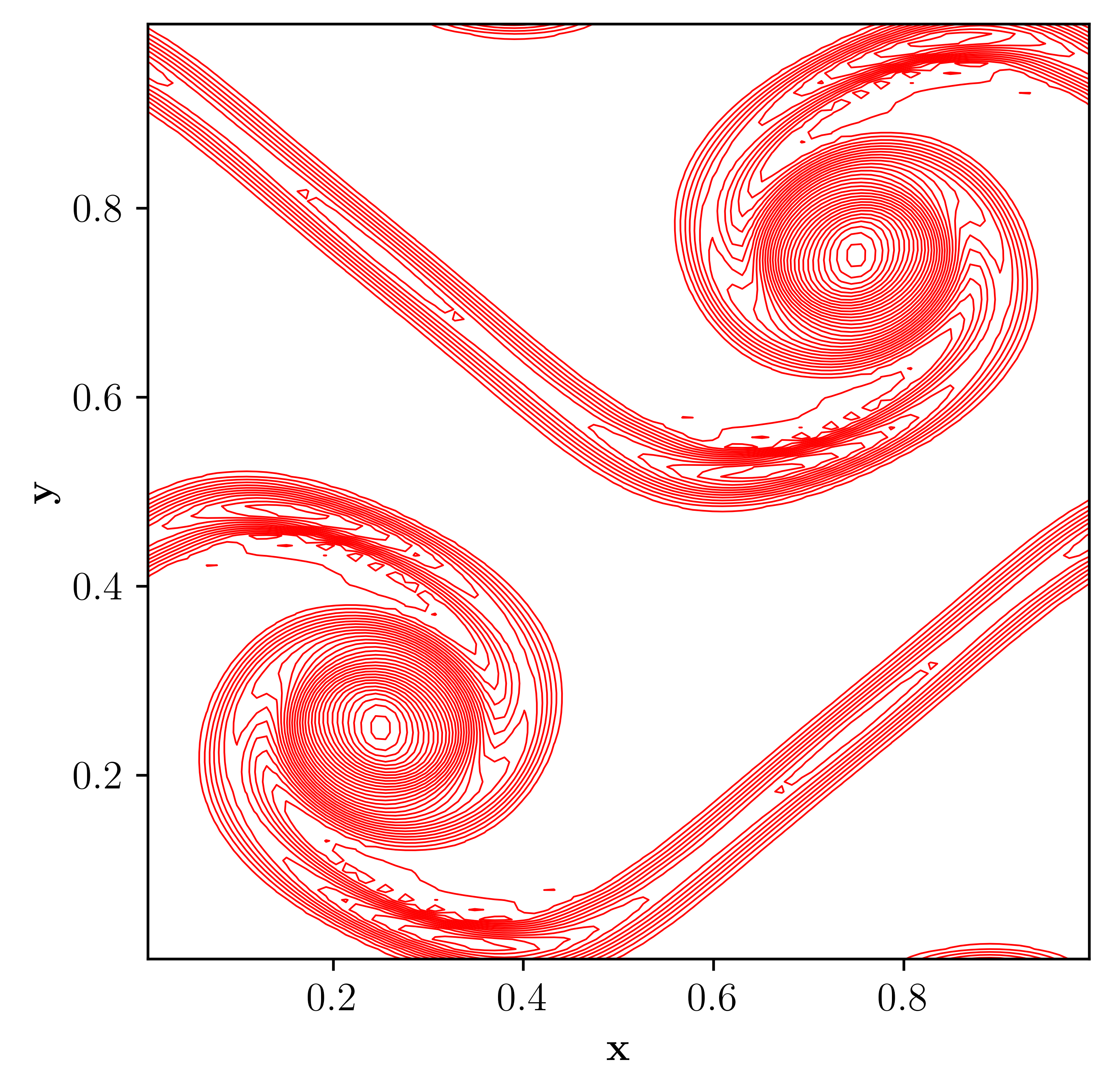}
\label{fig:34_dsl_80}}
\subfigure[Linear 5th/6th]{\includegraphics[width=0.3\textwidth]{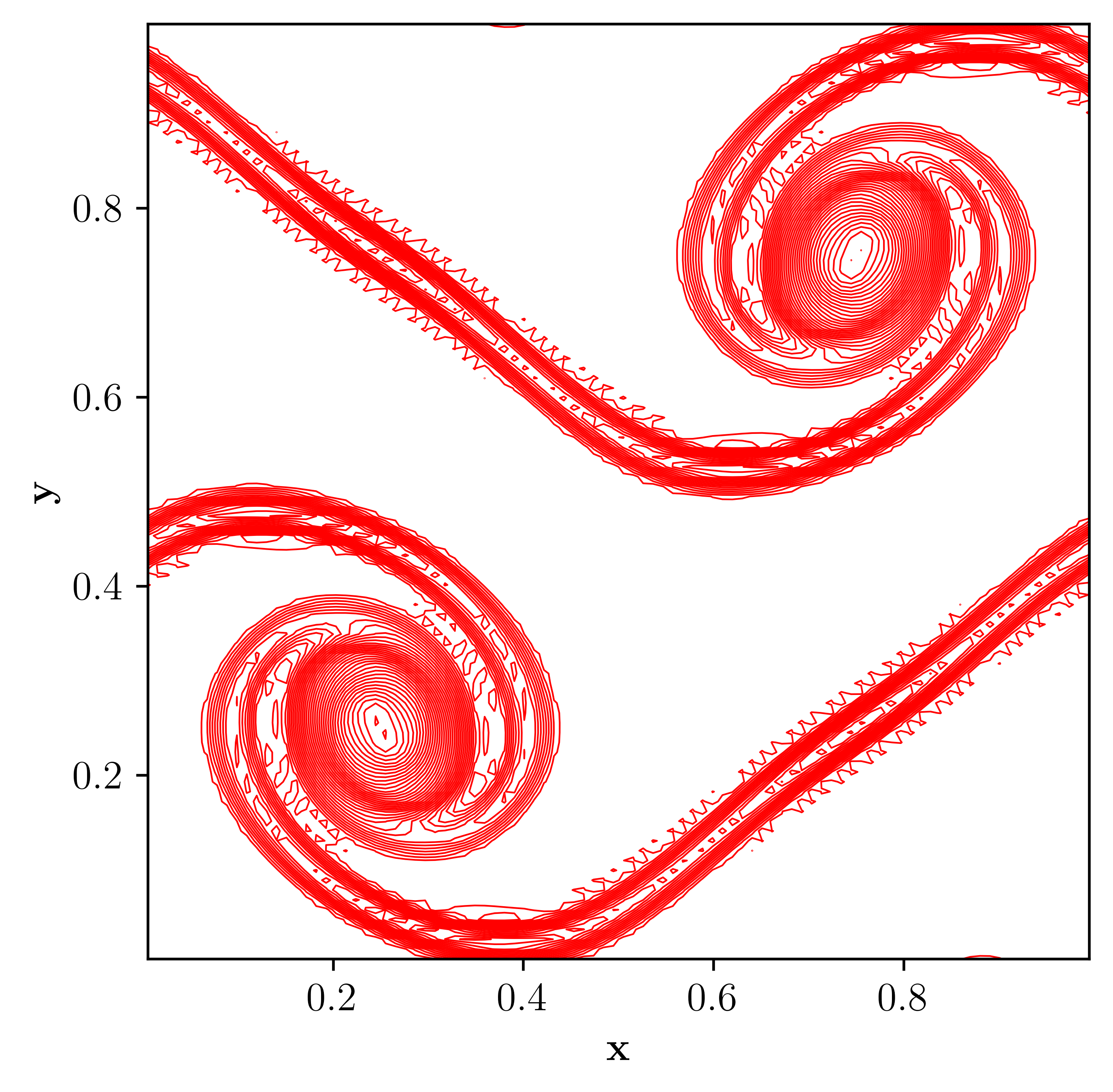}
\label{fig:56_dsl_80}}
    \caption{$z$-vorticity contours of the considered schemes using a grid size of $96^2$ for $\theta$ =80.}
    \label{fig:dpsl_80}
\end{figure}

Finally, Fig. \ref{fig:dpsl_80} displays the $z$-vorticity computed for MEG8-CC, linear third-fourth scheme and linear fifth-sixth-order scheme on a grid size of 96 $\times$ 96  for $\theta$=80. It can be observed that all the schemes are free of spurious vortices on such a coarse grid. These results show the advantages and uniqueness of the proposed algorithm. The algorithm also produced significantly better results than the kinetic energy preserving schemes considered in \cite{chamarthi2023role} on much coarser grids.

\begin{figure}[H]
\centering\offinterlineskip
\subfigure[MEG8-CC]{\includegraphics[width=0.3\textwidth]{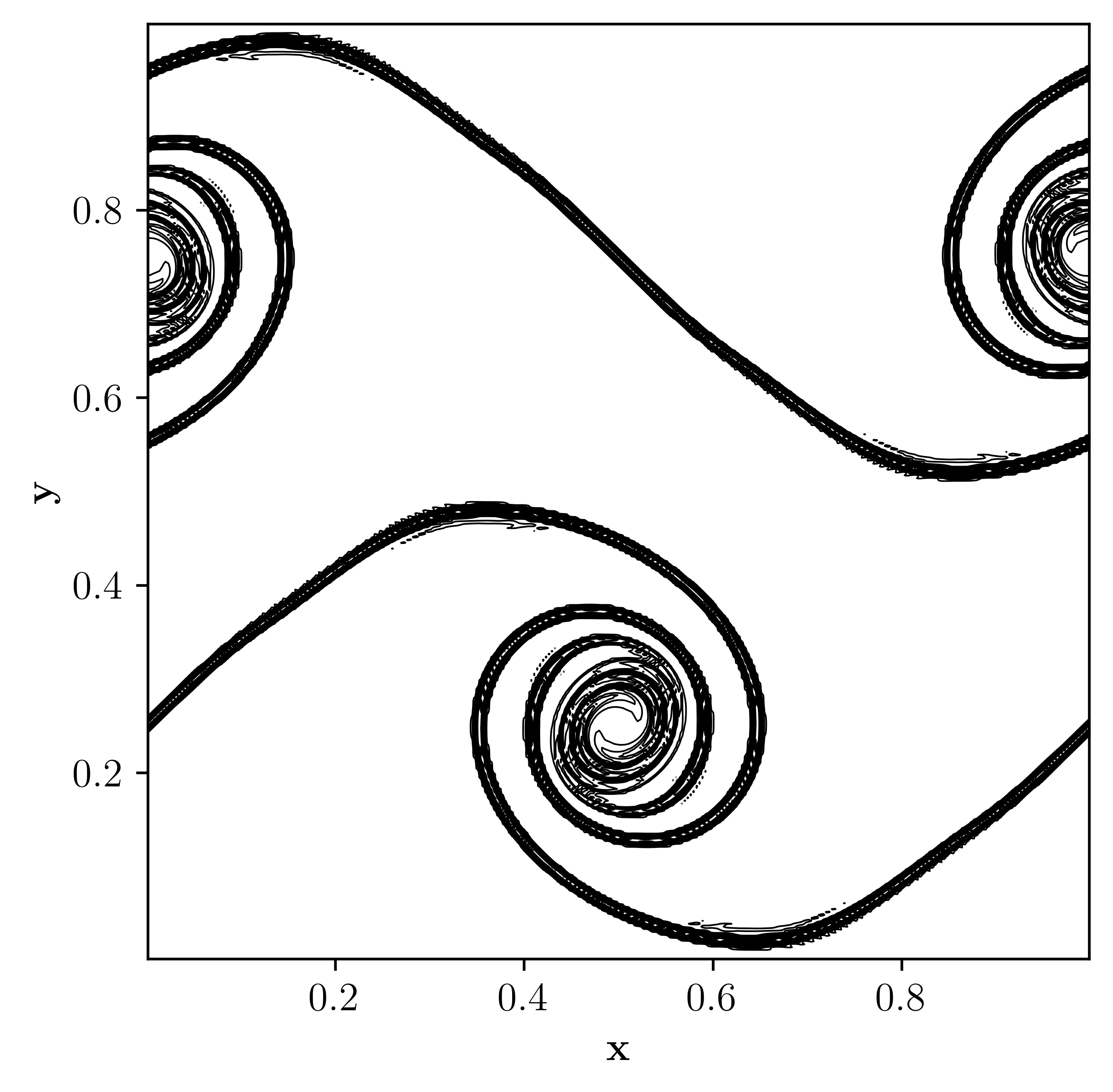}
\label{fig:mcc_dsl_80_inv}}
\subfigure[Linear 5th/6th]{\includegraphics[width=0.3\textwidth]{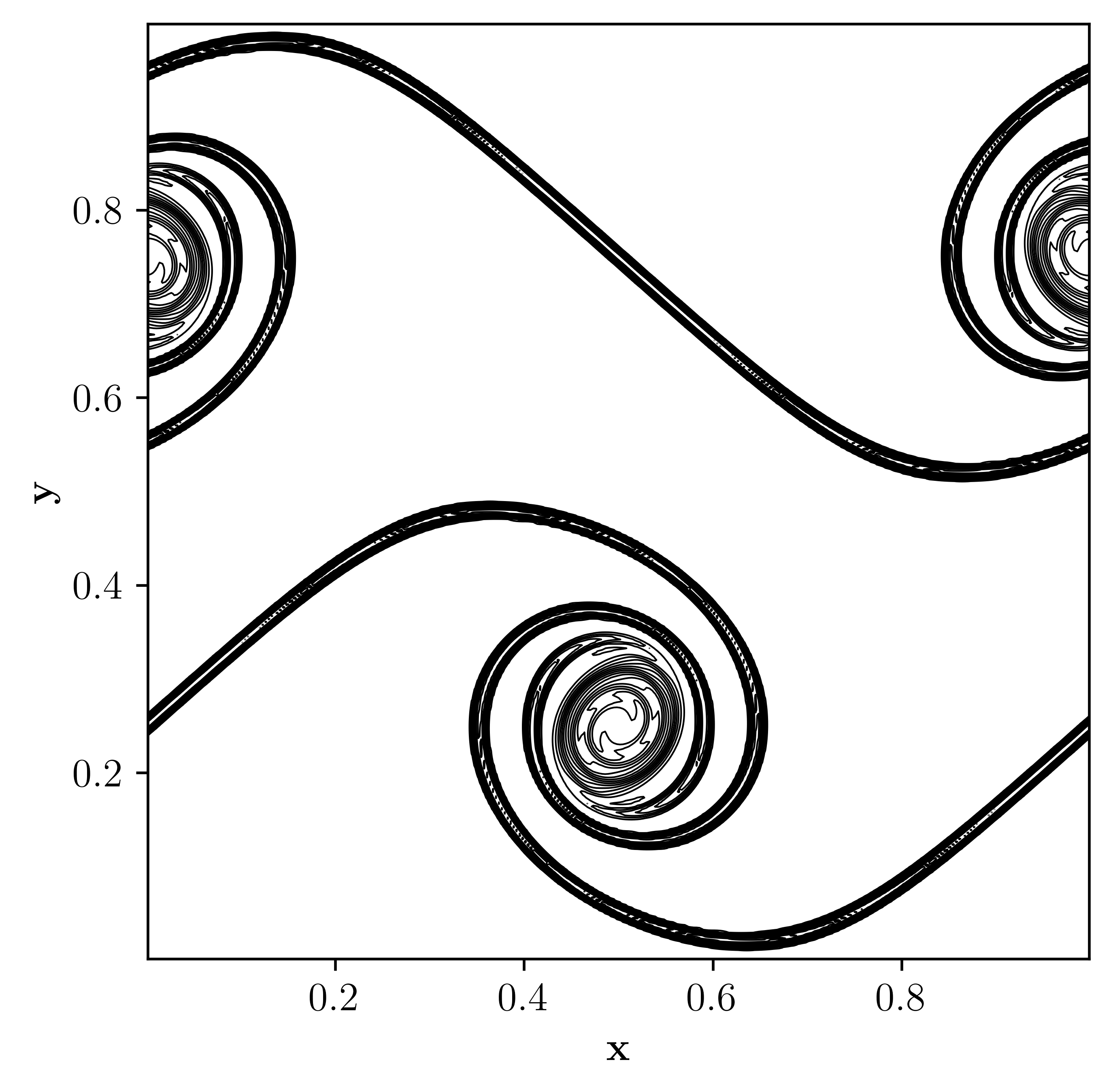}
\label{fig:56_dsl_80_inv}}
\subfigure[MP6-CC]{\includegraphics[width=0.3\textwidth]{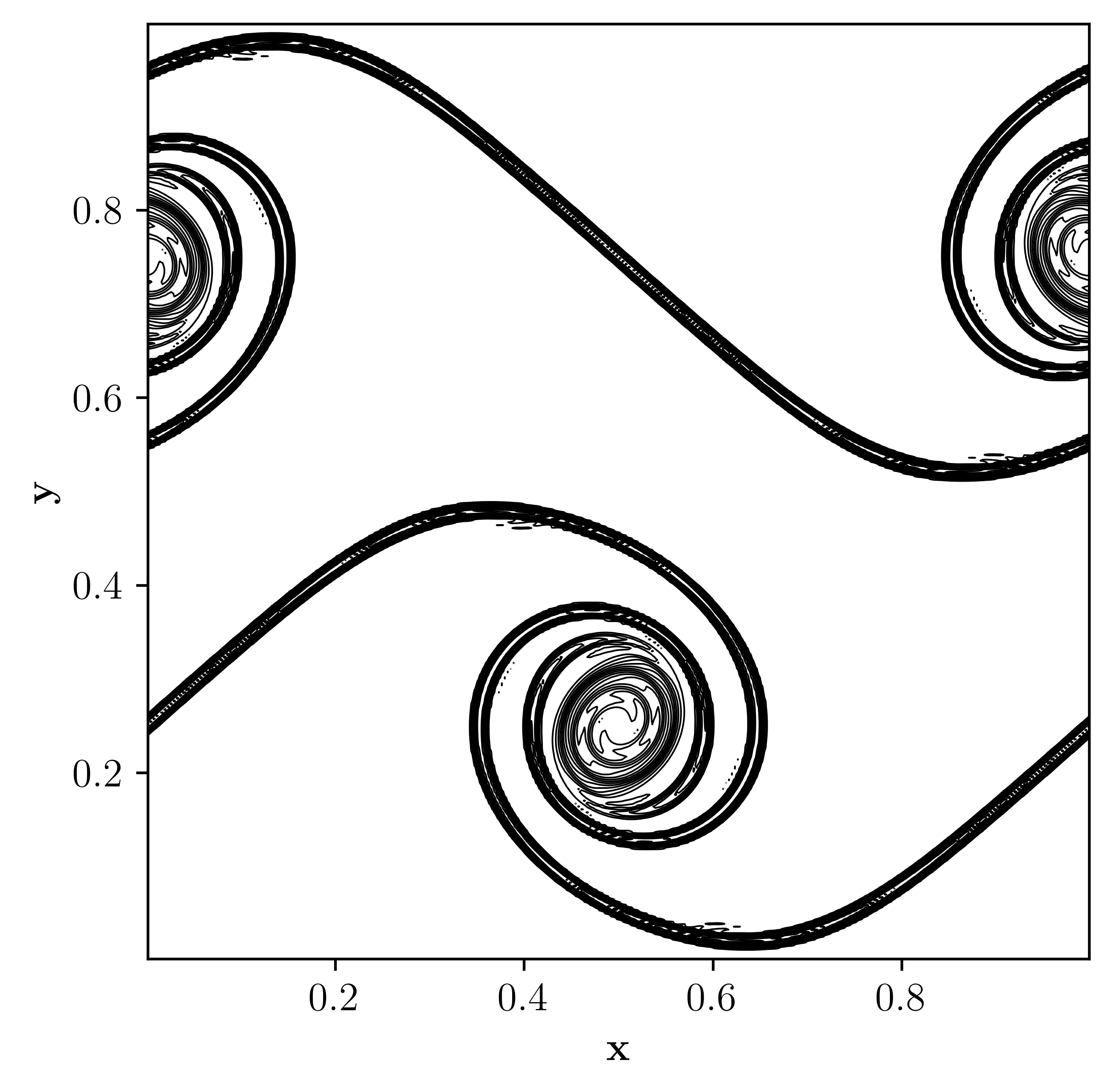}
\label{fig:mp6_dsl_80_inv}}
    \caption{$z$-vorticity contours of the considered schemes using a grid size of $320^2$ for $\theta$ =80 and $Re$=$\infty$.}
    \label{fig:dpsl_80_inv}
\end{figure}

Finally, simulation is also carried out for inviscid, $Re$=$\infty$, scenario  with $\theta$=80 for $v = 0.05 \sin \left[ 2 \pi(x) \right]$ in the initial conditions as in \cite{feng2024general}. Fig. \ref{fig:dpsl_80_inv} displays the $z$-vorticity computed for MEG8-CC, MP6-CC and the linear fifth-sixth-order scheme on a grid size of 320 $\times$ 320. In Ref. \cite{feng2024general}, authors have computed the simulation on a grid size of 512 $\times$ 512 for these initial conditions. With the present algorithm, even the inviscid case is free of spurious vortices on a grid size that is two and a half times smaller than \cite{feng2024general}. The fifth-sixth-order scheme has a slightly thicker shear layer width than MEG8-CC. The proposed algorithm performed better than the optimized TENO8 scheme considered in \cite{feng2024general} (see Fig. 13 of the concerned reference). All the schemes, both convective and viscous flux discretization, are only second-order accurate \cite{van2021towards}, as fluxes are not reconstructed directly, making these results even more significant.

\textbf{Possible reasons for lack of spurious vortices:}
\begin{enumerate}
	\item Contact discontinuities are characterized by the condition that no mass flows through the discontinuity (normal velocities are zero) and continuous pressure. This results in a jump in density and tangential velocity (page 137 of \cite{hirschvol2}). The tangential velocities are $v$ in the $x-$ direction and $u$ in the $y-$ direction. In the absence of contact discontinuity, the tangential velocities will be continuous. However, while tangential velocities are discontinuous for inviscid flows, they are not discontinuous for viscous flow simulations. The driving force behind interfacial instabilities is shear. In the interface between two different fluids, the conservation of momentum requires that the tangential viscous stresses be equal, leading to the continuity of the tangential velocity. The fluids would slip in the actual inviscid limit, resulting in discontinuous tangential velocity. However, in the presence of viscosity (\textit{even artificial viscosity due to upwind reconstructions}), the tangential velocity remains continuous across an interface according to Batchelor \cite{batchelor1967introduction}. In the present algorithm, the tangential velocities are deemed continuous if the MP criterion for density, and Ducros sensor for shocks are met and, therefore, reconstructed using a central scheme in conservative variable space - Equations (\ref{eqn:centralScheme_x}), (\ref{eqn:centralScheme_y}), and (\ref{eqn:centralScheme_z}) or wave appropriate centraliztion in characteristic space.
	\item The shear layer test case does not involve shocks, and the density is almost constant and therefore does not require any shock or contact discontinuity sensor and can be simulated using linear schemes itself, i.e. directly using the Equations (\ref{eqn:centralScheme_d}), (\ref{eqn:centralScheme_x}), (\ref{eqn:centralScheme_y}), and (\ref{eqn:centralScheme_z}). The flow is viscous, and the density is almost constant; therefore, the tangential velocities are also continuous, which is what the proposed algorithm is incidentally doing: centralizing the $\rho$v in $x-$ direction and $\rho$u in $y-$ direction.
	\item \textbf{Potential improvement of optimization of the numerical schemes:} In \cite{feng2024general} authors have optimized the parameter $\eta$ in the following (a similar one) equation:

\begin{equation}
\phi^{C}_{i+\frac{1}{2}} = \left( 1 - \eta \right) \phi^{L, Linear}_{i+\frac{1}{2}} + \eta  \phi^{R, Linear}_{i+\frac{1}{2}}
\end{equation}

The parameter $\eta$ is \textit{constant}, once the optimum value is found, for \textit{all the variables}. Despite the optimization procedure, the optimized schemes, including the Ducros sensor, lead to spurious vortices even on much larger grid sizes than the current paper (see Figure 13 of Ref. \cite{feng2024general} and the corresponding discussion). In this paper, the value of $\eta$ is either 0.5 or 1 and is different for different variables in various directions. If one has to optimize, then $\eta$ has to be an independent value for each variable in each direction. It implies that instead of optimizing one parameter $\eta$, there will be $\eta_1$ to $\eta_8$, eight parameters to optimize in a two-dimensional scenario (four variables in two directions)
	\item It is only a partial explanation because there is no clear understanding of why some variables are reconstructed using upwind reconstruction for this test case (it is necessary for test cases with discontinuities as in \cite{hoffmann2024centralized}). Nevertheless, there are no spurious vortices even with the linear schemes using the current central-upwind algorithm, which is a significant outcome.
\end{enumerate}

The shear layer test case does not have shock waves or contact discontinuities. The algorithm is tested for such cases in the following test cases, and its computational benefits and robustness are shown in the subsequent test cases.
\begin{example}\label{ex:kh}{Kelvin Helmholtz instability}
\end{example}

\begin{figure}[H]
\centering\offinterlineskip
\subfigure[TENO5]{\includegraphics[width=0.25\textheight]{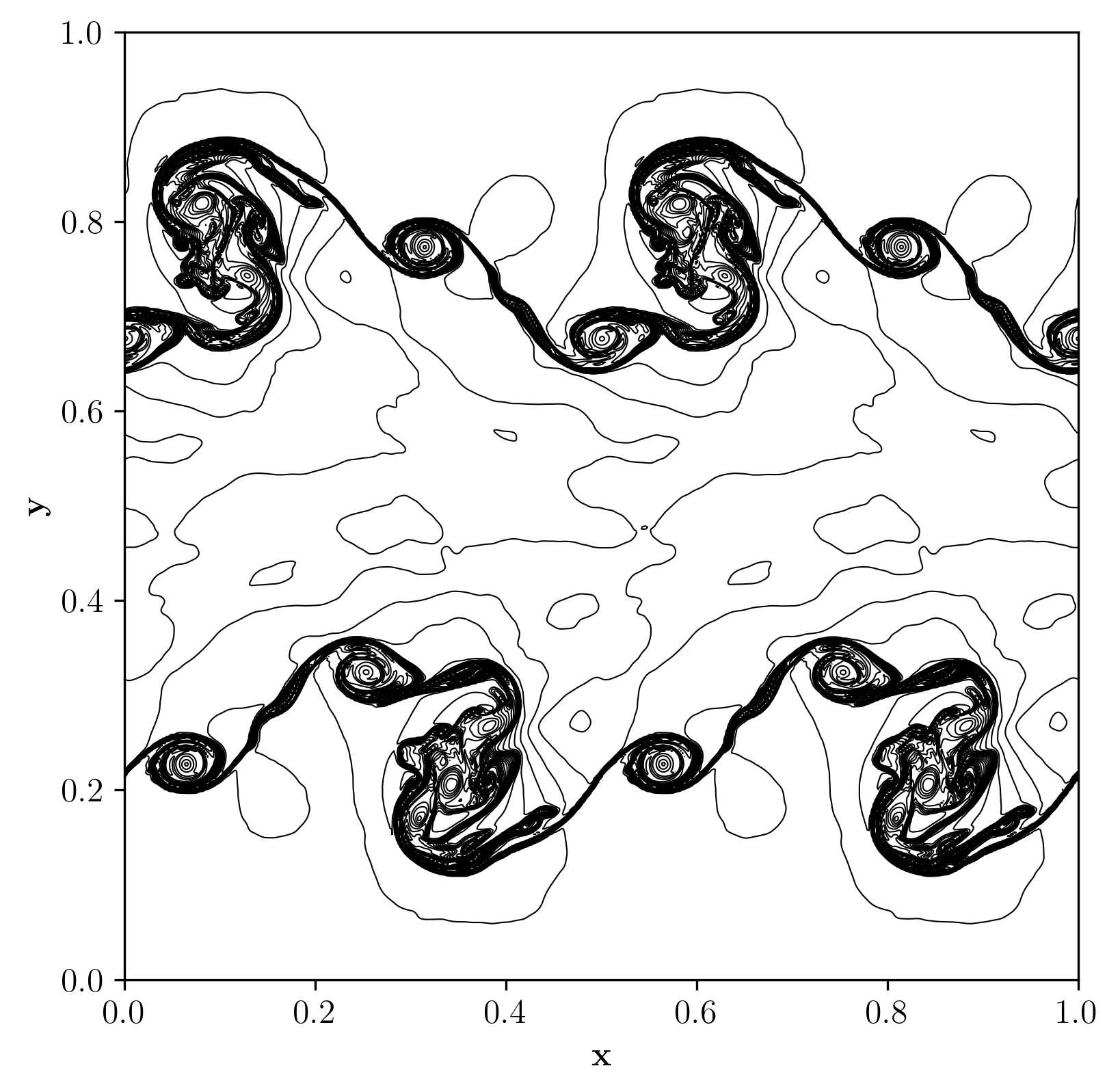}
\label{fig:teno_khi}}
\subfigure[MP6-CC]{\includegraphics[width=0.25\textheight]{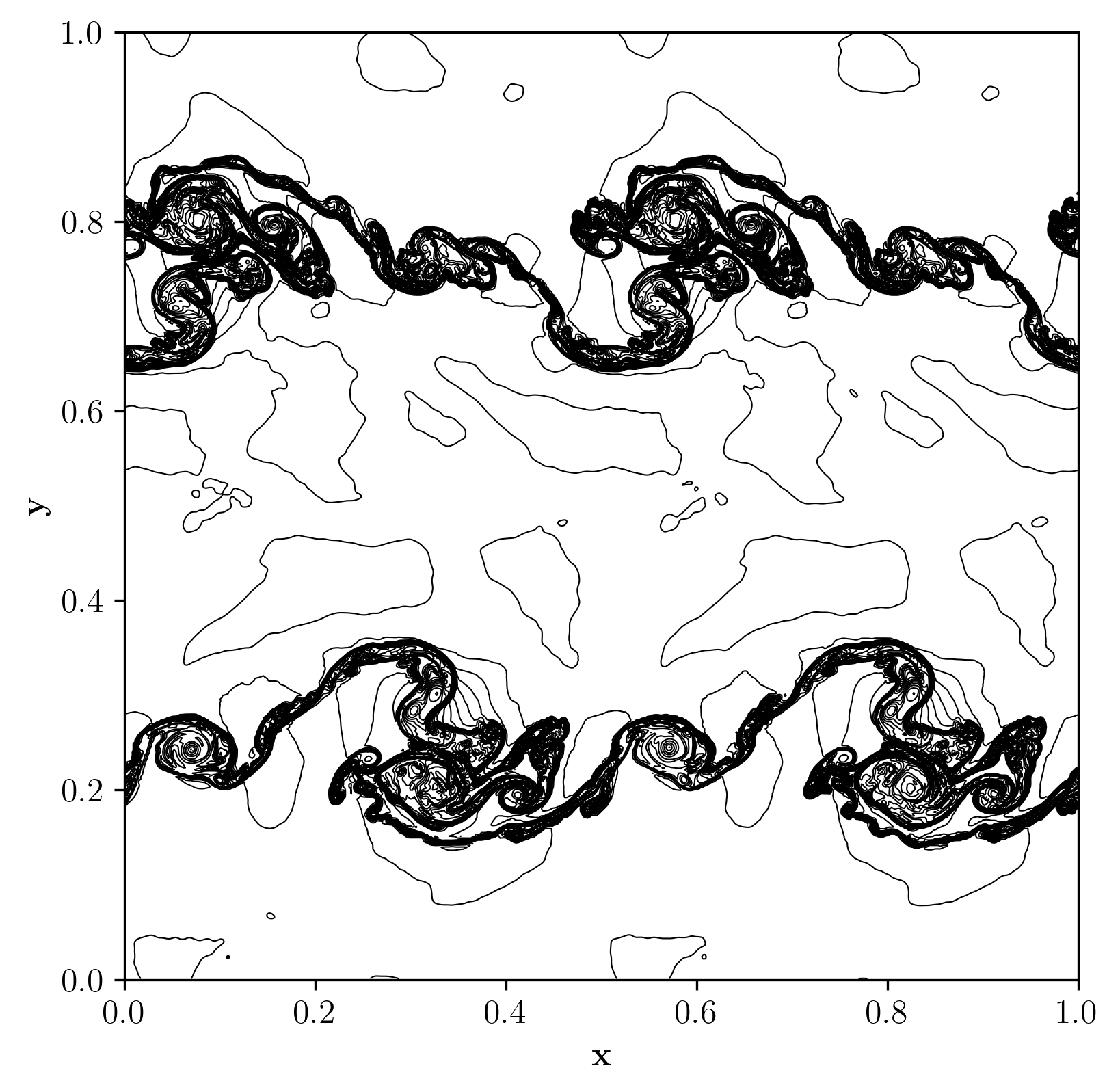}
\label{fig:mp6_khi}}
\subfigure[MEG8-C]{\includegraphics[width=0.25\textheight]{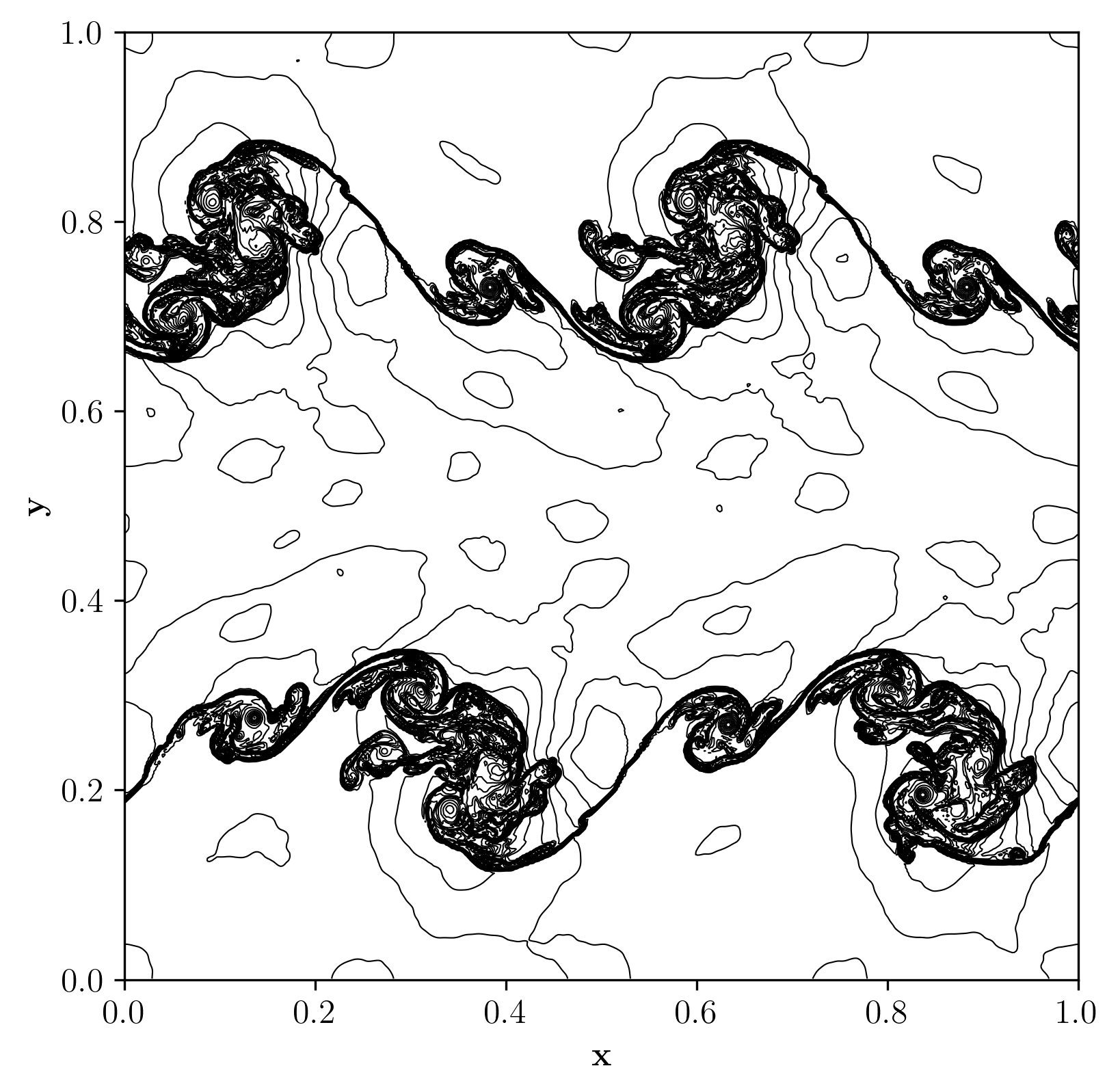}
\label{fig:meg8c_khi}}
\subfigure[MEG8-CC]{\includegraphics[width=0.25\textheight]{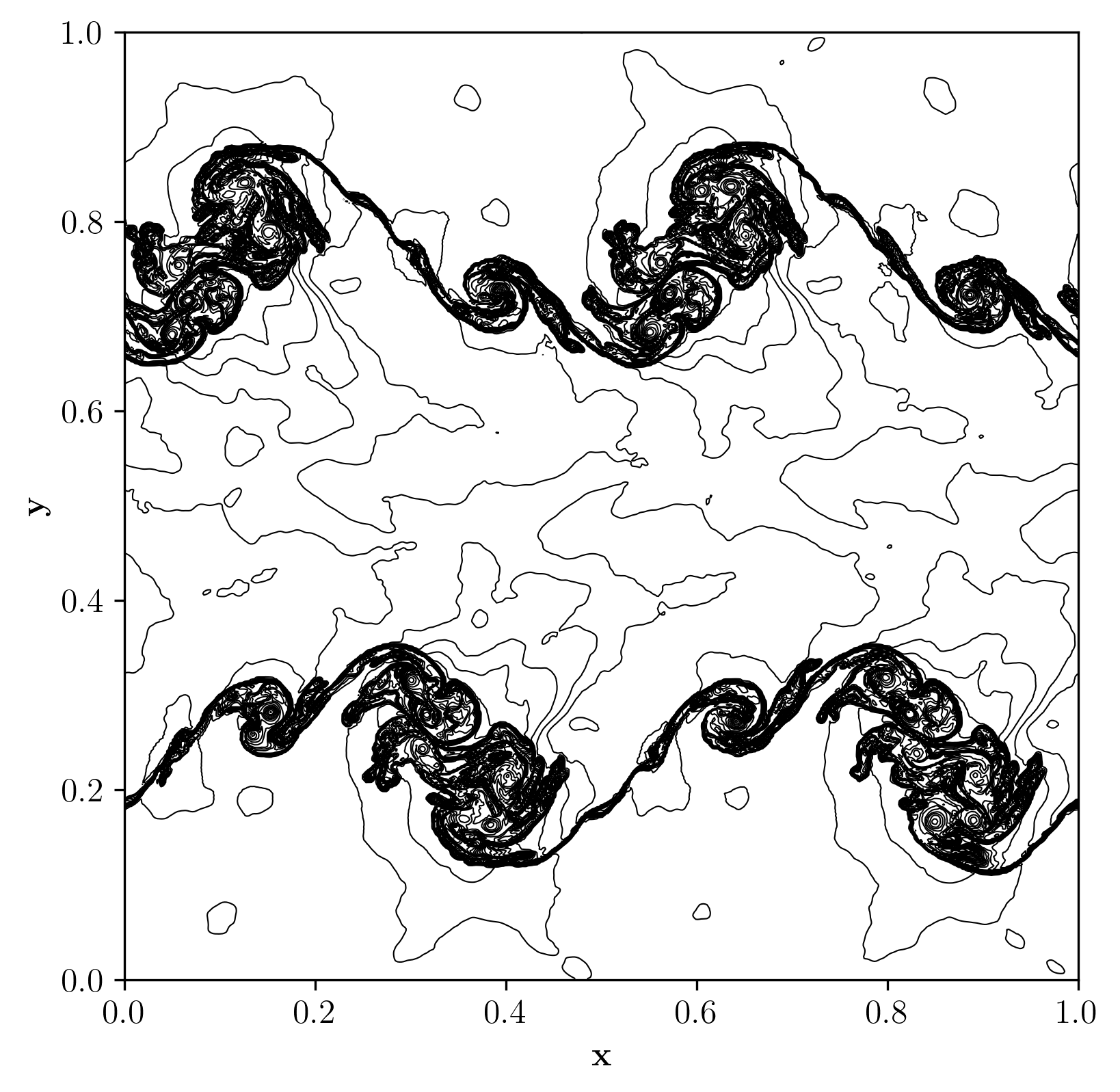}
\label{fig:meg8cc_khi}}
\caption{Numerical results of Kelvin-Helmholtz instability test case, Example \ref{ex:kh}, by different schemes on a grid size of 512 $\times$ 512.}
\label{fig_KH}
\end{figure}

First, the Kelvin-Helmholtz instability case is considered. This instability arises due to an unstable velocity gradient at the interface between the two fluids, leading to vortices and mixing patterns. The test case has the following initial conditions over a periodic domain of [0, 1] $\times$ [0, 1],

\begin{equation}
\begin{array}{l}
p=2.5, \quad \rho(x, y)=\left\{\begin{array}{l}
2, \text { If } 0.25<y \leq 0.75, \\
1, \text { else, }
\end{array}\right. \\
u(x, y)=\left\{\begin{array}{l}
0.5, \quad \text { If } 0.25<y \leq 0.75, \\
-0.5, \text { else },
\end{array}\right. \\
v(x, y)=0.1 \sin (4 \pi x)\left\{\exp \left[-\frac{(y-0.75)^{2}}{2 \sigma^{2}}\right]+\exp \left[-\frac{(y-0.25)^{2}}{2 \sigma^{2}}\right]\right\}, \text{where}\ \sigma = 0.05/\sqrt {2}.
\end{array}
\end{equation}
The computational domain is discretized with 512 cells in each direction, and the final time is $t$ = 0.8. The numerical solutions are computed using various schemes and are depicted in Fig. \ref{fig_KH}.  We can observe that all the schemes capture complex structures and small-scale vortices. Both MP6-CC and MEG8-CC are free of oscillations despite being reconstructed with the adaptive conservative-characteristic variable reconstruction. Despite sharing the same stencil, MP6-CC shows more vortical structures than the TENO5 scheme. 

Table \ref{tab:kh_cost} shows the schemes' computational times. MP6-CC is also 15$\%$ faster than the TENO5 scheme. For computational efficiency, the MEG8-CC scheme takes the same time as the TENO5 scheme with improved results. MEG8-CC is also 36 $\%$ faster than the MEG8-C presented in \cite{hoffmann2024centralized}, indicating the proposed approach's computational benefit.

\begin{table}[H]
    \centering
    \caption{Comparison of computational costs and efficiency for the evaluated schemes for Example \ref{ex:kh}.}
    \begin{tabular}{c c c c}
        \hline
        \hline
        MEG8-CC & MEG8-C & MP6-CC & TENO5 \\
        \hline
        1036 s (0.97) & 1614 s (1.51) & 895 s (0.84) & 1067 s (1.0) \\
        \hline
        \hline
    \end{tabular}
    \label{tab:kh_cost}
\end{table}

\begin{example}\label{ex:rt}{Rayleigh-Taylor instability}
\end{example}

The Rayleigh-Taylor instability is a hydrodynamic phenomenon that occurs when a dense fluid is above a less dense fluid in a gravitational field. This instability arises from an unstable density gradient, causing the denser fluid to accelerate downward and the lighter fluid to accelerate upward. As a result, complex and turbulent mixing patterns emerge between the two fluids. According to \cite{xu2005anti}, the initial conditions for the Rayleigh-Taylor instability are as follows:

\begin{equation}
\begin{aligned}
(\rho,u,v,p)=
\begin{cases}
&(2.0,\ 0,\ -0.025\sqrt{\frac{5p}{3\rho}\cos(8\pi x)},\ 2y+1.0),\quad 0\leq y< 0.5,\\
&(1.0,\ 0,\ -0.025\sqrt{\frac{5p}{3\rho}\cos(8\pi x)},\ 1y+1.5), \quad  0\leq y\leq 0.5,
\end{cases}
\end{aligned}
\label{eu2D_RT}
\end{equation}

\begin{figure}[H]
\centering\offinterlineskip
\subfigure[TENO5]{%
\includegraphics[width=0.15\textwidth]{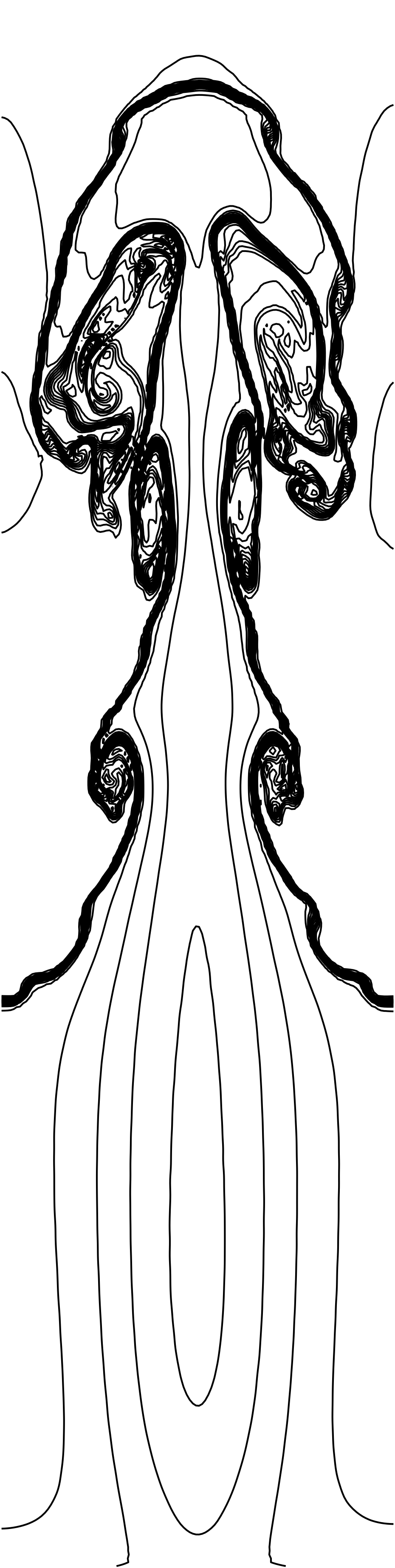}
\label{fig:teno_rt}}
\subfigure[MP6-CC]{%
\includegraphics[width=0.15\textwidth]{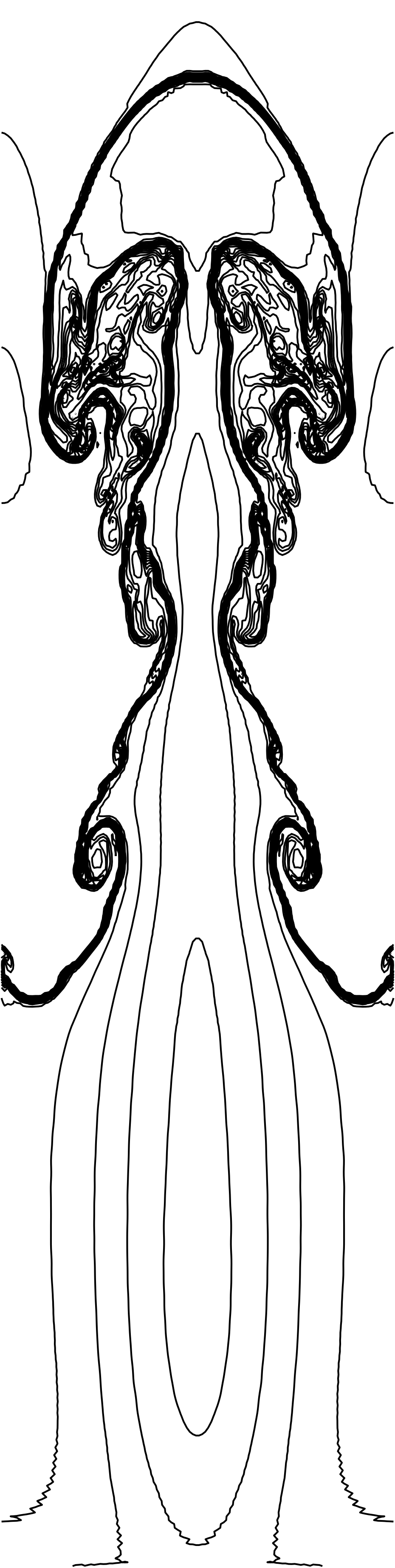}
\label{fig:mp6_rt}}
\subfigure[MEG8-C]{%
\includegraphics[width=0.15\textwidth]{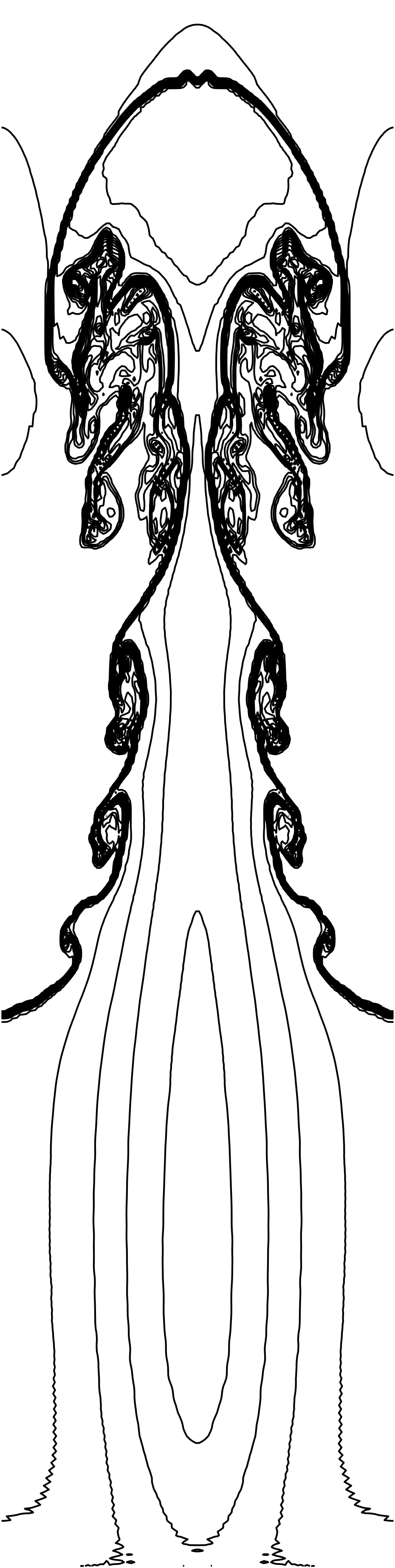}
\label{fig:meg8c_rt}}
\subfigure[MEG8-CC]{%
\includegraphics[width=0.15\textwidth]{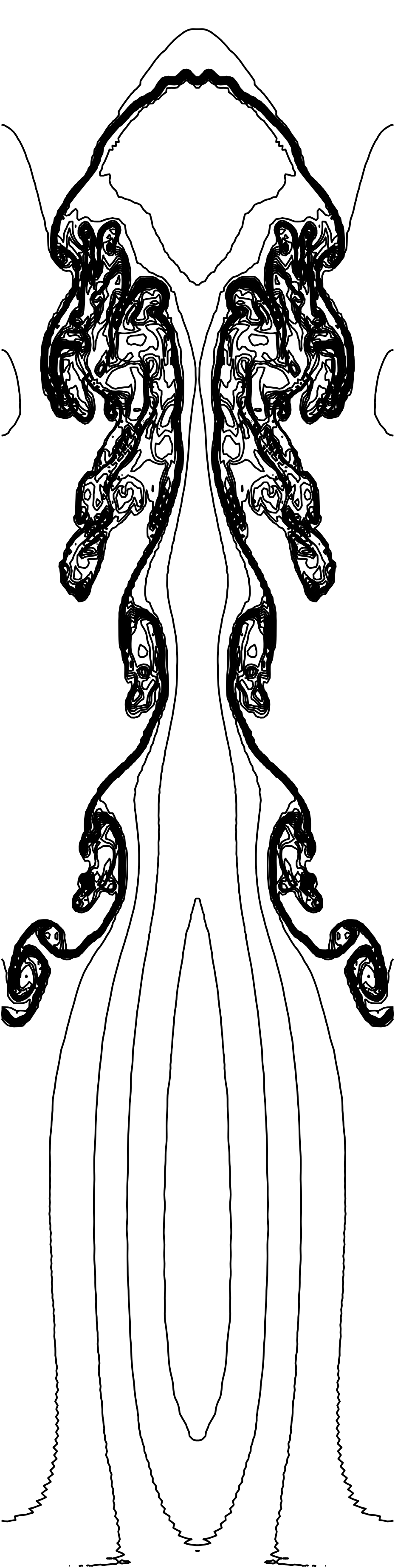}
\label{fig:meg8cc_rt}}
\caption{Comparison of density contours obtained by considered schemes on a grid size of 128 $\times$ 512 for the test case in Example \ref{ex:rt}.}
\label{fig:2d-RT}
\end{figure}

In this test case, simulations were conducted using a uniform mesh with a $128 \times 512$ resolution until $t=1.95$. The computational domain for the simulations was $[0, 1/4]\times [0,1]$, and the adiabatic constant $\gamma$ was set to $\frac{5}{3}$. The flow conditions were specified as $\rho=1$, $p=2.5$, and $u=v=0$ on the top boundary, and $\rho=2$, $p=1.0$, and $u=v=0$ on the bottom boundary. Source terms $S=(0,0,\rho, \rho v)$ were added to the Euler equations. The density distribution of the Rayleigh-Taylor instability problem is shown in Figs. \ref{fig:2d-RT}. The proposed approaches, MP6-CC and MEG8-CC, are free of oscillations as the MP criterion with density as the variable could detect contact discontinuity and prevent oscillations. The MP6-CC is again significantly faster, as shown in Table \ref{tab:rt_cost}, than the TENO5 scheme with the thinner material interface and more vortical structures. MEG8-CC is 35 $\%$ faster than the MEG8-C approach as it avoids the costly characteristic variable transformation and produces slightly more vortical structures.
\begin{table}[H]
    \centering
    \caption{Comparison of computational costs and efficiency for the evaluated schemes for Example \ref{ex:rt}.}
    \begin{tabular}{c c c c}
        \hline
        \hline
        MEG8-CC & MEG8-C & MP6-CC & TENO5 \\
        \hline
        404 s (0.825) & 670 s (1.36) & 373 s (0.761) & 490 s (1.0) \\
        \hline
        \hline
    \end{tabular}
    \label{tab:rt_cost}
\end{table}

\begin{example}\label{ex:dmr}{Double Mach Reflection}
\end{example}

In this test example, we will examine the proposed algorithm for the Double Mach Reflection (DMR) case originally presented by Woodward and Collela \cite{woodward1984numerical}. This scenario involves a Mach $10$ unsteady planar shock wave interacting with a 30-degree inclined surface, forming intricate flow patterns. The computational domain for this particular case spans $[x,y] = [0, 3] \times [0, 1]$, and our simulations will be conducted on a uniform grid of $768 \times 256$. The specified initial conditions are as follows:
\begin{equation}
(\rho, u, v, p)=\left\{\begin{array}{cc}
(1.4,0,0,1), & \text { if } y<1.732(x-0.1667) \\
(8,7.145,-4.125,116.8333), & \text { otherwise }
\end{array}\right.
\end{equation}

\begin{figure}[H]
%\begin{halfspacing}
\centering\offinterlineskip
\subfigure[TENO]{\includegraphics[width=0.42\textwidth]{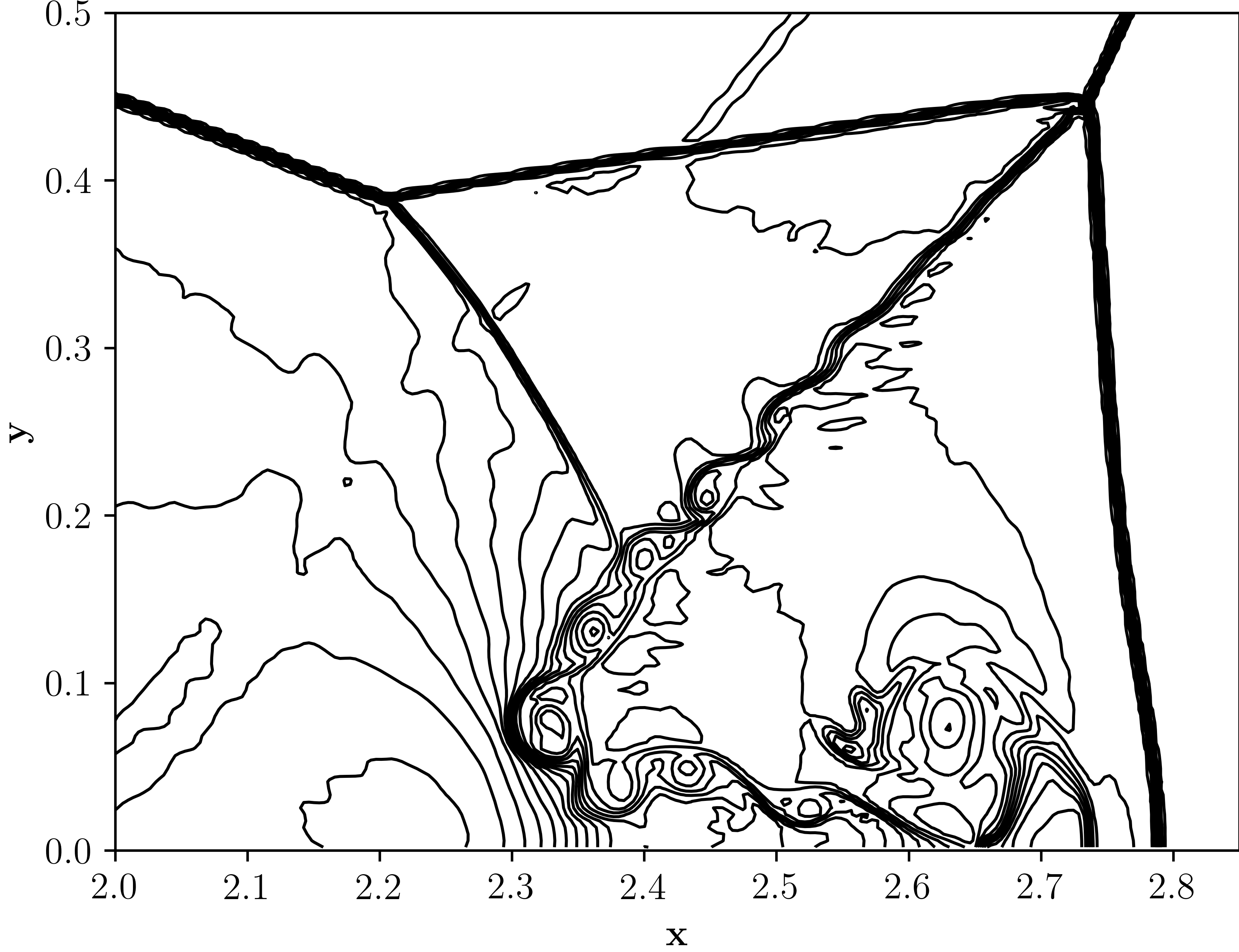}
\label{fig:teno_dmr}}
\subfigure[MP6-CC]{\includegraphics[width=0.42\textwidth]{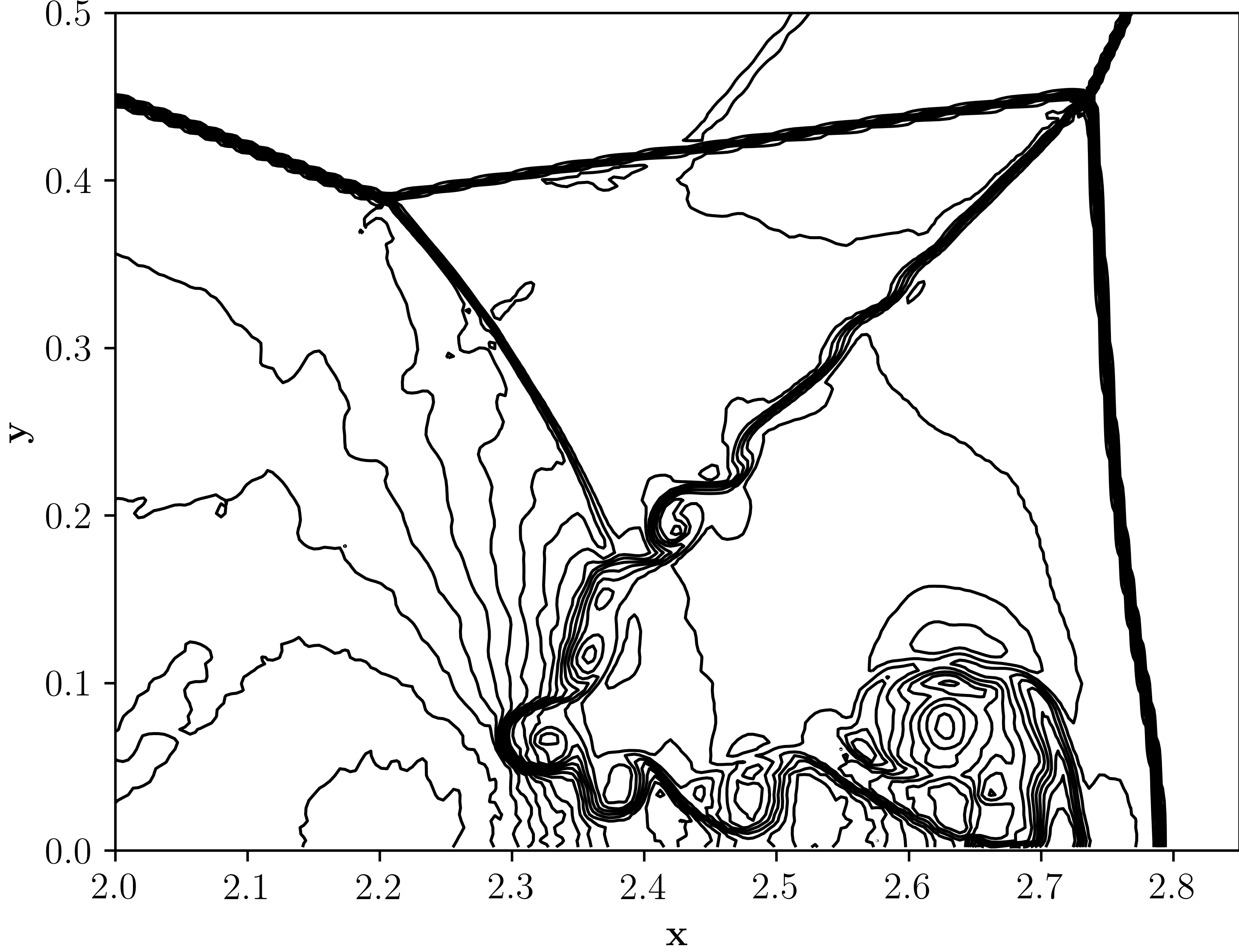}
\label{fig:migd_dmr}}
\subfigure[MEG8-C]{\includegraphics[width=0.42\textwidth]{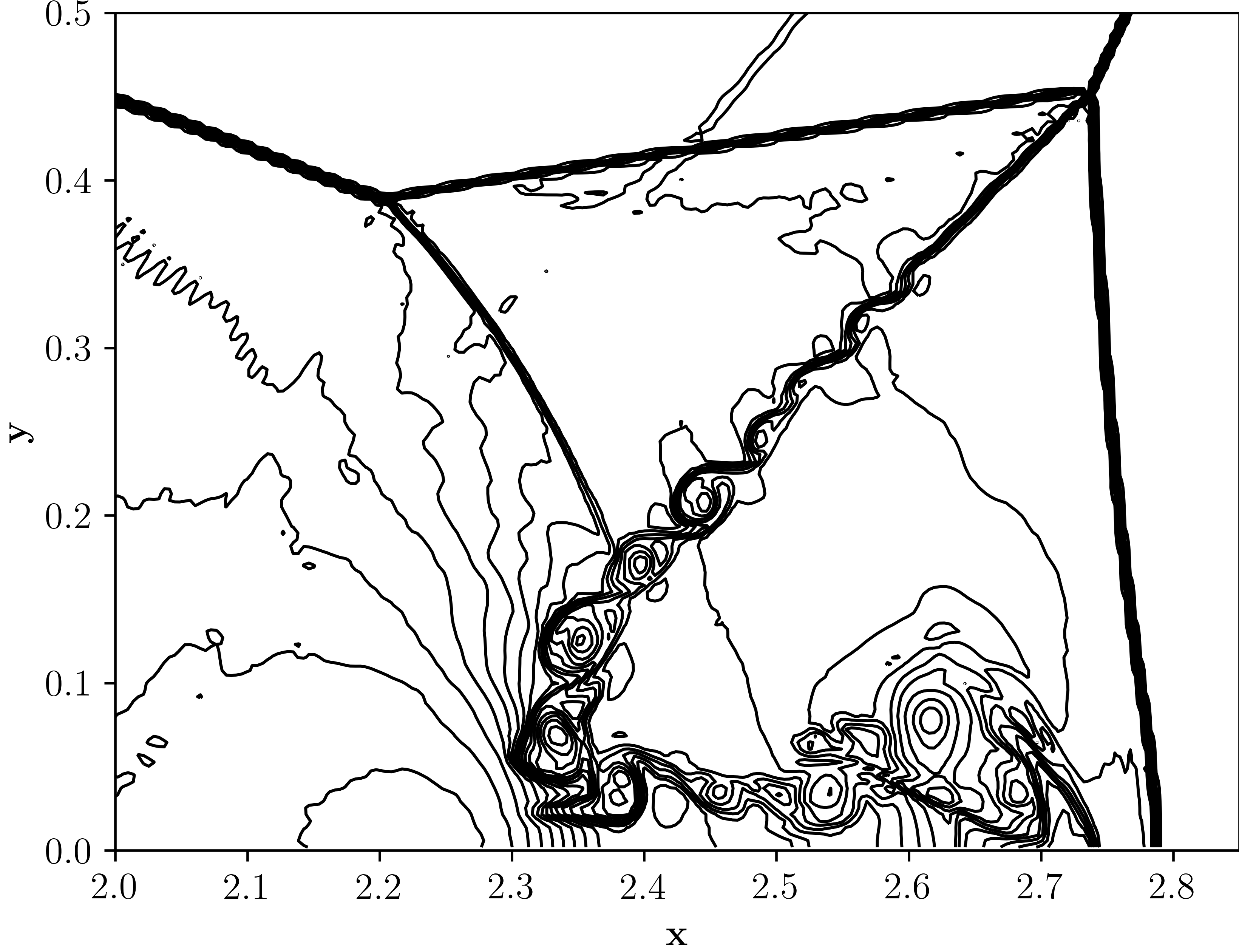}
\label{fig:migb_dmr}}
\subfigure[MEG8-CC]{\includegraphics[width=0.42\textwidth]{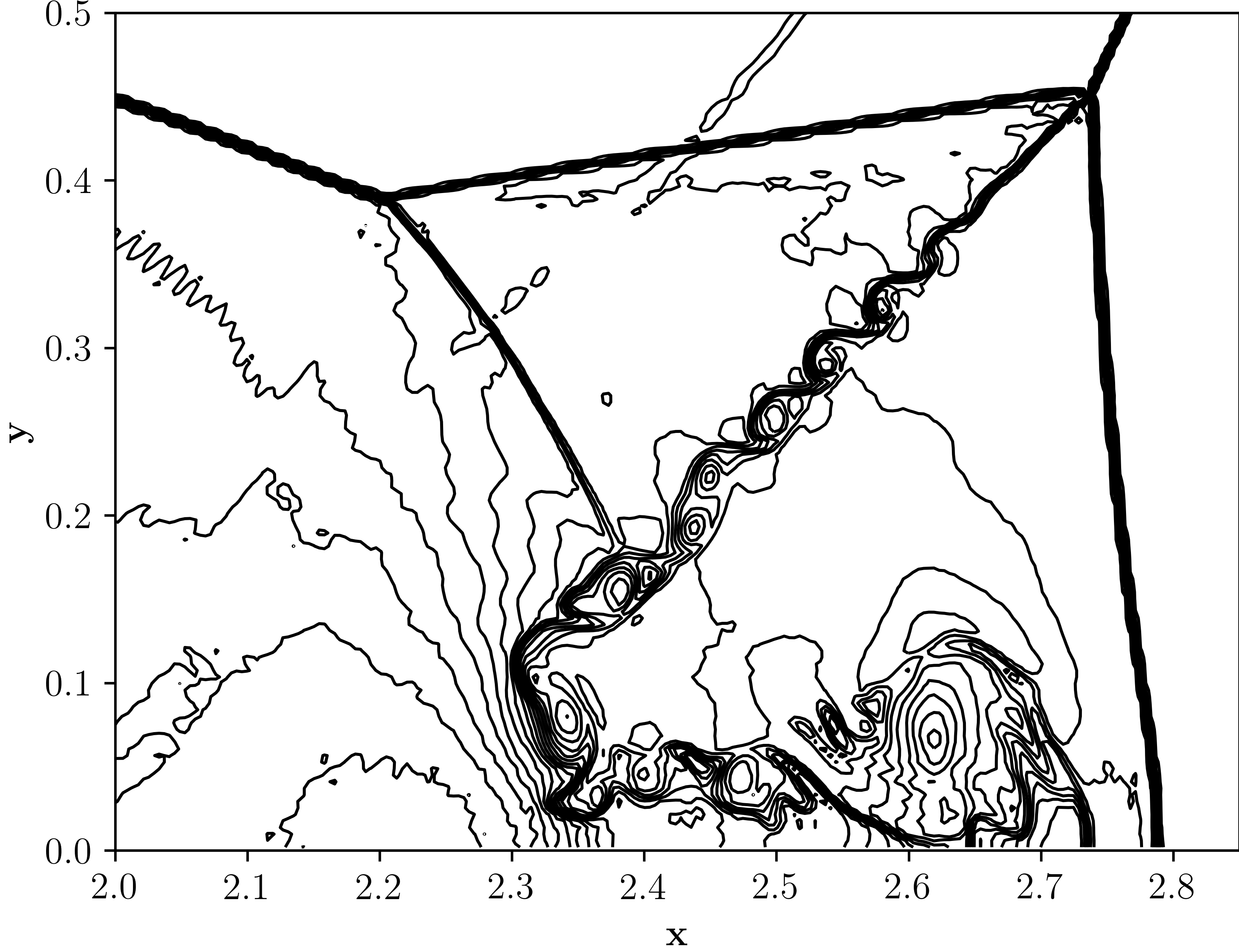}
\label{fig:migs_dmr}}
\caption{Computed density contours of the zoomed in Mach stem region of the Double Mach Reflection for the considered schemes. The figures are drawn with 38 contours.}
\label{fig_doublemach}
%\end{halfspacing}
\end{figure}

For this test case, we utilized the shock-stable cLLF Riemann solver from the reference \cite{fleischmann2019numerical} to avoid the ``carbuncle'' phenomenon. The simulations were conducted up to a final time of $t=0.3$. Reflecting wall conditions were applied to the bottom boundary for $0.1667 < x \leq 4.0$, and post-shock conditions for $0.0 \leq x \leq 0.1667$. The top boundary was set using the exact solution of the time-dependent oblique shock. Density contours from all the considered schemes are depicted in Figs. \ref{fig_doublemach}. Among the results, the TENO5 and MP6-CC schemes exhibited the highest dissipation and failed to resolve small-scale vortical structures. Due to their superior dispersion properties, the MEG schemes captured the shocks sharply compared to the TENO and MP6-CC schemes. Furthermore, within the MEG family of schemes, the MEG8-CC outperformed the MEG8-C scheme in capturing vortical structures and near-wall jets. Similar to the previous Example \ref{ex:rt}, it can be concluded that the MP detection criterion with the density as the variable and the Ducros sensor effectively identify discontinuities, aiding in the reconstruction of conservative variables. 
Once again, from Table \ref{tab:dmr_cost}, the computational cost of the MEG8-CC scheme is cheaper than the MEG8-C and TENO5 schemes. MP6-CC is the most efficient of all the schemes, and MEG8-CC is a close second.

\begin{table}[H]
    \centering
    \caption{Comparison of computational costs and efficiency for the evaluated schemes for Example \ref{ex:dmr}.}
    \begin{tabular}{c c c c}
        \hline
        \hline
        MEG8-CC & MEG8-C & MP6-CC & TENO5 \\
        \hline
        778 s (0.914) & 989 s (1.16) & 762 s (0.895) & 851 s (1.0) \\
        \hline
        \hline
    \end{tabular}
    \label{tab:dmr_cost}
\end{table}

\begin{example}\label{ex:rp}{Riemann Problem}
\end{example}

This example considers the two-dimensional Riemann problem of configuration 3 \cite{schulz1993numerical} for testing the proposed schemes. The initial conditions for the configuration considered are as follows:

\begin{equation}\label{ex:rp1}
(\rho, u, v, p)=\left\{\begin{array}{cc}
(1.5,0,0,1.5), & \text { if } x>0.5, y>0.5 \\
(0.5323,1.206,0,0.3), & \text { if } x<0.5, y>0.5 \\
(0.138,1.206,1.206,0.029), & \text { if } x<0.5, y<0.5 \\
(0.5323,0,1.206,0.3), & \text { if } x>0.5, y<0.5
\end{array}\right.
\end{equation}

The initial conditions described above result in four shocks at the interfaces of the four quadrants, and the Kelvin-Helmholtz instability along the slip lines gives rise to small-scale structures, often used as a benchmark for evaluating the numerical dissipation of a particular scheme. In this test case, simulations are conducted within a computational domain of $[x,y] = [0,1]\times [0,1]$, using a uniform grid of $1024 \times 1024$, until a final time, $t=0.3$.

\begin{figure}[H]
\begin{onehalfspacing}
\centering\offinterlineskip
\subfigure[TENO5]{\includegraphics[width=0.45\textwidth]{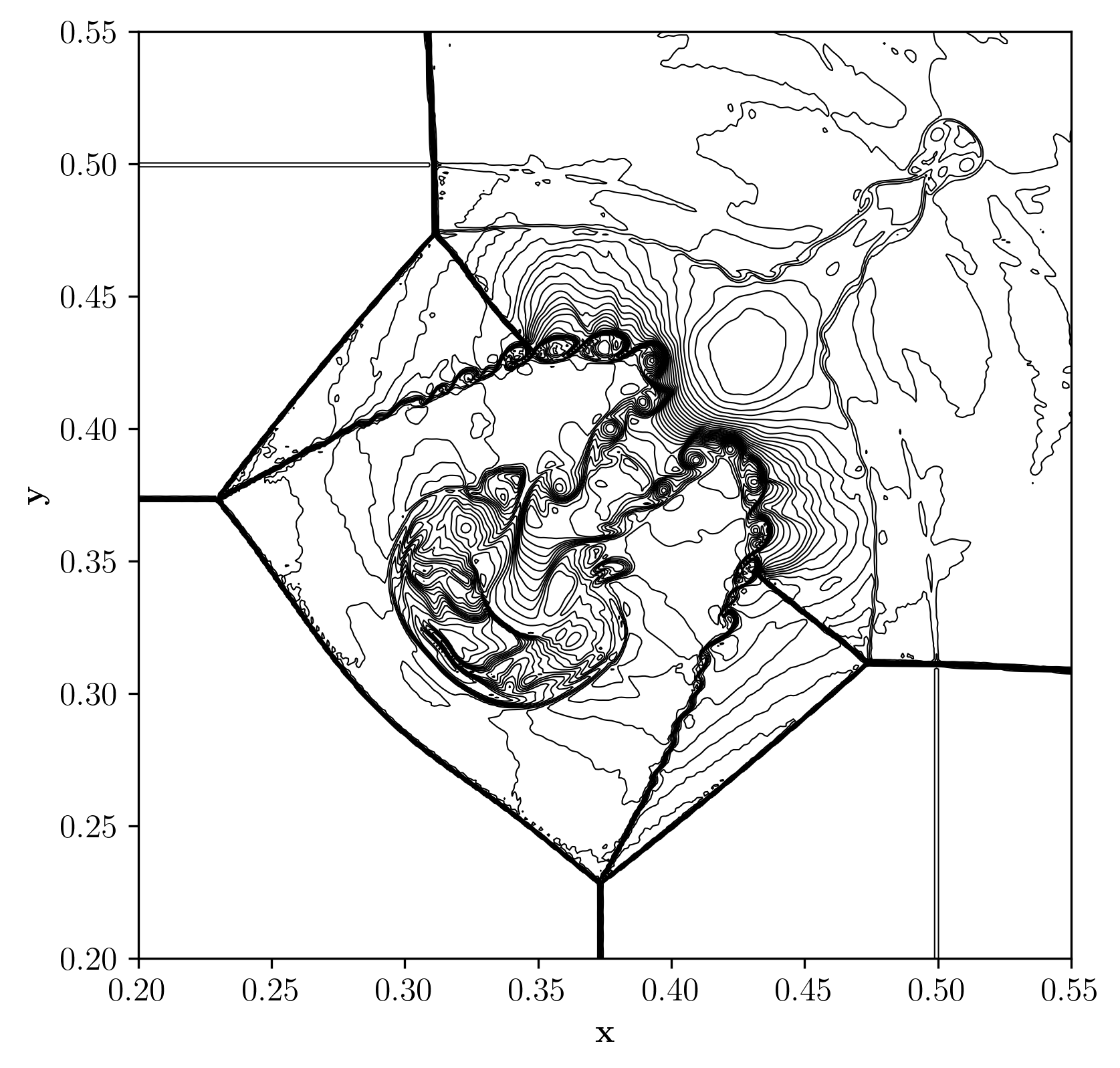}
\label{fig:TENO_r12}}
\subfigure[MP6-CC]{\includegraphics[width=0.45\textwidth]{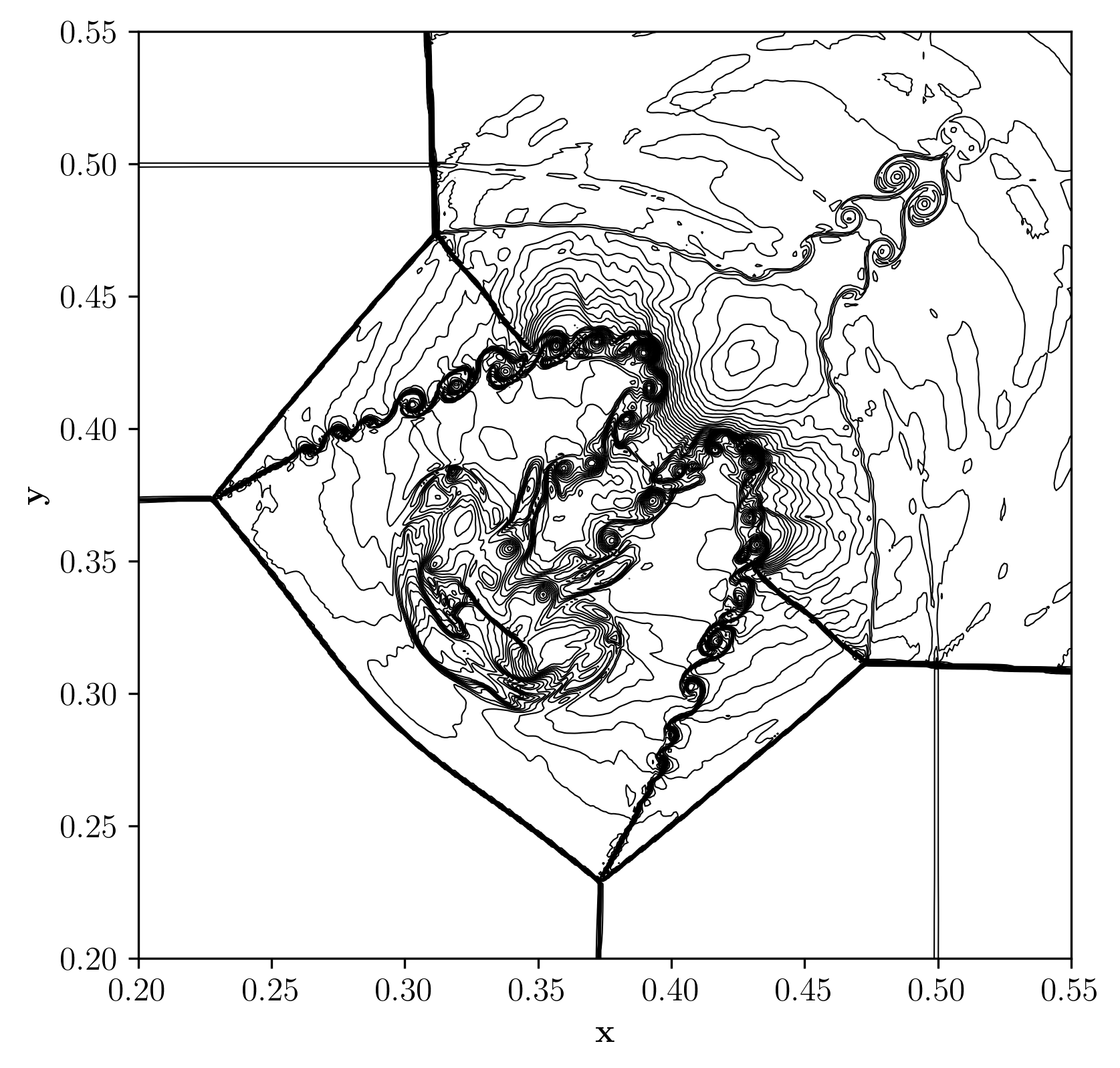}
\label{fig:mp6_r12}}
\subfigure[MEG8-C]{\includegraphics[width=0.45\textwidth]{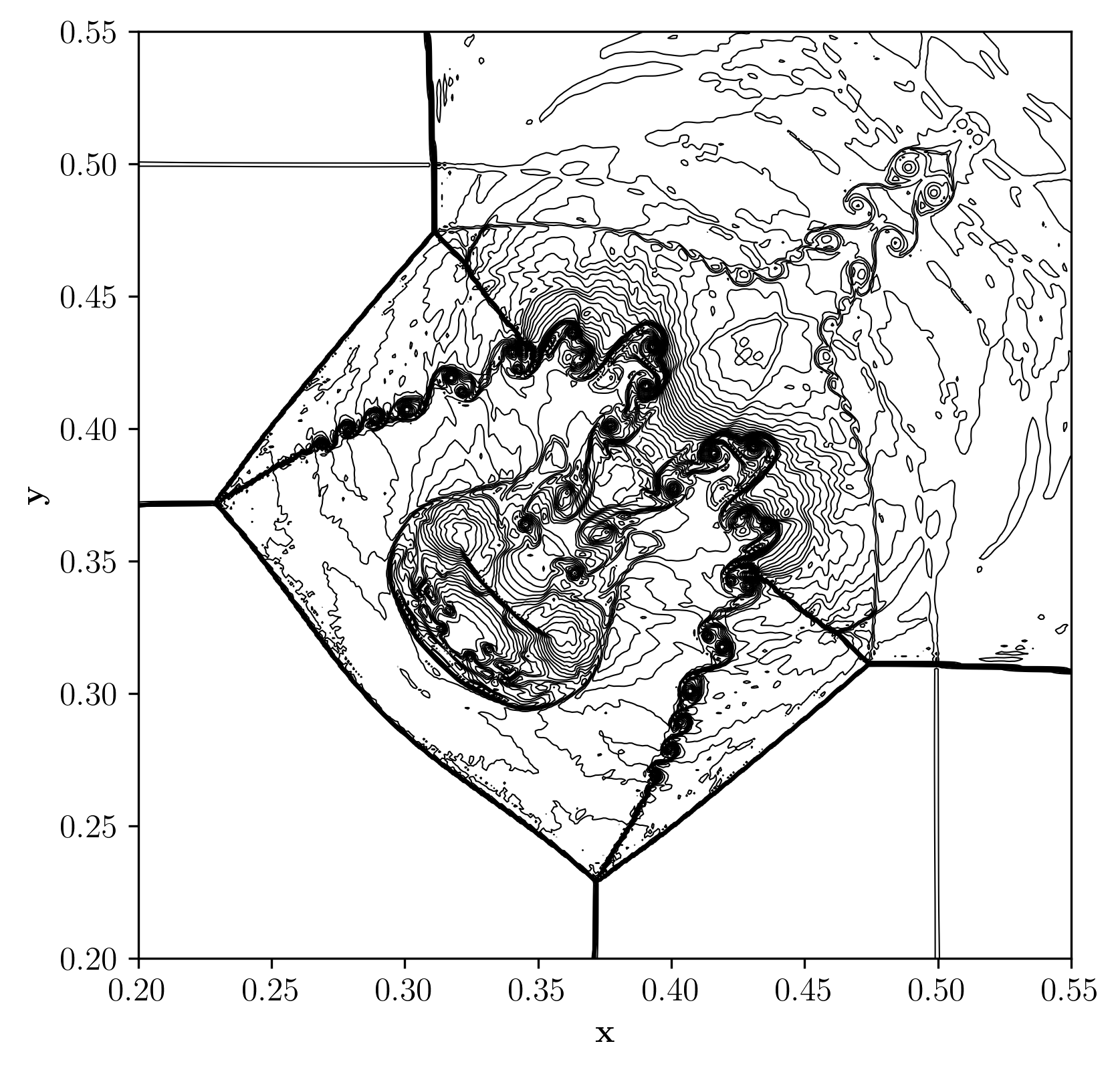}
\label{fig:Meg8c_r12}}
\subfigure[MEG8-CC]{\includegraphics[width=0.45\textwidth]{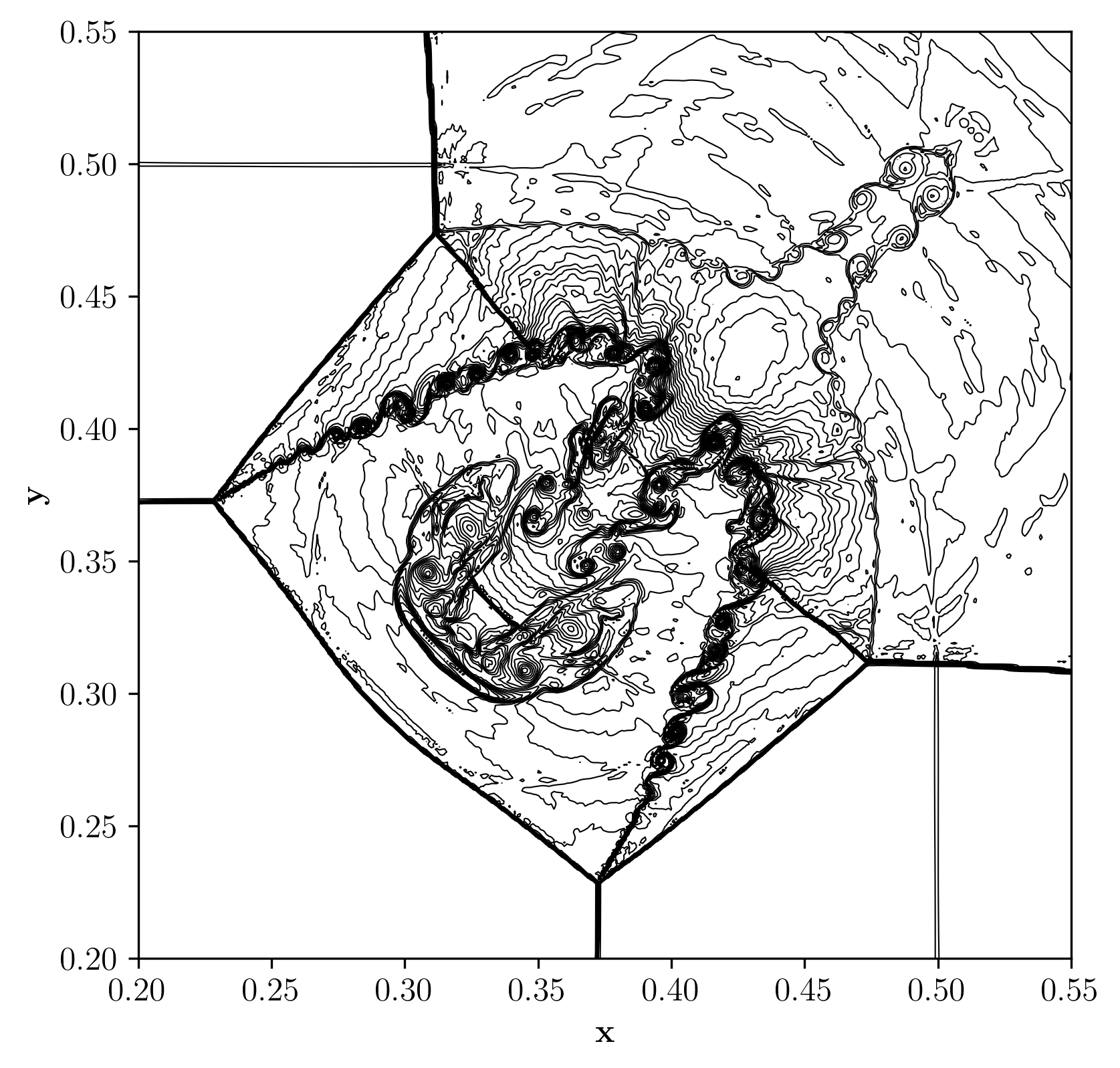}
\label{fig:Meg8cc_r12}}
\caption{Computed density contours of the Riemann problem described in Example \ref{ex:rp} for the considered schemes. The figures are drawn with 42 contours.}
\label{fig_riemann}
\end{onehalfspacing}
\end{figure}

The density contours obtained by the considered schemes are presented in Figs. \ref{fig_riemann}. The small-scale structures produced by the TENO scheme, Fig. \ref{fig:TENO_r12}, are dissipative compared to that of MP6-CC, Fig. \ref{fig:mp6_r12}, and they both share the same stencil for reconstruction. These results indicate that a properly designed scheme with limiters can also outperform the ENO-type schemes (considered standard and most popular schemes in the literature for compressible flow simulations). Furthermore, the small-scale vortices obtained by the MEG8-CC scheme are significantly better than the MEG8-C and the other schemes. 

It is important to note that this test case does not involve any contact discontinuities, and there is little difference between the MEG8-C and MEG8-CC schemes regarding computational cost, shown in Table \ref{tab:rp_cost}. MP6-CC is the cheapest of all the schemes, about 36$\%$ faster than the TENO5.

\begin{table}[H]
    \centering
    \caption{Comparison of computational costs and efficiency for the evaluated schemes for Example \ref{ex:rp}.}
    \begin{tabular}{c c c c}
        \hline
        \hline
        MEG8-CC & MEG8-C & MP6-CC & TENO5 \\
        \hline
        4286 s (1.07) & 4552 s (1.14) & 2535 s (0.636) & 3984 s (1.0) \\
        \hline
        \hline
    \end{tabular}
    \label{tab:rp_cost}
\end{table}

\begin{example}\label{ex:SB}{Shock-Bubble interaction}
\end{example}
In the sixth example, we focus on the interaction between a shock wave and a bubble at Mach 6, as described in Hu (2010). The shock wave and the helium bubble are assumed to behave as ideal gases. The helium bubble is positioned at coordinates $x = 0.25$, $y = 0$ within a domain spanning [0, 1] $\times$ [-0.5, 0.5]. The initial conditions for the problem are as follows:

 \begin{equation}
\begin{aligned}
(\rho,u,v,p)=
\begin{cases}
&(1.0,\ -3,\ 0,\ 1),~~~~~~~~~~~~~~~~~~~~~~~~~~~~~~~~ \rm{pre-shocked \ air},\\
&(216/41,\ (1645/286)-3,\ 0,\ 251/6) ~~~~~~~ \rm{post-shocked \ air},\\
&(0.138,\ -3,\ 0,\ 1)~~~~~~~~~~~~~~~~~~~~~~~~~~~~~~~ \rm{helium \ bubble}.
\end{cases}
\end{aligned}
\label{sb}
\end{equation}

\begin{figure}[H]
\centering\offinterlineskip
\subfigure[TENO5]{\includegraphics[width=0.48\textwidth]{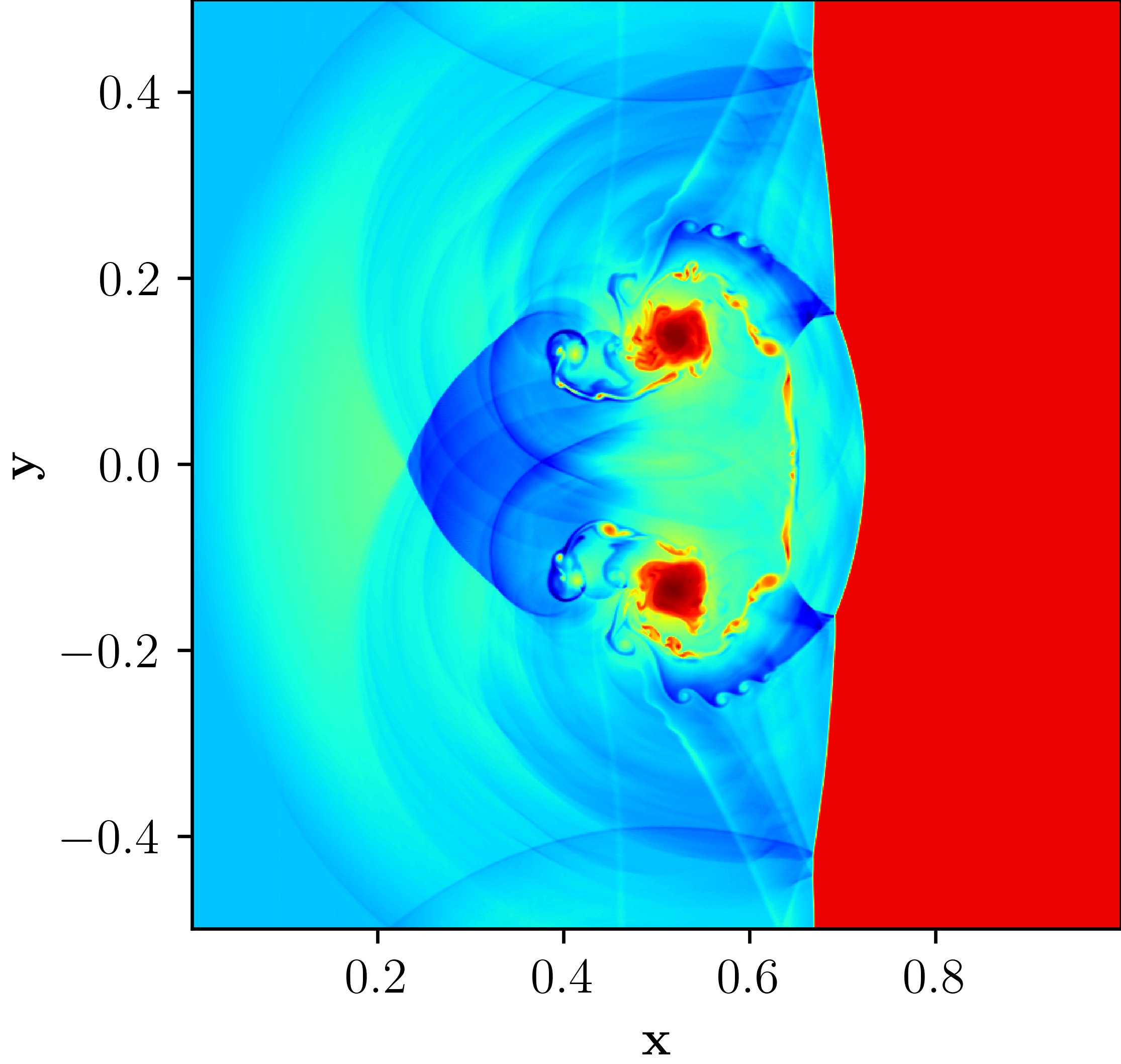}
\label{fig:teno_SB-LLFM}}
\subfigure[MP6-CC]{\includegraphics[width=0.48\textwidth]{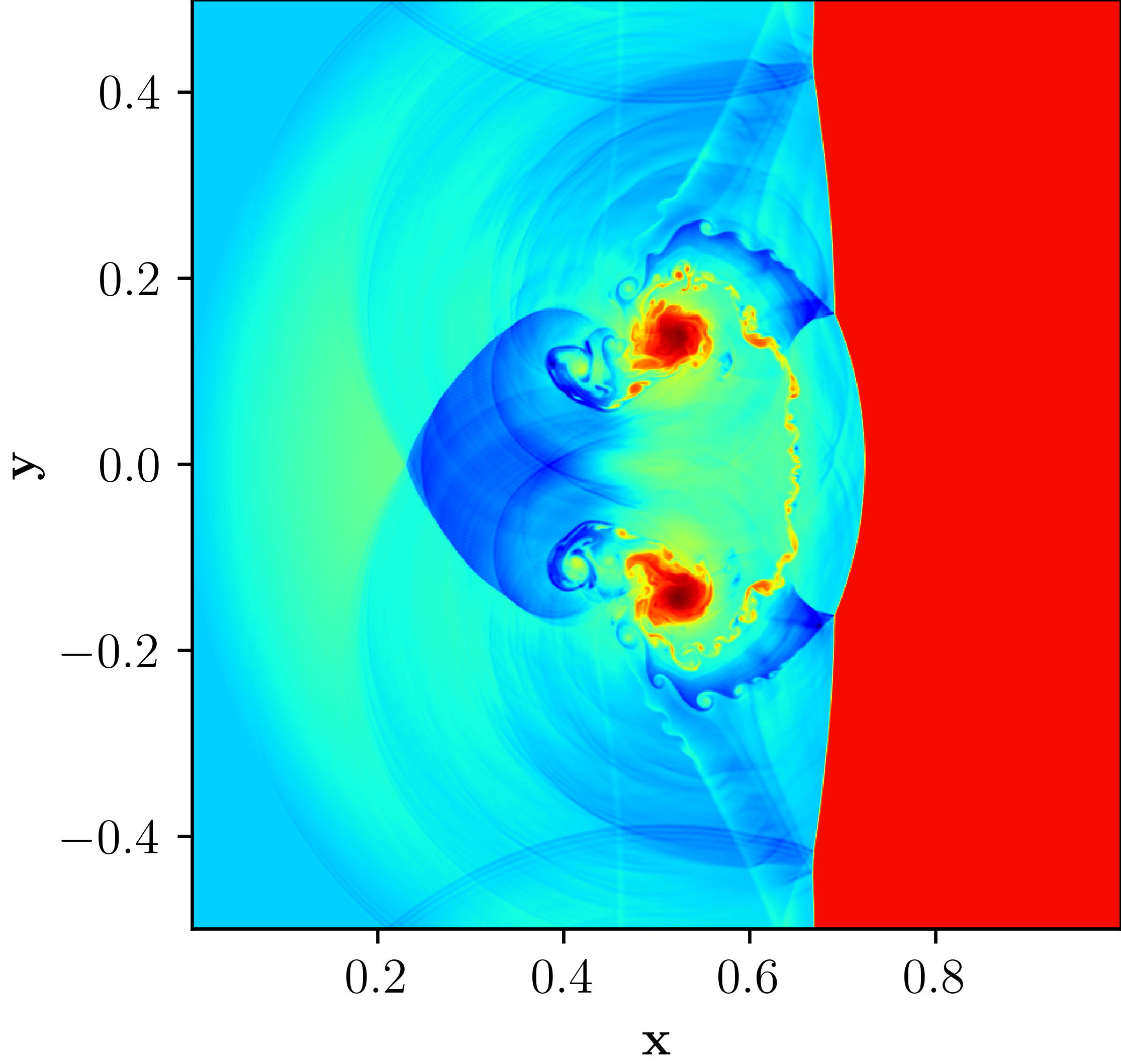}
\label{fig:mp6_SB-LLFM}}
\subfigure[MEG8-C]{\includegraphics[width=0.48\textwidth]{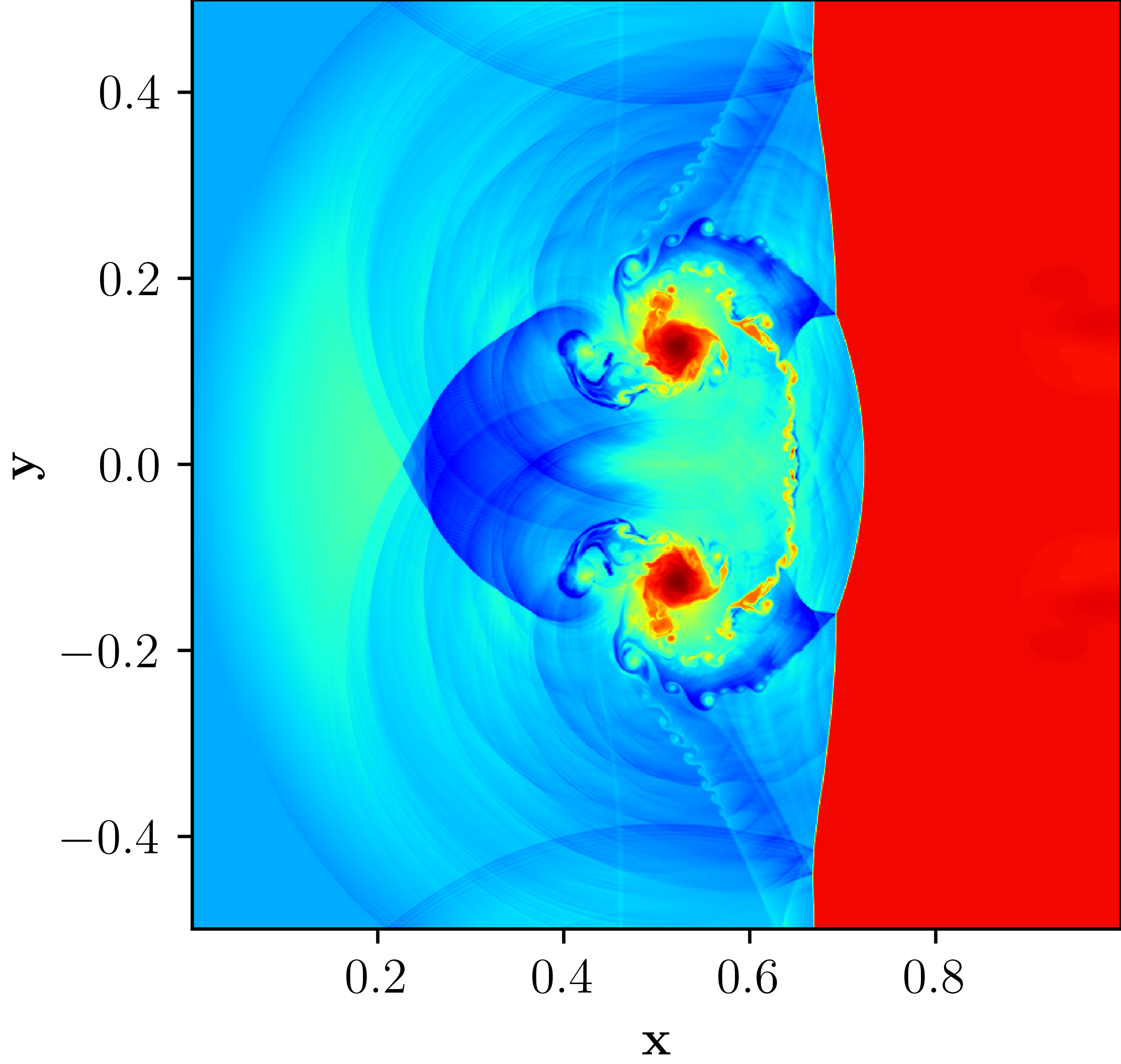}
\label{fig:Meg8c_base-GLF}}
\subfigure[MEG8-CC]{\includegraphics[width=0.48\textwidth]{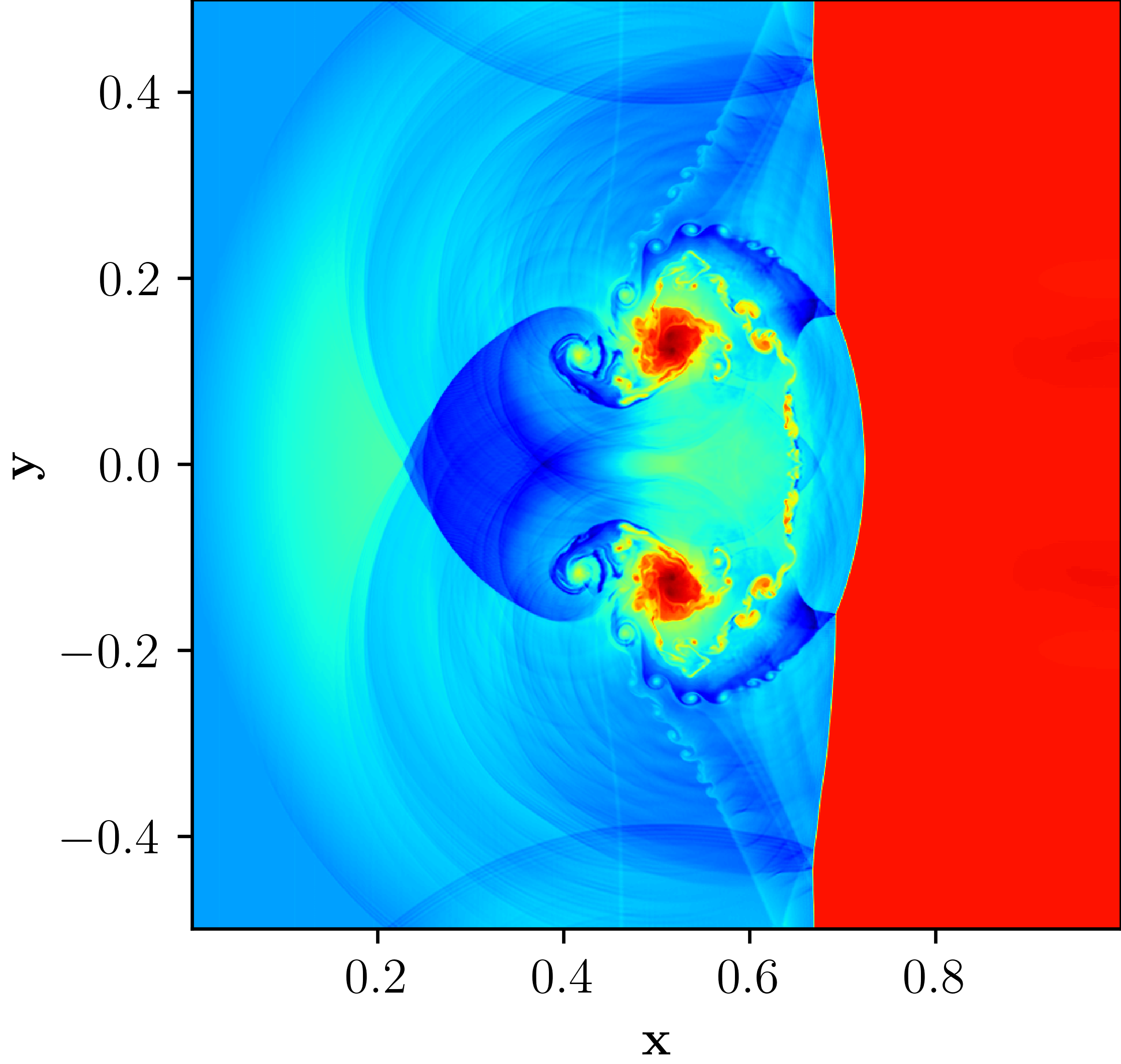}
\label{fig:mcc-Z_SB-LF}}
\caption{Shock-Bubble interaction for example \ref{ex:SB} using various schemes on a grid size of 800 $\times$ 800.}
\label{fig_allbubble}
\end{figure}

The initial radius of the bubble is 0.15, while the initial position of the shock front is at x = 0.05. Inflow and outflow conditions are enforced at the left and right boundaries. Neumann conditions with zero gradients for all quantities are applied at the remaining boundaries. The simulation resolution is set to 800 $\times$ 800. Simulations are also carried out with shock-stable cLLF Riemann solver \cite{fleischmann2019numerical} to prevent the ``carbuncle'' phenomenon. Once again, from Fig. \ref{fig_allbubble}, one can observe that the MEG8-CC scheme resolves more small-scale vortices than TENO5 and MP6-CC schemes. Results obtained by the MEG8-C and MEG8-CC are very similar, but the MEG8-CC scheme is 12$\%$ faster than the MEG8-C scheme.

\begin{table}[H]
    \centering
    \caption{Comparison of computational costs and efficiency for the evaluated schemes for Example \ref{ex:SB}.}
    \begin{tabular}{c c c c}
        \hline
        \hline
        MEG8-CC & MEG8-C & MP6-CC & TENO5 \\
        \hline
        5573 s (0.92) & 6339 s (1.04) & 4996 s (0.825) & 6061 s (1.0) \\
        \hline
        \hline
    \end{tabular}
    \label{tab:sb_cost}
\end{table}

\begin{example}\label{ex:TGV}{Inviscid Taylor-Green Vortex}
\end{example}

The Taylor-Green vortex problem is a common benchmark for evaluating the performance of the adaptive central-upwind schemes. It is a three-dimensional inviscid test case with the following initial conditions for the simulation:

\begin{equation}\label{itgv}
\begin{pmatrix}
\rho \\
u \\
v \\
w \\
p \\
\end{pmatrix}
=
\begin{pmatrix}
1 \\
\sin{x} \cos{y} \cos{z} \\
-\cos{x} \sin{y} \cos{z} \\
0 \\
100 + \frac{\left( \cos{(2z)} + 2 \right) \left( \cos{(2x)} + \cos{(2y)} \right) - 2}{16}
\end{pmatrix}.
\end{equation}

 The simulations are conducted until time $t=10$ on a grid size of $64^3$, with a specific heat ratio of $\gamma=5/3$ on a periodic domain of size $x,y,z \in [0,2\pi)$. The flow is incompressible since the mean pressure is significantly considerable. The study aims to evaluate the ability of the proposed schemes to preserve kinetic energy and the growth of enstrophy over time. Enstrophy is the integral of the square of the vorticity and is used as a measure of the scheme's ability to preserve vortical structures. The fourth-order compact scheme, Equation (\ref{eqn:firstDerivative}), computes the velocity derivatives required for the enstrophy computation for all the schemes, and the HLLC Riemann solver is used for this test case.

\begin{figure}[H]
\centering
\subfigure[Kinetic energy- MP6 and TENO]{\includegraphics[width=0.45\textwidth]{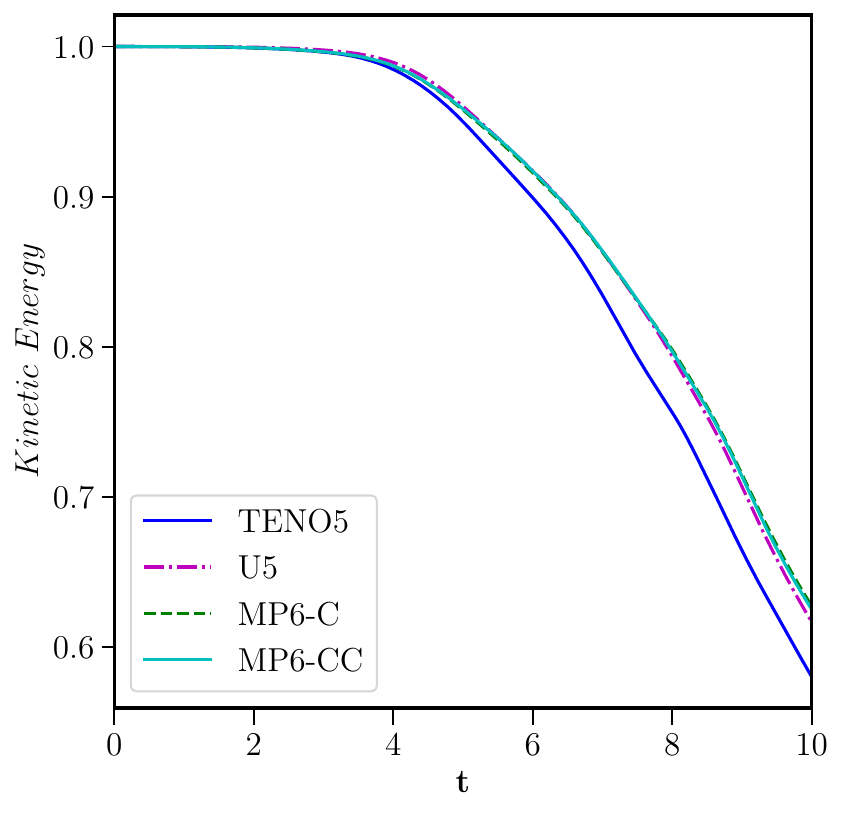}
\label{fig:TGV_KE_mp6}}
\subfigure[Kinetic energy- MEG8]{\includegraphics[width=0.46\textwidth]{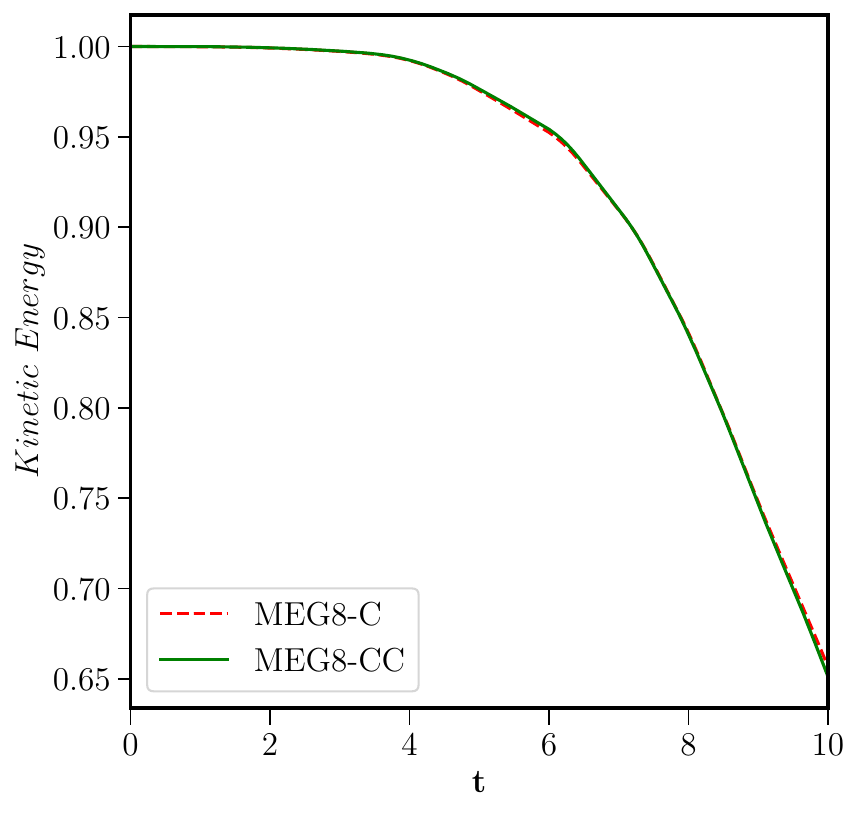}
\label{fig:TGV_ke_meg8}}
\caption{Normalised kinetic energy for different schemes presented in Example \ref{ex:TGV} on grid size of $64^3$. Solid blue line: TENO5; Dashed magenta line: U5; Solid cyan: MP6-CC; Dashed green line: MP6-C; Solid green line: MEG8-CC and dashed red line: MEG8-C.}
\label{fig_TGV_ke}
\end{figure}

\begin{figure}[H]
\centering
\subfigure[Enstrophy- MP6 and TENO]{\includegraphics[width=0.45\textwidth]{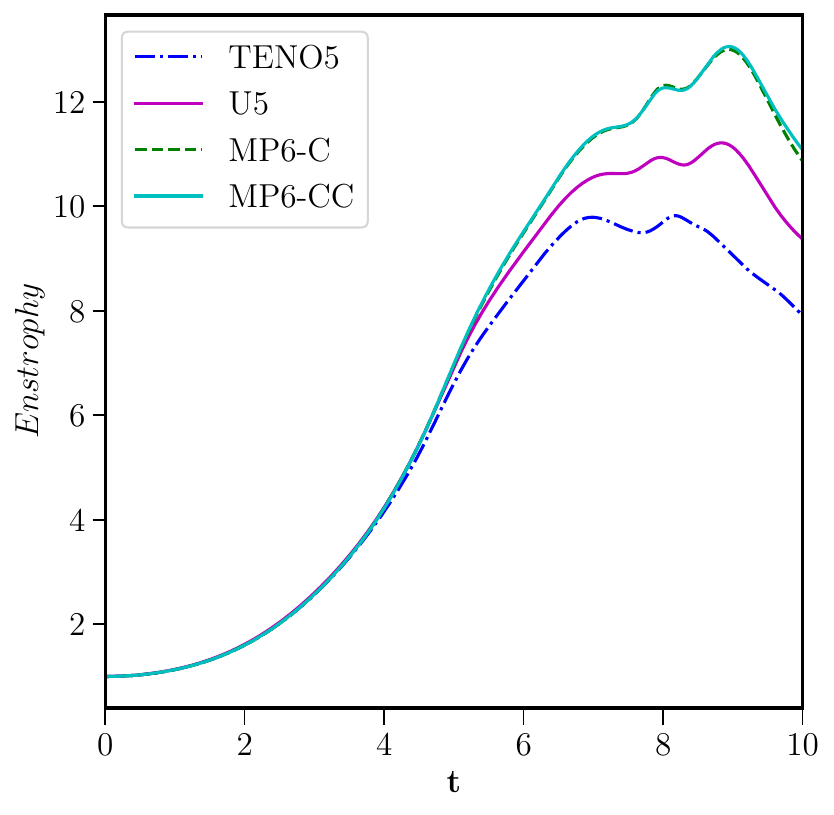}
\label{fig:TGV_ens_mp6}}
\subfigure[Enstrophy- MEG8]{\includegraphics[width=0.45\textwidth]{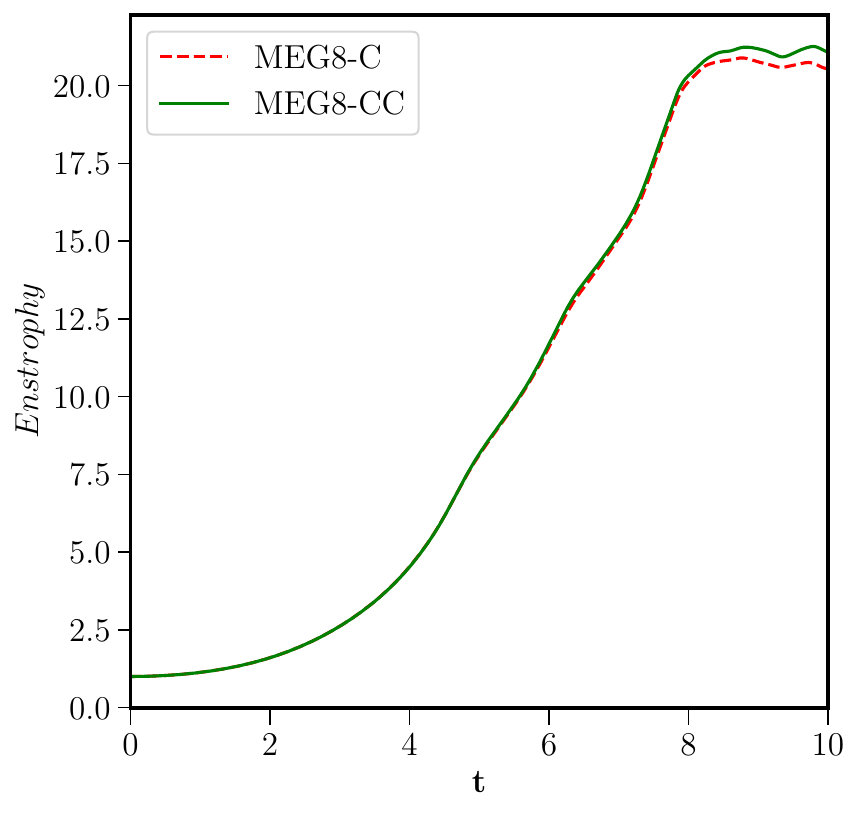}
\label{fig:TGV_ens_meg8}}
\caption{Normalised Enstrophy for different schemes presented in Example \ref{ex:TGV} on grid size of $64^3$. Solid blue line: TENO5; Dashed magenta line: U5; Solid cyan: MP6-CC; Dashed green line: MP6-C; Solid green line: MEG8-CC and dashed red line: MEG8-C.}
\label{fig_TGV_ens}
\end{figure}

The kinetic energy evolution of the numerical schemes MP6-CC, MP6-C, TENO5 and the linear fifth-order upwind scheme (U5) is shown in the Fig. \ref{fig:TGV_KE_mp6}, indicating that the MP6-CC scheme better preserves kinetic energy than the TENO5 scheme and even the linear U5 scheme. Furthermore, the MEG8-C and MEG8-CC schemes preserve the kinetic energy significantly better than all the other schemes, as shown in Fig. \ref{fig:TGV_ke_meg8}. There is little to no difference between the MEG8-C and MEG8-CC schemes, as this test case has no contact discontinuity or shockwaves. Figures \ref{fig:TGV_ens_mp6} and \ref{fig:TGV_ens_meg8} depict the computed enstrophy plots of the MP6-CC, MP6-C, TENO5, U5, MEG8-C and MEG8-CC schemes. The schemes with the proposed algorithm MP6-CC and MEG8-CC perform better than TENO and U5 schemes. There is a small difference between the MEG8-C and MEG-CC schemes for enstrophy values at late times, with MEG8-CC marginally better than MEG8-C. From Table \ref{tab:tgv_cost}, one can infer that the MP6-CC scheme is significantly cheaper than the TENO5 scheme and better preserves kinetic energy. While the computational costs of the TENO5 and MEG8-CC are almost the same, the MEG8-CC scheme outperforms the TENO5 scheme significantly in terms of both kinetic energy preservation and enstrophy values.

\begin{table}[H]
    \centering
    \caption{Comparison of computational costs and efficiency for the evaluated schemes for Example \ref{ex:TGV}.}
    \begin{tabular}{c c c c}
        \hline
        \hline
        MEG8-CC & MEG8-C & MP6-CC & TENO5 \\
        \hline
        4185 s (1.07) & 5271 s (1.14) & 3381 s (0.636) & 4217 s (1.0) \\
        \hline
        \hline
    \end{tabular}
    \label{tab:tgv_cost}
\end{table}

\section{Conclusions}\label{sec:conclusions}
This work proposed a general adaptive central-upwind scheme for the simulations of compressible flows with discontinuities in the flow field. The approach significantly improves over the centralized gradient-based reconstruction proposed by Hoffmann, Chamarthi and Frankel \cite{hoffmann2024centralized} with reduced computational cost where the reconstruction is carried out for characteristic variables. In the current approach, conservative variables are reconstructed if no discontinuities are detected by the Durcos shock sensor and the MP criterion-based contact discontinuity sensor; otherwise, the reconstruction is the same as that of \cite{hoffmann2024centralized}. While it is well known in the literature that reconstruction of the characteristic variables will result in \textit{cleaner and oscillation-free results}, the conversion of the conservative variables to characteristic variables will incur significant computational expense. The current approach overcomes this by reconstructing conservative variables in the regions with no discontinuities and characteristic variables in the areas of discontinuities, significantly reducing the computational cost and improving the results in some instances. The generalized approach also works with a standard fifth/sixth-order reconstruction scheme, MP6-CC. It has been shown that the MP6-CC scheme is 20-30$\%$ cheaper than the TENO5 scheme (which has the same stencil) and yet produces significantly superior results (judging by the vortical structures produced, which is the norm in the literature for assessing a low-dissipation scheme). The computational expense of the MEG8-CC scheme is the same as that of the TENO5 scheme, yet it produces superior results despite additional computations for the derivatives and Ducros sensor. 

The proposed approach did not produce spurious vortices for the periodic double shear layer test case with the proposed algorithm for all the schemes on coarse grids, unlike that of the TENO5 and the other existing schemes in the literature. It has been shown that the current algorithm (a combination of upwind and central schemes for appropriate variables in different directions )  can also prevent spurious vortices even with linear schemes (even the third/fourth scheme has prevented spurious vortices).

\section*{Inspiration and Acknowledgements:}

The work is inspired by the paper written by Van Leer \cite{van2006upwind}, where the author wrote - ``Presumably, the advent of high-performance computing and promise of massively parallel computing has quelled any drive toward a systematic modernization of CFD algorithms. I personally believe, though, that the next round of gains in CFD will not come from hardware improvement but, once again, from method development'', coupled with the lack of resources. A.S. thanks his wife and son for checking the code and the manuscript.

\bibliographystyle{elsarticle-num}
\bibliography{contact_ref}

\end{document}